\documentclass[preprint]{aastex631}
\usepackage{CJK}

\shorttitle{Molecules in 21\,$\mu$m sources}
\shortauthors{Qiu et al.}

\begin{document}
\begin{CJK*}{UTF8}{gkai}

\title{Gas-phase molecules in protoplanetary nebulae with the
21\,$\mu$m emission feature}

\correspondingauthor{Yong Zhang}
\email{zhangyong5@mail.sysu.edu.cn}

\author{Jian-Jie Qiu(邱建杰)}
\affiliation{School of Physics and Astronomy, Sun Yat-sen University, 2 Da Xue Road, Tangjia, Zhuhai 519000, Guangdong Province, PR China}
\affiliation{Key Laboratory of Modern Astronomy and Astrophysics (Nanjing University), Ministry of Education, Nanjing 210093, Jiangsu Province, PR China}
\affiliation{CSST Science Center for the Guangdong-Hongkong-Macau Greater Bay Area, Sun Yat-Sen University, Zhuhai, PR China}

\author{Yong Zhang(张泳)}
\affiliation{School of Physics and Astronomy, Sun Yat-sen University, 2 Da Xue Road, Tangjia, Zhuhai 519000, Guangdong Province, PR China}
\affiliation{CSST Science Center for the Guangdong-Hongkong-Macau Greater Bay Area, Sun Yat-Sen University, Zhuhai, PR China}
\affiliation{Xinjiang Astronomical Observatory, Chinese Academy of Sciences, Urumqi, 830011, Xinjiang Province, PR China}

\author{Jun-ichi Nakashima(中岛淳一)}
\affiliation{School of Physics and Astronomy, Sun Yat-sen University, 2 Da Xue Road, Tangjia, Zhuhai 519000, Guangdong Province, PR China}
\affiliation{CSST Science Center for the Guangdong-Hongkong-Macau Greater Bay Area, Sun Yat-Sen University, Zhuhai, PR China}

\author{Jiang-Shui Zhang(张江水)}
\affiliation{Center For Astrophysics, Guangzhou University, 230 Wai Huan Xi Road, Guangzhou Higher Education Mega Center, Guangzhou 510006, Guangdong Province, PR China}
                 
\author{Fei Li(李菲)}
\affiliation{Key Laboratory of Modern Astronomy and Astrophysics (Nanjing University), Ministry of Education, Nanjing 210093, Jiangsu Province, PR China}
\affiliation{School of Astronomy and Space Science, Nanjing University, 163 Xianlin Avenue, Nanjing 210023, Jiangsu Province, PR China}
                 
\author{Deng-Rong Lu(逯登荣)}
\affiliation{Purple Mountain Observatory, Chinese Academy of Sciences, Nanjing 210034, Jiangsu Province, PR China}
                 
\author{Xin-Di Tang(汤新弟)}
\affiliation{Xinjiang Astronomical Observatory, Chinese Academy of Sciences, Urumqi, 830011, Xinjiang Province, PR China}

\author{Xiao-Ling Yu(俞效龄)}
\affiliation{College of Physics and Electronic Engineering, Qujing Normal University, Qujing 655011, Yunnan Province, PR China}

\author{Lan-Wei Jia(贾兰伟)}
\affiliation{School of Physics and Materials Science, Guangzhou University, 230 Wai Huan Xi Road, Guangzhou Higher Education Mega Center, Guangzhou 510006, Guangdong Province, PR China}

\begin{abstract}

It has been more than 30 years since the enigmatic 21\,$\mu$m emission feature was first discovered in protoplanetary nebulae (PPNs). 
Although dozens of different dust carrier candidates have been proposed, there is as yet no widely accepted one. 
We present the results of molecular observations toward 21\,$\mu$m objects using the 10\,m Submillimeter Telescope of Arizona Radio Observatory at the 1.3\,mm band 
and the 13.7\,m telescope of Purple Mountain Observatory at the 3\,mm band, 
aiming to investigate whether the gas-phase environments of these unusual sources have some peculiarities compared to normal PPNs. 
We detect 31 emission lines belonging to seven different molecular species, most of which are the first detection in 21\,$\mu$m PPNs.
The observations provide clues on the identification of the 21\,$\mu$m feature. 
We report a correlation study between the fractional abundance of gas-phase molecules and the strengths of the 21\,$\mu$m emission.
Our study shows that given the small sample size, the 21\,$\mu$m feature has weak or no correlations with the gas-phase molecules. 
Future radio observations of high spatial and spectral resolution toward a large sample are desirable to elucidate the 21\,$\mu$m emission phenomena.

\end{abstract}


\section{Introduction}

Low- and intermediate-mass (0.8--8.0\,$M_{\odot}$) stars eject material into the interstellar medium (ISM) during their evolution 
from the asymptotic giant branch (AGB) to the white-dwarf phase, creating circumstellar envelopes \citep[CSEs, see, e.g.,][]{Hofner18}. 
CSEs can serve as a natural chemical laboratory and play a significant role on cosmic gas/dust cycle. 
However, we are still far from fully understanding circumstellar chemistry \citep[see, e.g.,][]{Ziurys06, Cernicharo11}.
The chemical processes in the protoplanetary nebula (PPN) phase that is a transition phase from the AGB to white-dwarf stages 
with a timescale of about $\sim$10$^{3}$\,yr \citep{Herwig05} are particularly unclear. 
A PPN consists of a post-AGB star with a effective temperature of less than $3\times10^{4}$\,K and a separated shell. 
The central star is not hot enough to ionize the shell, 
but the growing ultraviolet (UV) radiation field could process the dust and gas in PPNs and stimulate photochemistry.

Protoplanetary nebule exhibit various infrared (IR) unidentified features, among which
the so-called 21\,$\mu$m emission feature is one of the most intriguing enigmas \citep{Kwok21}. 
This feature was first discovered in four PPNs by \cite{Kwok89} 
from the $Infrared$ $Astronomical$ $Satellite$ $(IRAS)$ Low-Resolution Spectrometer (LRS) database. 
Subsequent studies of \cite{Volk99} expanded the sample of 21\,$\mu$m feature objects and show that 
this feature has a consistent line profile from source to source and actually peaks at 20.1\,$\mu$m. 
The objects with the 21\,$\mu$m feature are rare, 
suggesting that this feature may arise from a transient molecule \citep{Kwok89}. 
So far, this feature has been unambiguously detected only in 18 and 9 PPNs 
in the Galaxy and the Large/Small Magellanic Clouds, respectively \citep{Cerrigone11, Volk11}, 
suggesting that it must be a transient phenomenon. 
Accurate identification of its carrier is crucial for understanding circumstellar chemistry, 
but is a subject of considerable debate (see \citealt{Volk20} for a recent review).

Proposed carriers of the 21\,$\mu$m feature include 
SiS$_{2}$, Fe$_{\rm m}$O$_{\rm n}$, TiC, thiourea, hydrogenated fullerenes, polycyclic aromatic hydrocarbons, 
hydrogenated amorphous carbon, nano-diamonds, etc. \citep[see, e.g.,][and references therein]{Volk20}. 
No definite consensus has been reached. 
\citet{Zhang09} conducted a quantitative investigation for nine kinds of inorganic candidates, 
including TiC nanoclusters, fullerenes coordinated with Ti atoms, SiS$_{2}$, doped-SiC, SiO$_{2}$-coated SiC dust, 
carbon and silicon mixtures, nano FeO, Fe$_{2}$O$_{3}$, and Fe$_{3}$O$_{4}$ dust, and concluded that the cold FeO dust, 
which was first suggested by \cite{Posch04}, is a feasible candidate for the 21\,$\mu$m emission feature carrier. 
However, the laboratory studies of \cite{Koike17} suggested that the IR spectra of FeO vary depending on defects and disorders and 
the synthesis of FeO is sensitive to surrounding conditions, disfavoring FeO as the carrier of the 21\,$\mu$m feature. 
Fullerenes have recently been discovered in circumstellar environments \citep{cami10}, 
which has renewed interest in the idea that hydrogenated fullerenes might be the 21\,$\mu$m feature carrier \citep{Zhang20b,Zhang20c,Zhang20a}. 
These fullerene derivatives may partly survive, be transferred into more complex compounds during the dispersion of CSEs, 
and finally contribute to other unidentified spectral features such as anomalous microwave emission \citep{Iglesias06} 
and diffuse interstellar bands \citep{Omont16}. 
According to \cite{Mishra16}, there is a strong correlation between the mass-loss rates in the AGB phase and the flux of the 21\,$\mu$m feature. 
They proposed that its carrier has been formed in the AGB phase but is unable to be excited by sufficient UV photons until the PPN phase.

It is established that the gas-phase chemistry and the dust processes depend on the physical conditions and the evolutionary stages of CSEs. 
Different astronomical environments are characterized by different gas-phase molecular patterns \citep[e.g.,][]{Cernicharo11, Aladro15, Jorgensen20, Taniguchi23}. 
Therefore, it is instructive to statistically investigate the gas-phase molecules in 21\,$\mu$m sources, which could presumably provide some clues on the origin of the 21\,$\mu$m feature. 
Such studies have been rather scarce because the molecular lines in 21\,$\mu$m sources are typically very faint. 
The majority of earlier investigation were focused on CO only (e.g., \citealt{Woodsworth90, Hrivnak05, Nakashima12}).
\cite{Zhang20b} reported a line survey of the 21\,$\mu$m source IRAS\,22272+5435 at 3 and 1.3 mm bands, 
which resulted in the detection of ten molecules and three isotopologues. 
A large sample is still required to obtain statistically meaningful results.

In this paper, we present observations of gas-phase molecules in six 21\,$\mu$m sources, 
aiming to investigate the relation between the 21\,$\mu$m feature and the gas-phase molecules. 
The paper is organized as follows. 
In Sect.~\ref{Section2}, we present the sample, observations, and data reduction. 
In Sect.~\ref{Section3}, we present the observation results as well as the calculations of column densities, fractional abundances, and isotopic ratios. 
In Sect.~\ref{Section4}, we discuss the correlations between the strengths of the 21\,$\mu$m emission feature and the measurements of the molecular lines in our sample.

\section{Sample, Observations, and Data Reduction} 
\label{Section2}

\subsection{Description of the sample}

The sample of 21\,$\mu$m sources is listed in Table~\ref{Table1}. 
They were chosen for observation because of their comparatively strong CO emission. 
The properties of the sample are described below. 

{\it \bf IRAS\,06530-0213:} 
Its central star is an F5\,I spectral type supergiant with an effective temperature of 6\,900\,K \citep{Hrivnak03}. 
The star is strongly carbon-enriched with a C/O abundance ratio of 2.8 and has a sub-solar metallicity with [Fe/H] =~$-0.5$ \citep{Reyniers04}. 
The significantly enhanced $s$-process elements indicate that the star has experienced a strong third dredge-up \citep{Reyniers04}. 
The optical image taken by the Hubble Space Telescope (HST) shows an elliptical nebula ($\sim$2$\arcsec$ $\times$ 1$\arcsec$) 
with limb-brightened bipolar lobes \citep{Ueta00}.

{\it \bf IRAS\,07134+1005:} 
This PPN, also named HD\,56126 or SAO\,96709, has an elliptical morphology with an extended circular reflection nebula \citep{Ueta00}. 
It is the biggest 21\,$\mu$m source, with an optical size of $4\arcsec$, according to the HST image. 
A two-peaked, limb-brightened dust torus is plainly seen in the mid-IR (MIR) image \citep{Meixner97}. 
Its central supergiant star has a variability period of 12.1 days with an effective temperature of 7\,250\,K and a spectral class of F5\,Iab \citep{Bakker97,Ueta00}. 
The [Fe/H] and C/O ratio are $-1.0$ and 1.2, respectively \citep{Reddy02,Reyniers04}.
It is one of the four sources in which the 21\,$\mu$m emission feature was first discovered \citep{Kwok89}.

{\it \bf IRAS\,19500+1709:} 
This PPN, also named HD\,187885 or SAO\,163075, has a central star with an effective temperature of 8\,000\,K 
and an optical spectral type of F2--6\,Ia, a metallicity of [Fe/H] =~$-0.7$ to $-0.6$, and a C/O ratio of 1.1 \citep{Hrivnak89,Winckel00,Reddy02,Reyniers04}. 
It is an intrinsically polarized bipolar reflection nebula with a size of 3$\rlap{.}\arcsec$6 $\times$ 2$\rlap{.}\arcsec$2 at the near-IR (NIR) $J-$band,  
consisting of two lobes divided by a dark lane that is perpendicular to the major axis \citep{Gledhill01}. 
The 21\,$\mu$m emission feature in IRAS\,19500+1709 was first detected by \cite{Justtanont96} using the United Kingdom Infrared Telescope (UKIRT).

{\it \bf IRAS\,22272+5435:} 
This PPN, also named HD\,235858 or SAO\,34504, 
is one of the four sources in which the 21\,$\mu$m emission feature was first reported \citep{Kwok89}. 
It has a G5\,Ia central star with an effective temperature of 6\,500\,K. 
Optical and IR  observations reveal an elliptically elongated nebula \citep{Ueta00, Gledhill01} 
with a dust torus (1$\rlap{.}\arcsec$6 diameter at 11.8\,$\mu$m, \citealt{Meixner97}). 
The [Fe/H] and C/O ratio of this PPN is $-0.8$ to $-0.5$ and 1.6--1.8, respectively \citep{Reddy02,Zacs16}.

{\it \bf IRAS\,23304+6147:} 
This PPN is one of the four 21\,$\mu$m objects first reported by \cite{Kwok89}. 
It is a quadrupolar shaped nebula (about 5$\rlap{.}\arcsec$5 at the NIR band) with two pairs of unevenly spaced lobes \citep{Sahai07}. 
The [Fe/H] and C/O ratio is $-0.8$ and 2.9--3.2, respectively \citep{Winckel00, Reddy02, Reyniers04}.
The central star is G2\,Ia type and has an effective temperature of 6\,750\,K.

{\it \bf IRAS\,Z02229+6208:} 
The HST NIR image of this PPN reveals an elliptical nebula with a size of about 2$\arcsec$ $\times$ 1$\arcsec$ \citep{Ueta05}. 
Its 21\,$\mu$m feature was first discovered by \cite{Volk99} based on the Infrared Space Observatory (ISO) data. 
The central star is assigned as G8-K0\,0-Ia type \citep{Hrivnak99} and has an effective temperature of 5\,500\,K \citep{Reddy99}. 
It has a sub-solar metalicity ([Fe/H] =~$-0.5$) and is overabundant with C, N, and $s$-process elements \citep{Reddy99}. 
Based on the Na\,{\sc i} absorption spectra, \cite{Reddy99} proposed that IRAS\,Z02229+6208 has an intermediate-mass progenitor star ($>$3\,$M_\odot$).

\subsection{SMT 10\,m Observations}

Five 21\,$\mu$m sources were observed in the 1.3\,mm window with the 
Arizona Radio Observatory 10\,m Submillimeter Telescope\footnote{https://aro.as.arizona.edu/?q=facilities/submillimeter-telescope} (SMT) at Mt.~Graham 
on 2014 April 21--26 and 2015 April 15--21. 
The study is complemented by the 1.3\,mm spectra of IRAS\,22272+5435 reported by \cite{Zhang20b}.
We used the 2048-channel acousto-optical spectrometer and the 1024-channel Forbes Filterbanks simultaneously 
with a frequency channel spacing of 500\,kHz and 1\,MHz, respectively, 
to cover the frequency range from 216--273\,GHz with several gaps. 
The selections of these frequency scans are based on the relatively strong lines discovered in IRC+10216 and CRL\,2688 \citep{He08,Zhang13}. 
As the 21\,$\mu$m sources are much fainter, it is reasonable to suppose that no line in the frequency gaps is detectable with our instrumental sensitivity. 
The half-power beam widths (HPBW) are in the range of 27$\rlap{.}\arcsec$7--34$\rlap{.}\arcsec$8, 
which are larger than the source sizes derived from their mid-IR images. 
The standard wobbler switching mode was carried out with an azimuth beam throw of $\pm$2$\arcmin$. 
The on-source times of each band are about 1--3 hours. 
The dual-channel superconductor-insulator-superconductor (SIS) receivers in single-sideband dual polarization mode were applied. 
The typical system temperatures are about 200--500\,K. 
Pointing and focus were checked every 2 hours with nearby planets.

\subsection{PMO 13.7\,m Observations}

We also observed the the $J =1 \rightarrow 0$ lines from CO, $^{13}$CO, and C$^{18}$O toward the six 21\,$\mu$m sources by using 
the Purple Mountain Observatory (PMO) 13.7\,m millimeter telescope\footnote{http://www.dlh.pmo.cas.cn/wyjsbjs/hmbwyj} at Delingha in China. 
The observations were performed during May 11--20 2022. 
A supplementary observation toward IRAS~06530-0213 was carried out on November 9 2023. 
The on-source time of each band is 0.5--1 hour. 
The receiver used is a two-sideband SIS mixer. 
The spectrometer backends for each sideband utilize a Fast Fourier Transform Spectrometer (FFTS) 
with a bandwidth of about 1\,GHz including 16384 channels and a frequency resolution of 61\,kHz. 
The corresponding velocity resolution is 0.16\,km\,s$^{-1}$ for the CO $J = 1 \rightarrow 0$ line at the rest frequency of 115\,GHz. 
The two-sideband SIS receiver was designed to cover the CO $J = 1 \rightarrow 0$ line at the upper sideband 
and the $J = 1 \rightarrow 0$ transitions of $^{13}$CO and C$^{18}$O at the lower sideband simultaneously with a specific local oscillator (LO) frequency \citep{Shan12}. 
The observations were made using a beam switch method. 
Regular pointing checks were made with a precision better than 5$\arcsec$. 
The HPBWs of this telescope are about 55$\arcsec$ and 52$\arcsec$ at 110 and 115\,GHz, respectively. 
All sources are thus unresolved.

\subsection{Data reduction}

The spectral data were reduced by using the Grenoble Image and Line Data Analysis Software 
(GILDAS) of the Continuum and Line Analysis Single-dish Software (CLASS) package\footnote{https://www.iram.fr/IRAMFR/GILDAS/} \citep{Pety05}. 
Bad scans caused by bad atmospheric conditions and receiver instabilities were discarded. 
The artificial features at the edge of each band were removed. 
A low-order polynomial baseline subtraction was carried out for each scan. 
The scans were co-added by using the inverse noise as the weight. 
The antenna temperature ($T_{A}^{*}$) is converted to the main-beam temperature ($T_{\rm mb}$) following the expression of $T_{\rm mb} = T_{A}^{*} / B_{\rm eff}$, 
where $B_{\rm eff}$ is the main beam efficiency of 0.7 and 0.6 for the SMT 10\,m and PMO 13.7\,m telescopes, respectively. 
To further increase the signal-to-noise ratio, we smoothed the spectra over several channels using the `boxcar' method. 
The resultant root mean square (RMS) levels and velocity resolutions of the transitions are listed in Tables~\ref{Table2}--\ref{Table7}.

\section{Results} 
\label{Section3}

\begin{figure}[htb]
\centering
\includegraphics[angle=0,scale=.21]{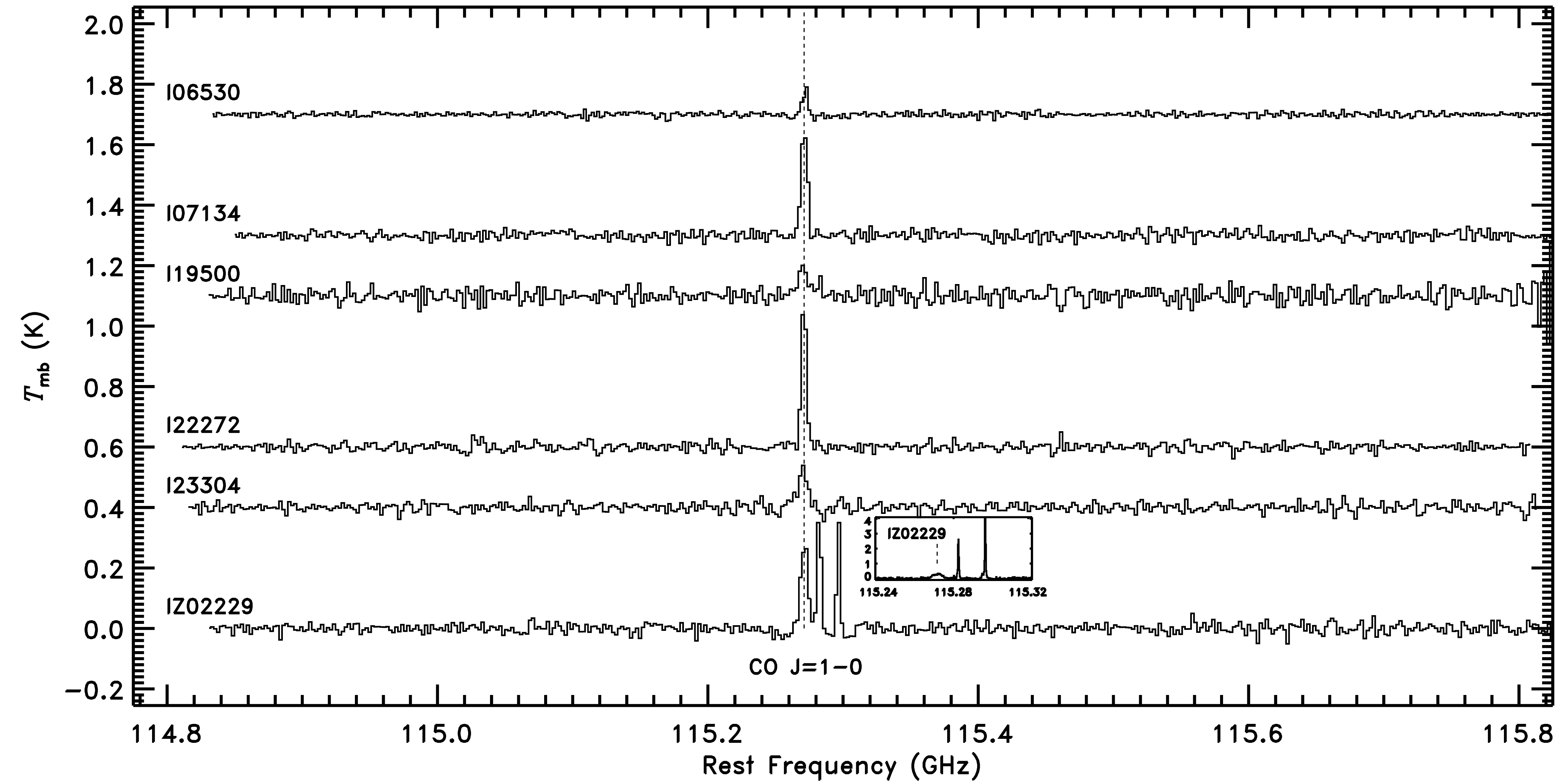}
\caption{
Spectra of the 21\,$\mu$m sources obtained with the PMO 13.7\,m telescope. 
The dotted vertical lines mark the strong lines in the spectrum of CRL\,2688. 
To bring out the lines against the noise, 36 adjacent channels have been binned, 
resulting in a frequency resolutions of about 2\,MHz. 
To avoid smearing we cut off part of the spectra of IRAS\,Z0229+6208, and the zoom-out spectra are shown in the insets.
}
\label{Figure1}
\end{figure}

The spectra for the whole frequency range covered by our observations are shown in Figs.~\ref{Figure1} and \ref{Figure2}, where for a better view of the detected lines against the noise we have binned several adjacent channels. 
Thirty-one emission lines assigned to seven molecular species and isotopologues, including CO, $^{13}$CO, HCN, H$^{13}$CN, HNC, CN, and HC$_{3}$N, are detected. 
Figs.~\ref{Figure3}--\ref{Figure8} display the profile of each line separately. 
SiS and SiO are marginally detected in a few sources. 
The density of the detected lines is about one line per GHz. 
Except for IRAS\,22272+5435, none of the 21\,$\mu$m sources display SiC$_{2}$, 
a compound that is frequently detected in C-rich CSE. 
In previous work, CO, $^{13}$CO, and HCN were the only molecules detected in these sources. 
The velocity-integrated intensities of strong lines are obtained through the stellar-shell model fitting using the CLASS package \citep{Pety05}.
The fitting function of the stellar-shell model is 
\begin{equation}
f(\nu) = \frac{A}{\Delta\nu} \frac{1+4H[(\nu-\nu_{0})/\Delta\nu]}{1+H/3}, 
\label{Equation1}
\end{equation}  
where $A$ is the integrated intensity in K\,MHz, $\nu_{0}$ and $\Delta\nu$ are the middle frequency and full width at zero level in MHz. 
$H$ dictates the Horn/Center parameter, which represents the edge-to-center intensity ratio, with the expression of $f(\Delta\nu/2)/f(0)$ = $1 + H$. 
As a result, we can estimate the expanding velocity by $V_{\rm {exp}}$ = $c\Delta\nu/(2\nu_{0})$ with $c$ the light speed. 
The faint lines and the blended CN $N = 2 \rightarrow 1$ fine structures cannot be fitted by the stellar-shell model. 
Tables~\ref{Table2}--\ref{Table7} present the fitting results, 
where the rest frequency, molecular species, upper and lower energy levels, expanding velocity ($V_{\rm exp}$), Horn/Center parameter ($H$), intensity at the line center ($T_{\rm c}$), velocity-integrated intensity ($\int T_{\rm mb}{\rm d}V$), RMS of the baseline, line center velocity ($V_{\rm LSR}$), velocity resolution (D$_{v}$), HPBW of the telescope at the corresponding frequency are listed.
In the following, the observational results of each source are reported.

\subsection{\texorpdfstring{IRAS\,06530-0213}{}}

The CO $J = 2 \rightarrow 1$ line in this object was previously reported by \cite{Hrivnak05} using the 10\,m diameter single dish Heinrich Hertz Submillimeter Telescope (HHT), 
which exhibits a slightly asymmetric double-peak profile. 
In this work, we detect five emission lines from gas-phase molecules,
including the CO $J = 1 \rightarrow 0$ and $^{13}$CO $J = 2 \rightarrow 1$ lines, 
the $J = 3 \rightarrow 2$ transitions of HCN and HNC, and the CN $N = 2 \rightarrow 1$ line, 
as shown in Fig.~\ref{Figure3}. 
A narrow feature with a width of 2\,km\,s$^{-1}$ is visible at the blue side of the CO $J = 1 \rightarrow 0$ line. 
Based on the Herschel archive data and CO map of the Milk Way \citep{Dame01}, 
it is likely to be a contamination from a nearby interstellar cloud lying within the telescope beam. 
The broad CO $J = 1 \rightarrow 0$ feature and the HCN $J = 3 \rightarrow 2$ line show a parabolic profile, 
suggesting that they are  optically thick. 
The CO $J = 1 \rightarrow 0$ and $^{13}$CO $J = 2 \rightarrow 1$ lines exhibit a prominent absorption at about 45\,km\,s$^{-1}$, 
which has also been seen in the CO $J = 2 \rightarrow 1$ line \citep{Hrivnak05}. 
The $^{13}$CO $J = 2 \rightarrow 1$ line shows an emission feature at 55--70\,km\,s$^{-1}$, 
which is not detected in the other transitions. 
The mean line width of CO $J = 1 \rightarrow 0$, $^{13}$CO $J = 2 \rightarrow 1$, and the $J = 3 \rightarrow 2$ transitions of HCN and HNC indicate an expansion velocity of 9.7 $\pm$ 0.5 km\,s$^{-1}$. 
The HNC $J = 3 \rightarrow 2$ line shows an asymmetric double-peak profile with a stronger red peak. 

The HC$_{3}$N species are marginally detected through the $J = 24 \rightarrow 23$ transition.
The $J = 1 \rightarrow 0$ transitions of $^{13}$CO and C$^{18}$O, SO $N_{J} = 5_{6} \rightarrow 4_{5}$, SiO $J = 5 \rightarrow 4$, 
SiS $J = 12 \rightarrow 11$ and $15 \rightarrow 14$, and SiC$_{2}$ $J_{K+,K-} = 10_{0,10} \rightarrow 9_{0,9}$ lines 
lie within our frequency coverage but are not seen.

\subsection{\texorpdfstring{IRAS\,07134+1005}{}}

The first detection of CO in this object was made by \cite{Zuckerman86} 
using the Five College Radio Astronomy Observatory (FCRAO) 14\,m telescope through the $J = 1 \rightarrow 0$ transition. 
The CO $J = 1 \rightarrow 0$, $2 \rightarrow 1$, and $4 \rightarrow 3$, 
$^{13}$CO $J = 2 \rightarrow 1$, and HCN $J = 1 \rightarrow 0$ transitions were previously detected in this object 
by Caltech Submillimeter Observatory (CSO) 10.4\,m telescope, Institut de Radioastronomie Millim{\'e}trique (IRAM) 30\,m telescope, and HHT \citep{Bujarrabal92, Omont93, Knapp98, Knapp00, Hrivnak05}. 
The HCO$^{+}$ $J = 1 \rightarrow 0$ line was possibly detected by \cite{Bujarrabal92} using the IRAM 30\,m telescope. 
As shown in Fig.~\ref{Figure4}, our observations clearly show the presence of 
CO $J = 1 \rightarrow 0$, $^{13}$CO $J = 2 \rightarrow 1$, HCN $J = 3 \rightarrow 2$, and CN $N = 2 \rightarrow 1$ lines. 
The profiles of the CO $J = 1 \rightarrow 0$ and $^{13}$CO $J = 2 \rightarrow 1$ transitions show 
a smoothed parabolic and a double-peaked profile, 
corresponding to optically thick and thin cases, respectively. 
\cite{Bujarrabal92} found that the HCN $J = 1 \rightarrow 0$ line, in spite of the poor signal-to-noise ratio, 
seems to be shifted to the red side, which can be ascribed to self-absorption. 
Our higher signal-to-noise ratio observations show that the HCN $J = 3 \rightarrow 2$ line clearly has a stronger emission at the red side, 
supporting the result of \cite{Bujarrabal92}. 
Based on an interferometric observation of the Berkeley-Illinois-Maryland-Association (BIMA) millimeter array, 
\cite{Meixner04} found that IRAS\,07134+1005 is a low-mass clumpy PPN with a size extending to about 7$\arcsec$ in radius. 
\cite{Nakashima09} constructed a morphokinematic model based on the Submillimeter Array (SMA) interferometric observations of the CO $J = 3 \rightarrow 2$ line, 
which indicates  the presence of a central torus with an inner and outer radius of 1$\rlap{.}\arcsec$2 and 3$\rlap{.}\arcsec$0, respectively. 
The mean width of the CO $J = 1 \rightarrow 0$, $^{13}$CO $J = 2 \rightarrow 1$, and HCN $J = 3 \rightarrow 2$ lines in our spectra 
suggests an expansion velocity of 9.9 $\pm$ 0.2\,km\,s$^{-1}$.

The HNC $J = 3 \rightarrow 2$ and HC$_{3}$N $J = 24 \rightarrow 23$ lines are marginally detected. 
The $J = 1 \rightarrow 0$ transitions of $^{13}$CO and C$^{18}$O, H$^{13}$CN $J = 3 \rightarrow 2$, SO $N_{J} = 5_{6} \rightarrow 4_{5}$, 
SiO $J = 5 \rightarrow 4$, SiS $J = 12 \rightarrow 11$ and $15 \rightarrow 14$, SiC$_{2}$ $J_{K+,K-} = 10_{0,10} \rightarrow 9_{0,9}$, 
and HC$_{3}$N $J = 25 \rightarrow 24$ lines, 
which are frequently detected in CSEs and lie within our observational frequency range, are not detected. 
The non-detection of SiO has also been reported by \cite{Bujarrabal92}.

\subsection{\texorpdfstring{IRAS\,19500-1709}{}}

The first detection of CO species in IRAS\,19500-1709 was made by \cite{Likkel87} in a survey of the CO $J = 1 \rightarrow 0$ line 
toward the coldest IRAS objects performed with the IRAM 30\,m telescope. 
They found that IRAS\,19500-1709 has the lowest circumstellar expansion velocity ($\sim$11\,km\,s$^{-1}$) among their sample. 
Subsequent observations of higher-$J$ CO transitions indicate the existence of a speedy outflow with a velocity of $\sim$30\,km\,s$^{-1}$ \citep{Knapp89, Likkel91, Bujarrabal92, Omont93, Hrivnak05}. 
The $^{13}$CO and HCN species have also been detected in previous observations \citep{Bujarrabal92}. 
The CO $J = 1 \rightarrow 0$, $^{13}$CO $J = 2 \rightarrow 1$, HCN $J = 3 \rightarrow 2$, and CN $N = 2 \rightarrow 1$ transitions, 
are visible in our spectra (Fig.~\ref{Figure5}). 
The CO $J = 1 \rightarrow 0$ and $^{13}$CO $J = 2 \rightarrow 1$ lines show broad and narrow components. 
Their peak velocities show a small offset (5.2\,km\,s$^{-1}$ for CO and 4.2\,km\,s$^{-1}$ for $^{13}$CO). 
This is primarily due to the rather sensitive dependence of the fitting on the faint broad wing, 
and cannot be considered as real. 
Nevertheless, the broad component suggests a outflow velocity of 29--33\,km\,s$^{-1}$, 
which is consistent with previous results. 
The two components are not distinct for the HCN $J = 3 \rightarrow 2$ line due to low signal-to-noise ratio. 
An inspection of the Fig.~\ref{Figure5} clearly indicates that the stellar-shell fitting for the HCN line is unreliable. 
We thus mark it as an uncertain measurement, and assume a 50$\%$ uncertainty for the analysis. 
\cite{Bujarrabal92} discovered a CO $J = 2 \rightarrow 1$ absorption at 13\,km\,s$^{-1}$, 
and ascribed this to a slow envelope. 
Such an absorption feature is also detected in our $^{13}$CO $J = 2 \rightarrow 1$ spectrum. 
Differing from \cite{Bujarrabal92}, we discover a $^{13}$CO emission component at $\sim$10\,km\,s$^{-1}$. 
The mean line widths suggest the expansion velocities of the broad and narrow components to be 9.2 $\pm$ 0.5 and 30.0 $\pm$ 1.8\,km\,s$^{-1}$, respectively.

The transitions of HNC $J = 3 \rightarrow 2$, SiO $J = 5 \rightarrow 4$, and SiS $J = 12 \rightarrow 11$, $15 \rightarrow 14$ are marginally detected. 
The $^{13}$CO and C$^{18}$O $J = 1 \rightarrow 0$, H$^{13}$CN $J = 3 \rightarrow 2$, SO $N_{J} = 5_{6} \rightarrow 4_{5}$, 
SiC$_{2}$ $J_{K+,K-} = 10_{0,10} \rightarrow 9_{0,9}$, and HC$_{3}$N $J = 24 \rightarrow 23$ and $25 \rightarrow 24$ transitions 
lie within the frequency range, but are not visible.

\subsection{\texorpdfstring{IRAS\,22272+5435}{}}

The detection of CO in IRAS\,22272+5435 was first reported by \cite{Woodsworth90} through the $J = 2 \rightarrow 1$ transition using the JCMT.  
\cite{Bujarrabal01} observed the $J = 1 \rightarrow 0$ and $2 \rightarrow 1$ lines of CO and $^{13}$CO using the IRMA 30\,m telescope. 
The CO $J = 2 \rightarrow 1$ and $4 \rightarrow 3$ lines were detected by \cite{Hrivnak05} using the HHT. 
We observe the $J = 1 \rightarrow 0$ transitions of CO and $^{13}$CO,
which yield an expansion velocity of 7.5 $\pm$ 0.7\,km\,s$^{-1}$.
As shown in Fig.~\ref{Figure6}, their line profiles are smoothed parabolic and double-peaked, respectively, 
corresponding to optically thick and thin cases, respectively. 
Interferometric observations toward IRAS\,22272+5435 have been carried out by \cite{Fong06} using the BIMA array in the CO $J = 1\rightarrow 0$ line 
and by \cite{Nakashima12} using the Combined Array for Research in Millimeter-wave Astronomy (CARMA) in the CO $J = 2\rightarrow 1$ line. 
The morphologies of the CO $J = 2 \rightarrow 1$ and $J = 1\rightarrow 0 $ transitions appear to be different. 
The CO $J = 1\rightarrow 0$ mapping shows a roughly spherical envelope 
with a size of about 20$\arcsec$ and some unusual protrusions. 
The CO $J = 2\rightarrow 1$ image clearly resolves the central structure, 
which exhibits two intensity peaks with a distance of about 1$\rlap{.}\arcsec$4 
and that their connecting line is almost in the vertical direction of the two 12.5\,$\mu$m peaks of \citet{Ueta01}. 
Thus far, the detected molecules in this object include 
C$_{2}$H, C$_{4}$H, HC$_{3}$N, SiC$_{2}$, CS, C$^{34}$S, CO, $^{13}$CO, CN, CH$_{3}$CN, HCN, H$^{13}$CN, and HNC 
(see the details in \citealt{Zhang20b}).

\subsection{\texorpdfstring{IRAS\,23304+6147}{}}

The first detection of CO in IRAS\,23304+6147 was made by \cite{Woodsworth90} using the JCMT through the $J = 2 \rightarrow 1$ transition. 
\cite{Likkel91} and \cite{Hrivnak05} observed the CO $J = 1 \rightarrow 0$ and $4 \rightarrow 3$ transitions. 
\cite{Omont93} tentatively detected the CS $J = 3 \rightarrow 2$ transition using the IRAM 30\,m telescope.
Our observations result in the detection of six lines, 
including the CO $J = 1 \rightarrow 0$, $^{13}$CO $J = 2 \rightarrow 1$, $J = 3 \rightarrow 2$ transitions of HCN, HCN, and H$^{13}$CN, and CN $N = 2 \rightarrow 1$, 
as shown in Fig.~\ref{Figure7}. 
The $^{13}$CO $J = 2 \rightarrow 1$ and HCN $J = 3 \rightarrow 2$ lines are composed of narrow and broad components. 
Similar to the CO $J = 1 \rightarrow 0$ line in IRAS\,19500-1709, the two components of the HCN $J = 3 \rightarrow 2$ line show slightly different peak velocities 
with an offset of 3.2\,km\,s$^{-1}$.
It ought to be regarded as a fitting  error
because the fitting of the broad component is substantially influenced by its faint wing. 
The CO $J = 1 \rightarrow 0$ line shows an asymmetric shape with stronger emission at the blue side. 
A similar profile has been reported by \cite{Hrivnak05} for the CO $J = 2 \rightarrow 1$ line. 
We observe a narrow emission feature at the velocity of about 0 km\,s$^{-1}$ for the CO $J = 1 \rightarrow 0$ line, 
which may stem from the interstellar medium close to the Galactic plane. 
The mean line widths result in expansion velocities of 10.3 $\pm$ 0.9 and 22 $\pm$ 0.8\,km\,s$^{-1}$ for the narrow and broad components, respectively.

The HC$_{3}$N $J = 25 \rightarrow 24$ transition is marginally detected. 
The $^{13}$CO and C$^{18}$O $J = 1 \rightarrow 0$, SO $N_{J} = 5_{6} \rightarrow 4_{5}$, SiO $J = 5 \rightarrow 4$, 
SiS $J = 12 \rightarrow 11$ and $J = 15 \rightarrow 14$, SiC$_{2}$ $J_{K+,K-} = 10_{0,10} \rightarrow 9_{0,9}$, 
and HC$_{3}$N $J = 24 \rightarrow 23$ lines are not seen.

\subsection{\texorpdfstring{IRAS\,Z02229+6208}{}}

Prior to this work, CO was the only gas-phase molecule detected in this object, whose $J = 2 \rightarrow 1$ and $3 \rightarrow 2$, $4 \rightarrow 3$ transitions were observed by \cite{Hrivnak99} and \cite{Hrivnak05}. 
The $J = 1 \rightarrow 0$ transitions of CO and $^{13}$CO, 
the $J = 3 \rightarrow 2$ transitions of HCN, H$^{13}$CN, and HNC,
CN $N = 2 \rightarrow 1$, and HC$_{3}$N $J = 24 \rightarrow 23$, $25 \rightarrow 24$, and $28 \rightarrow 27$ lines are detected in
our observations. 
HCN, H$^{13}$CN, HNC, and HC$_{3}$N are newly discovered in this object. 
As shown in Fig.~\ref{Figure8}, the $J = 1 \rightarrow 0$ transitions of CO and $^{13}$CO and the $J = 3 \rightarrow 2$ transitions of HCN and HNC show both broad and narrow components. 
The broad component is likely to originate from fast outflows. 
There is an emission feature at the velocity of about -5\,km\,s$^{-1}$ for CO $J = 1 \rightarrow 0$ and $^{13}$CO $J = 1 \rightarrow 0$ and $J = 2 \rightarrow 1$ lines; 
we infer that it may stem from the interstellar medium. 
The mean line widths indicate expansion velocities of 10.3 $\pm$ 0.9 and 22 $\pm$ 0.8\,km\,s$^{-1}$ for the broad and narrow components, respectively.

The SiS $J = 12 \rightarrow 11$ line is marginally detected. 
The C$^{18}$O $J = 1 \rightarrow 0$, SO $N_{J} = 5_{6} \rightarrow 4_{5}$ and $9_{8} \rightarrow 8_{8}$, SiO $J = 5 \rightarrow 4$, 
SiS $J = 15 \rightarrow 14$, and SiC$_{2}$ $J_{K+,K-} = 10_{0,10} \rightarrow 9_{0,9}$ transitions lie within our frequency coverage and are not detected.

\subsection{Column densities and fractional abundances}
\label{abundace}

Three HC$_{3}$N transitions are clearly detected in IRAS\,Z02229+6208 (Fig.~\ref{Figure7}), 
allowing us to perform a rotation-diagram analysis to derive the excitation temperature ($T_{\rm ex}$) and total column density ($N$) of 
HC$_{3}$N in this PPN. 
Based on the assumptions of local thermodynamical equilibrium (LTE) and optical thinness, 
the level populations can be expressed in the function of 
\begin{equation}
{\rm ln} \frac{N_{\rm u}}{g_{\rm u}} = {\rm ln} \frac{8\pi k \, \nu^{2} \, \int T_{\rm S} \, {\rm d}{V}}{hc^{3} \, A_{\rm ul} \, g_{\rm u}}\,  =  {\rm ln} \frac{N}{Q(T_{\rm ex})} - \frac{E_{\rm u}}{kT_{\rm ex}}, 
\label{Equation2}
\end{equation}  
where $k$ and $h$ represent the Boltzmann constant and Planck constant, respectively. 
The rest frequency $\nu$ (in Hz), upper-level energy $E_{\rm u}/k$ (in K), 
partition function $Q(T_{\rm ex})$, spontaneous emission coefficient $A_{\rm ul}$, 
and upper state degeneracy $g_{\rm u}$ of each transition are taken from the CDMS catalog \citep{Muller01, Muller05}. 
Based on the assumption of both the source brightness distribution and the antenna beam being Gaussian profiles, 
the line intensity $T_{\rm mb}$ in the scale of main beam temperature is converted to the source brightness temperature $T_{\rm S}$ 
by the expression of $T_{\rm S}=T_{\rm mb}(\theta_{\rm b}^{2} + \theta_{\rm s}^{2})/\theta_{\rm s}^{2}$. 
The $\theta_{\rm b}$ value is equal to the HPBW of each transition, as listed in Table~\ref{Table7}. 
The source size $\theta_{\rm s}$ is simply assumed to be 10$\rlap{.}\arcsec$4, 
a value that has been used to estimate the mass loss rate of IRAS\,Z02229+6208 by \cite{Mishra16}. 
$\int T_{\rm S}\,{\rm d}V$ is the velocity-integrated intensity of each line. 
We derive $T_{\rm ex}$ =~88.4 $\pm$ 27.7\,K 
and $N$ =~(1.37 $\pm$ 0.76) $\times$ 10$^{13}$\,cm$^{-2}$ for HC$_{3}$N in IRAS\,Z02229+6208. 
The resultant rotation diagram is shown in Fig.~\ref{Figure9}. 
Assuming $T_{\rm ex} = 80$\,K and taking the source sizes  from \cite{Mishra16} (see Table~\ref{Table1}), 
we can estimate the column densities of other molecules through Eq.~\ref{Equation2}. 
The analysis results are presented in Table~\ref{Table8}. 
The assumption of a constant $T_{\rm ex}$ is only a rough approximation,
which however does not cause significant errors
under the condition of $E_{\rm u}/k < 80$\,K.

To determine the fractional abundances with respect of H$_{2}$ ($f_{\rm X}$),
we utilize the equation proposed by \cite{Olofsson96}, 
\begin{equation}
f_{\rm X} = 1.7 \times 10^{-28} \frac{V_{\rm exp} \theta_{\rm b}D}{\dot{M}_{\rm H_2}} \frac{Q(T_{\rm ex})\nu^{2}}{g_{\rm u}A_{\rm ul}} \frac{e^{E_{\rm l}/kT_{\rm ex}}\int T_{\rm mb}{\rm d}V}{\int^{x_e}_{x_i}e^{-4x^2{\rm ln2}}{\rm d}x},
\label{Equation4}
\end{equation}  
where $A_{\rm ul}$ is in s$^{-1}$,
$V_{\rm exp}$ is the expansion velocity in km\,s$^{-1}$, 
$\theta_{\rm b}$ is the beam size in arcsecond, 
$D$ is the distance in pc, 
$\nu$ is the rest frequency in GHz, 
$\int T_{\rm mb}\,{\rm d}V$ is the velocity-integrated intensity in K\,km\,s$^{-1}$, 
$E_{\rm l}/k$ is the energy of the lower level in K, 
$\dot{M}_{\rm H_{2}}$ is the H$_{2}$ mass loss rate in $M_{\odot}/{\rm yr}$, 
and $x_{i}$ and $x_{e}$ represent the ratios of the inner ($R_{i}$) and outer ($R_{e}$) radii over the beam size, 
that is, $x_{e,i} = R_{e,i}/(\theta_{\rm b}D)$
(see Table~\ref{Table1} for these values).
If more than one transition from a molecule is detected, 
we adopt the average abundance. 
The obtained $f_{\rm X}$ values are listed in Table~\ref{Table8}. 
The indicated uncertainties come only from the measurement uncertainties, 
and may be significantly larger if considering the uncertainties associated with the distance estimation 
and the assumed gas-to-dust ratio. 
Nevertheless, the relative abundances between different molecules are free from the uncertainties of
the distances and gas-to-dust ratios. 
It should be noted that if the lines are optically thick, 
the $f_{\rm X}$ values should be considered as the lower limits.

\cite{Ramstedt14} investigated 19 C stars and found the fractional abundance of $^{13}$CO to be 
(0.1--5.5) $\times$ 10$^{-4}$ with a mean value of 1.3 $\times$ 10$^{-4}$, 
systematically higher than the $f_{\rm ^{13}CO}$ values of 
21\,$\mu$m sources of (1--2) $\times$ 10$^{-5}$. 
The CSE expansion leads to a decreased optical depth, 
and hence the shielding of the CO against the ambient interstellar radiation field is declining. 
As a result, more $^{13}$CO may be photo-dissociated. 
On the other hand, the more extensive penetration of UV photons, 
either from the center star (or a companion) or the external radiation field, 
may ionize more molecules and drive photochemistry during the evolution from the AGB to PPN phase.

Table~\ref{Table9} compares the molecular abundances of the 21\,$\mu$m sources and those of the prototypical C-star IRC+10216 and PPN CRL\,2688. 
We find that despite having molecular abundances more than one order of magnitude lower than the two prototype objects, 
the 21\,$\mu$m sources' abundance pattern more closely resembles CRL\,2688 than IRC+10216. 
Therefore, it seems that molecular spectra could serve as valuable means for classifying CSEs.

\subsection{\texorpdfstring{$^{12}$C/$^{13}$C abundance ratios}{}}

Isotopic abundance ratios can shed some insights on the stellar nucleosynthesis. 
As an intermediate reaction product of the CNO cycle, 
$^{13}$C can be brought to the surface of a red giant through the first dredge-up process. 
$^{12}$C is mainly generated by the triple-$\alpha$ reaction in AGB phase \citep{Herwig05, Karakas14}. 
The $^{12}$C/$^{13}$C ratios are estimated through CO and HCN lines and their $^{13}$C isotopic lines. 
As no C$^{18}$O lines are discovered, we estimate the lower limits of the $^{16}$O/$^{18}$O ratios based on the observational sensitivity. 
The results are presented in Table~\ref{Table10}. 
Because the CO and HCN are optically thicker than their isotopic lines, the $^{12}$C/$^{13}$C ratios listed in Table~\ref{Table10} should be treated as lower limits. 
For IRAS\,Z02229+6208, the $^{12}$C/$^{13}$C ratios obtained from the broad and narrow line components are listed. 
Their divergent values are most likely the cause of the two components having different optical depths. 
The lower limits of the $^{12}$C/$^{13}$C ratios in all 21\,$\mu$m sources 
are much lower than the solar value \citep[$\sim$89.3,][]{Lodders03}, 
which cannot be solely attributed to the influence of optical thickness. 
Therefore, the 21\,$\mu$m sources exhibit a $^{13}$C enhancements probably as a result of nonstandard mixing processes \citep{Karakas10}. 
We come to the conclusion that the isotopic ratios of 21\,$\mu$m sources are representative of evolved stars in general. 
We also note that the extremely low $^{12}$C/$^{13}$C ratios ($<$15) is a characteristic of J-type carbon stars 
that might be associated with binary evolution.

\section{Discussion and conclusion}
\label{Section4}

The CN $N = 2 \rightarrow 1$ transition consists of eleven hyperfine components, 
which are divided into two groups located at frequency ranges of 226.6--226.7 and 226.8--226.9\,GHz, respectively, 
as shown in Figs.~\ref{Figure3}--\ref{Figure8}. 
Their intensity ratios may differ from intrinsic values due to the influence of optical depth. 
However, the observed intensity ratios between the 226.8 and the 226.6\,GHz groups are lower than 1.2, 
which is significantly less than the intrinsic strength ratio ($\sim$1.8). 
To replicate the huge deviation, unreasonably large optical depths are needed. 
The same circumstances have been discovered for other CSEs \citep{Bachiller97,Zhang20b}. 
The pumping of CN lines by optical and/or near-IR radiation provides a more likely explanation for the strength abnormalities.

The interferometric observations of IRAS\,07134+1005 show a geometrically thick expanding torus with an expanding velocity of 8\,km\,s$^{-1}$ \citep{Nakashima09}. 
Based on a comparison of two sources, \cite{Zhang20b} finds that the 21\,$\mu$m source IRAS\,22272+5435 may have a relatively narrow line width, probably tracing a slowly expanding torus. 
The narrow line profile is also shown in our spectra of all the 21\,$\mu$m sources. 
It is reasonable to speculate that the presence of a torus might provide a favorable condition for the formation of the 21\,$\mu$m carrier. 
However, apart from the narrow component, the lines of IRAS\,19500-1709, IRAS\,23304+6147, and IRAS\,Z02229+6208 exhibit a broad wing with a width of 20--30\,km\,s$^{-1}$, 
tracing velocity-enhanced components. 
The CO observations with the Atacama Large Millimetre/submillimetre Array (ALMA) reveal that 
IRAS 07134+1005 has a hollow shell with a slight velocity enhancement at the poles, 
while another 21\,$\mu$m source, IRAS\,16594-4646 (not part of the current sample), 
does indeed exhibit a dense torus \citep{Ueta18}. 
Therefore, the 21\,$\mu$m sources seem to have a range of morphologies. 
This could also be inferred from the mid-IR observations of \citet{Ueta00}. 
Follow-up interferometric observations of a larger sample are required to elucidate the details of the three-dimensional structures and dynamics of 21\,$\mu$m sources.

As shown in Table~\ref{Table9}, the relative molecular abundances exhibit a similar degree of dispersion as the strengths of the 21\,$\mu$m feature. 
To establish a link between the 21\,$\mu$m emission feature and the gas-phase molecular species, 
we investigate the correlations between the relative intensities of the 21\,$\mu$m feature respective to total IR emission (F$_{\rm 21\mu m}$/F$_{\rm IR}$) 
and molecular fractional abundance f$_{\rm X}$ for the six 21\,$\mu$m feature sources, as shown in Fig.~\ref{Figure10}. 
The F$_{\rm 21\mu m}$/F$_{\rm IR}$ values equal to 4.0$\%$, 8.1$\%$, 1.0$\%$, 2.3$\%$, 4.7$\%$, and 1.3$\%$ 
for IRAS\,06530-0213, IRAS\,07134+1005, IRAS\,19500+1709, IRAS\,23304+6147, IRAS\,22272+5435, and IRAS\,Z02229+6208, respectively. 
We note that different authors have reported different intensities of the 21\,$\mu$m feature 
because of different spectral decomposition approaches \citep{Zhang10,Mishra15,Mishra16}. 
For the purpose of consistency, the values and uncertainties presented in \cite{Mishra16} are used in our work. 
It must bear in mind that the adopted uncertainties of the F$_{\rm 21\mu m}$/F$_{\rm IR}$ ratio in the analysis 
are likely to be underestimated. 
We fit the data with a linearized power law of the form 
${\rm log}(f_{\rm X}) = {\rm a} + {\rm b} \times {\rm log}[F_{\rm 21\mu m}/F_{\rm IR}]$, 
where f$_{\rm X}$ is the fractional abundances of CO, $^{13}$CO, HCN, H$^{13}$CN, HNC, CN, SiS, and HC$_{3}$N, 
as shown in Fig.~\ref{Figure10}. 
The fittings are performed using the LINMIX Bayesian regression package \cite{Kelly07}, 
which takes into account the uncertainties in x- and y-directions and is able to handle the upper limits of the y values. 
The fitting results are presented in Table~\ref{Table11}. 
The correlation coefficients are too low to have a strong statistical significance. 
However, some of the correlations would be enhanced if the points for the feeblest 21\,$\mu$m source IRAS\,19500+1709 are excluded.

In Fig.~\ref{Figure10}, we also examine the correlation between the 21\,$\mu$m feature and the HNC/HCN and CN/HCN abundance ratios. 
[HNC]/[HCN] has been found to decrease with increasing gas kinetic temperature (\citealp[and the references therein]{Graninger14}) 
and thus has been taken as a thermometer of astronomical environments (e.g., \citealt{Jin15, Hacar20, Long21}). 
We find a strong inverse correlation between [HNC]/[HCN] and F$_{\rm 21\mu m}$/F$_{\rm IR}$, 
suggesting that a warm environment may favor the formation of the 21\,$\mu$m carrier. 
The modelling of \cite{Huggins82} suggests that CN in C-rich CSEs is the direct photodissociation product of HCN. 
This is supported by the observations that CN is enhanced with respect to HCN during the PPN--PN evolution \citep{Bachiller97}. 
If one excludes the points for IRAS\,07134+1005 and IRAS\,19500+1709, 
which have the strongest and weakest relative intensities of the 21\,$\mu$m feature in our sample, 
the [CN]/[HCN] values lie within a narrow range. 
If we assume that [CN]/[HCN] is an indication of a change in the UV radiation field, 
this might suggest that these sources have a similar radiation environment. 
The rough trend shown in Fig.~\ref{Figure10} can be tentatively explained 
in terms of a scheme in which the 21\,$\mu$m carrier is being 
formed/excited with increasing UV radiation and is destroyed 
once the radiation strength reaches a certain level. 
This seems to be in line with the notion of \citet{Kwok89} 
that the 21\,$\mu$m  emission is a transient phenomenon. 
However, the interpretation is highly speculative and 
can be further complicated by other factors, such as the properties of the central star and the dust-to-gas ratio.

The intensity ratio between the $^{13}$CO 
$J = 2 \rightarrow 1$ line and the total IR emission can be taken as a proxy of the gas-to-dust ratio. 
If we assume a constant mass-loss ratio ($\dot{M}_{\rm H_{2}}$) and expansion velocity ($V_{\rm exp}$),  
the $\dot{M}_{\rm H_{2}}/V_{\rm exp}$ could be used to trace the average density of the CSE. 
A correlation study may shed new insights into the physical conditions of synthesizing the 21\,$\mu$m carrier. 
However, we cannot find any strong correlation, as shown in Fig.~\ref{Figure10}. 
A large sample study is needed to draw meaningful conclusions.

\begin{acknowledgments} 
We are grateful to the anonymous referee for constructive comments that contributed to improving the manuscript. 
This work was supported by the National Science Foundation of China (NSFC, grant No.\,11973099 and No.\,12333005). 
J.J.Q. acknowledges NSFC funding support (No.\,12003080), 
the Guangdong Basic and Applied Basic Research Foundation (No.\,2019A1515110588), 
and the Fundamental Research Funds for the Central Universities (Sun Yat-sen University, No.\,22qntd3101).
Y.Z. acknowledges Xinjiang Uygur Autonomous Region of China for 
support from the Tianchi Talent Program and 
the science research grants from the China Manned Space Project with Nos.\,CMS-CSST-2021-A10, CMS-CSST-2021-A09, etc. 
J.S.Z. acknowledges NSFC funding support (No.\,12041302). 
F.L. acknowledges the National Natural Science Foundation of China grant (12103024) and the fellowship of China Postdoctoral Science Foundation 2021M691531.
X.D.T. acknowledges the support of the Chinese Academy of Sciences (CAS) “Light of West China” Program under grant No.\,xbzg-zdsys-202212, the Natural Science Foundation of Xinjiang Uygur Autonomous Region under grant No.\,2022D01E06, and the Tianshan Talent Program of Xinjiang Uygur Autonomous Region under grant No.\,2022TSYCLJ0005.
We wish to express our gratitude to the staff at the SMT 10\,m and PMO 13.7\,m telescopes for their kind help and support during our observations.
\end{acknowledgments}

\vspace{5mm}
\facilities{SMT: 10\,m, PMO: 13.7\,m}

\software{CLASS, 
          }

\clearpage

\renewcommand{\thefigure}{\arabic{figure} (Cont.)}
\addtocounter{figure}{-1}

\begin{figure}[htb]
    \centering
\includegraphics[angle=0,scale=.21]{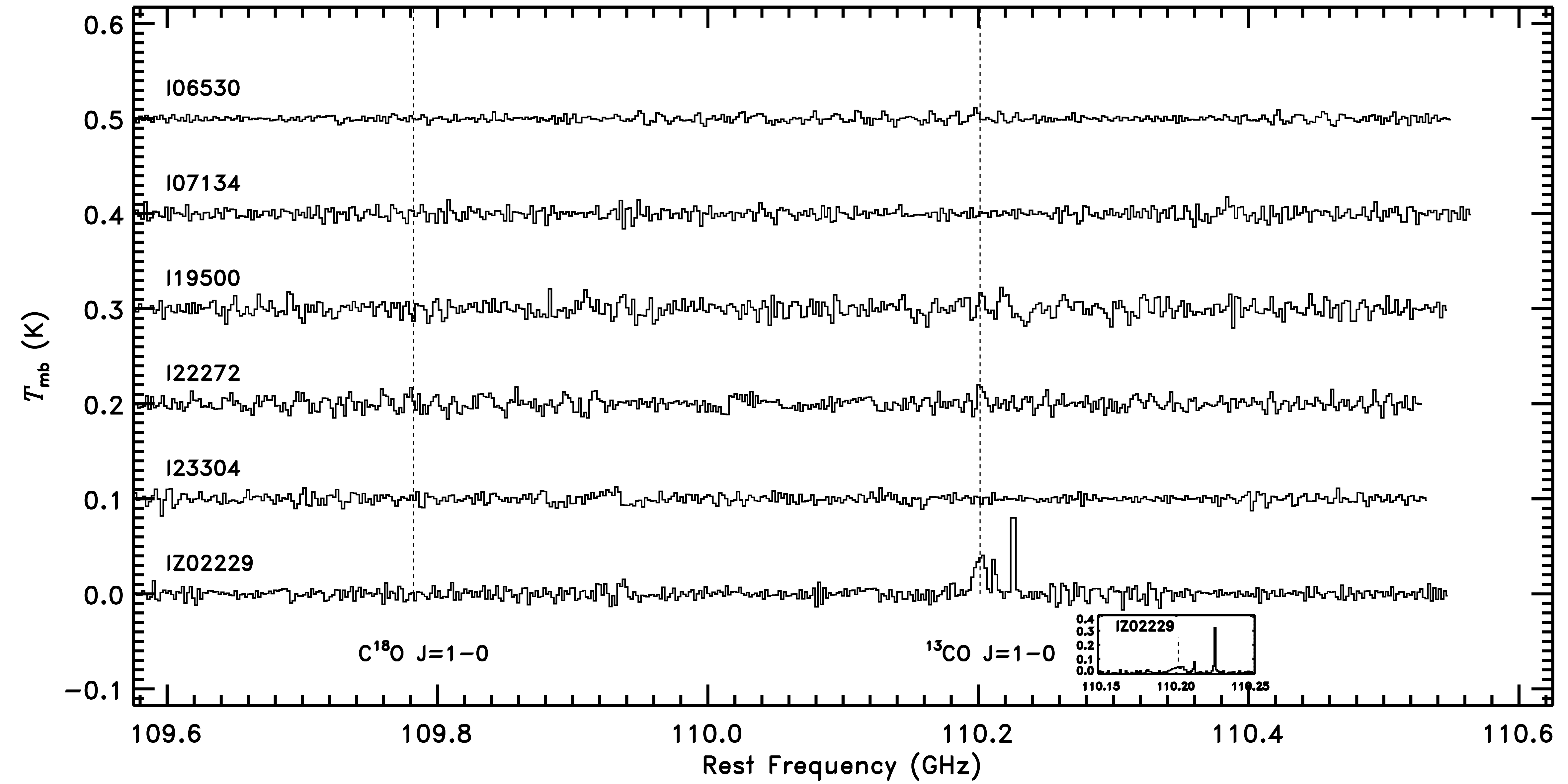}
\caption{}
\end{figure}

\renewcommand{\thefigure}{\arabic{figure}}

\begin{figure}[htb]
\centering
\includegraphics[angle=0,scale=.21]{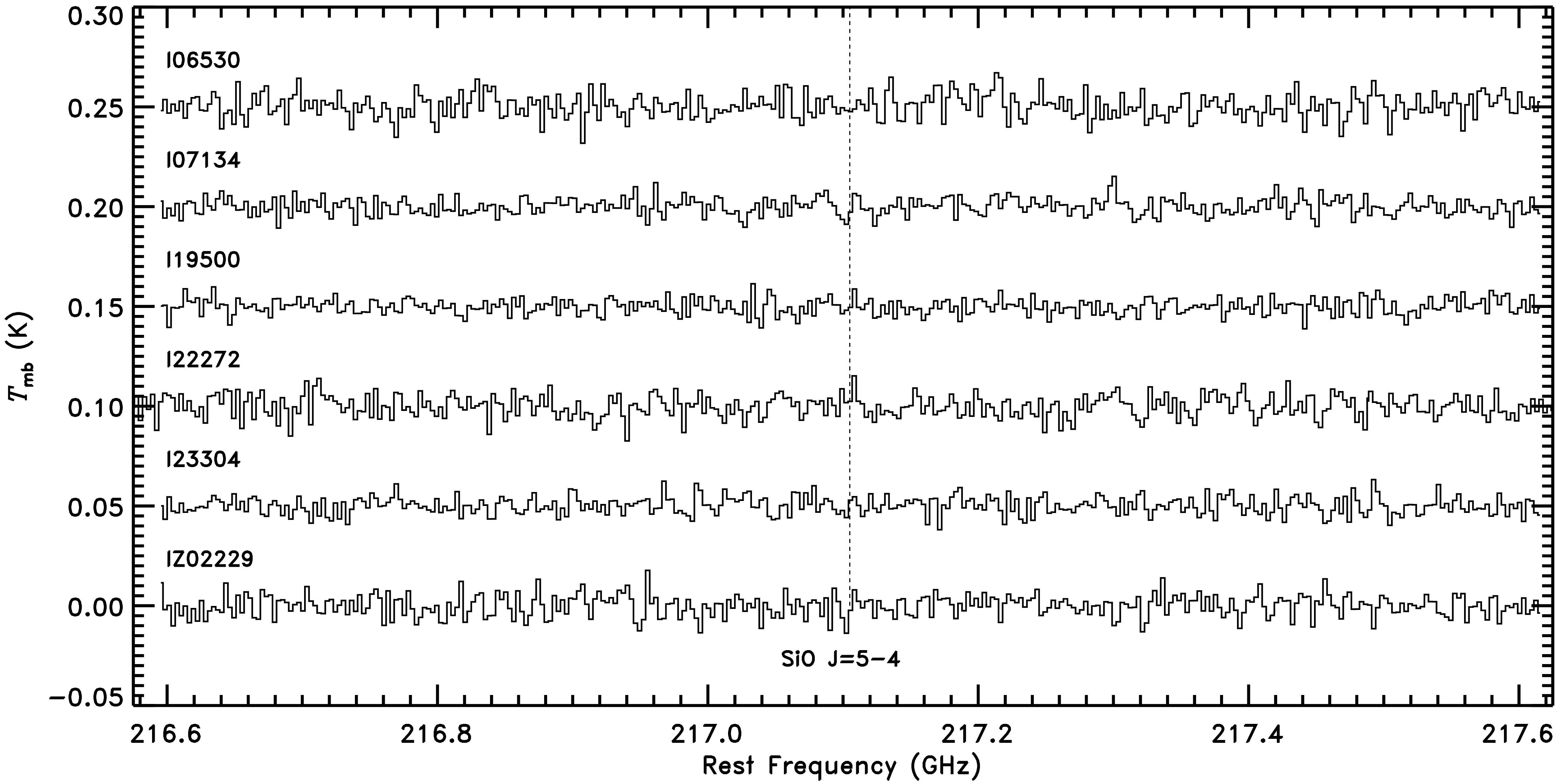}
\caption{Spectra of the 21\,$\mu$m sources obtained with the SMT 10\,m telescope. 
The dotted vertical lines mark the strong lines in the spectrum of CRL\,2688. 
To bring out the lines against the noise, three adjacent channels have been binned, 
resulting in a frequency resolutions of about 3\,MHz.
}
\label{Figure2}
\end{figure}

\renewcommand{\thefigure}{\arabic{figure} (Cont.)}
\addtocounter{figure}{-1}

\begin{figure}[htb]
\centering
\includegraphics[angle=0,scale=.21]{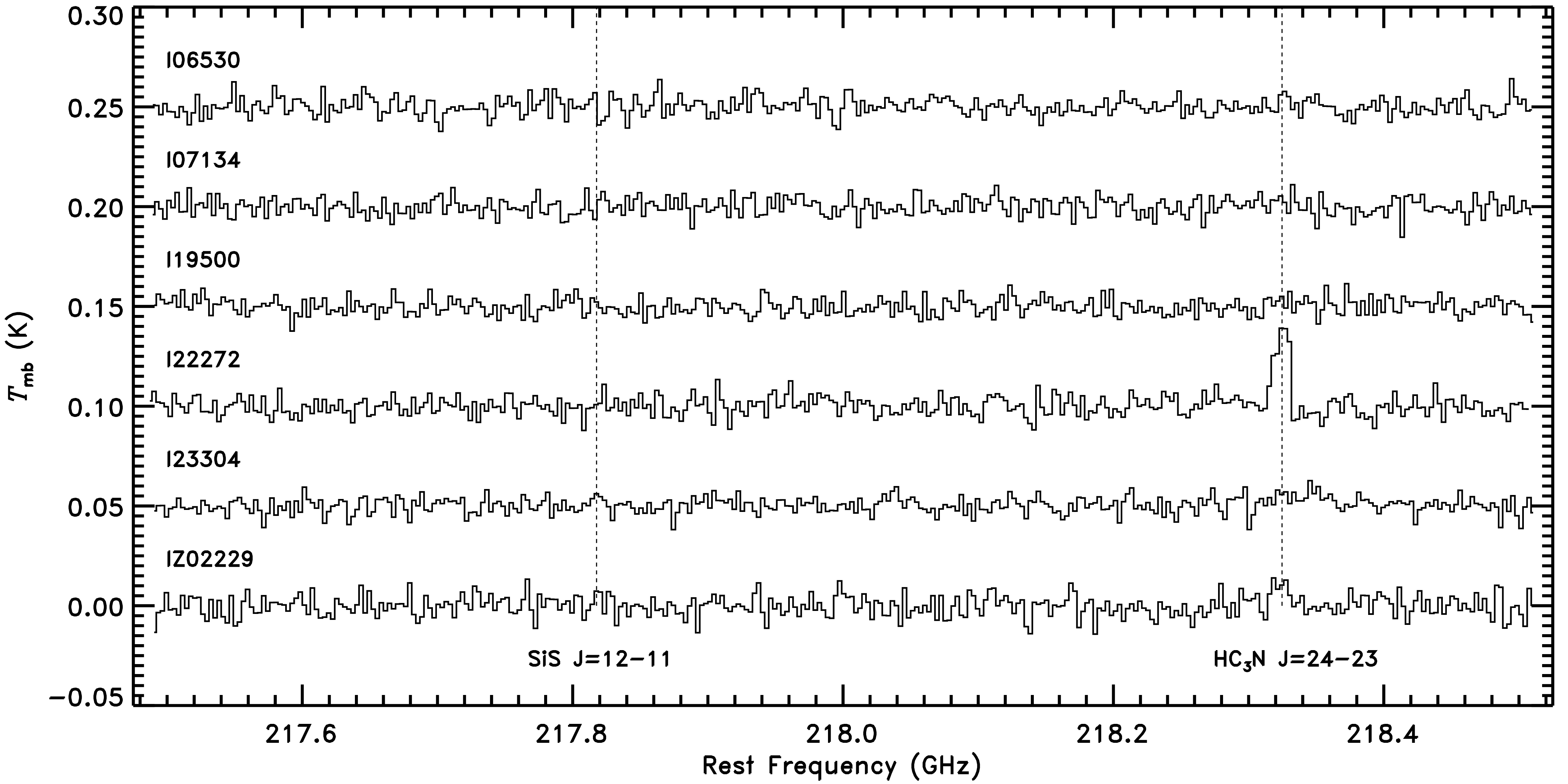}
\caption{}
\end{figure}

\renewcommand{\thefigure}{\arabic{figure} (Cont.)}
\addtocounter{figure}{-1}

\begin{figure}[htb]
\centering
\includegraphics[angle=0,scale=.21]{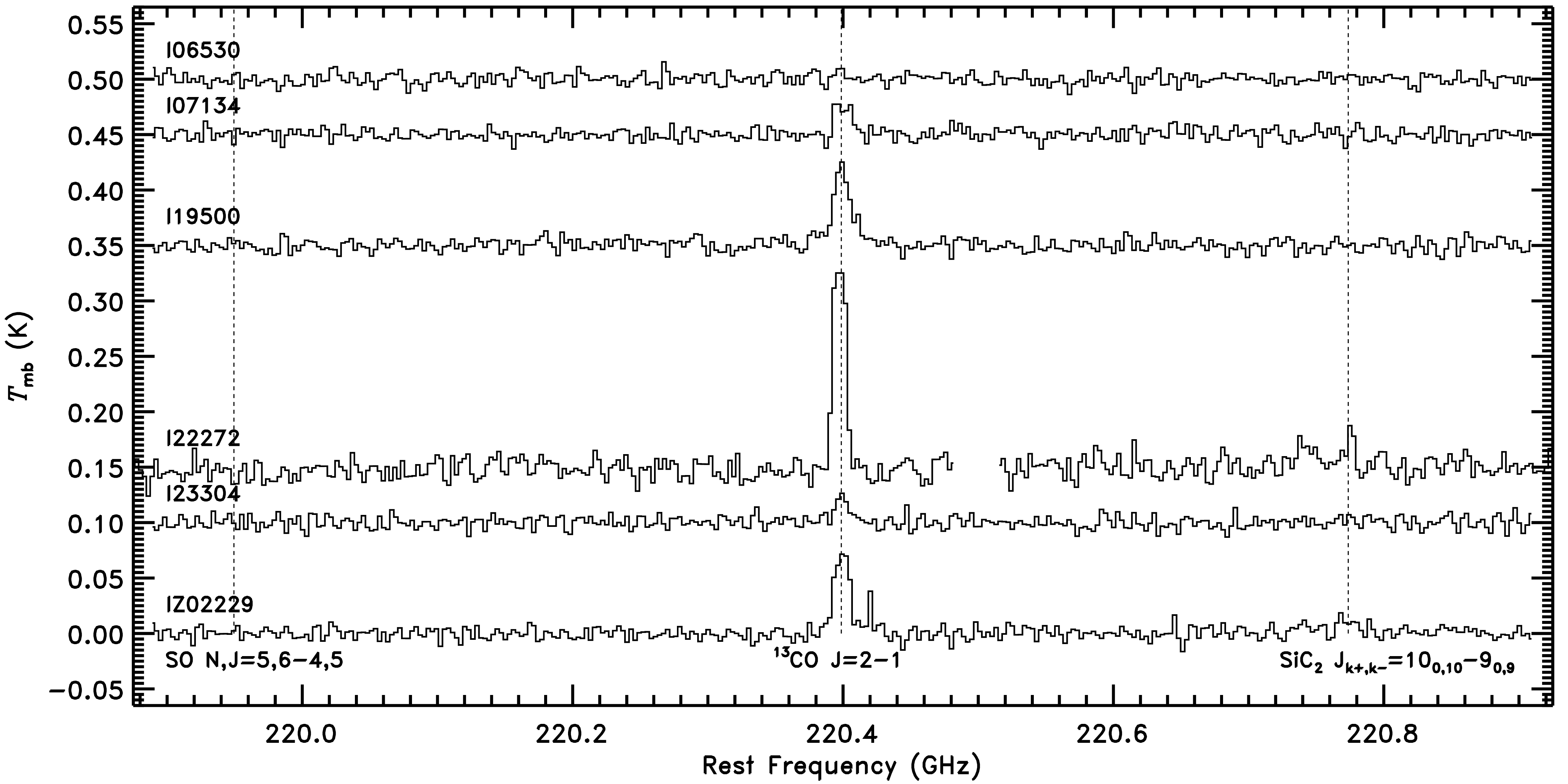}
\caption{}
\end{figure}

\renewcommand{\thefigure}{\arabic{figure} (Cont.)}
\addtocounter{figure}{-1}

\begin{figure}[htb]
\centering
\includegraphics[angle=0,scale=.21]{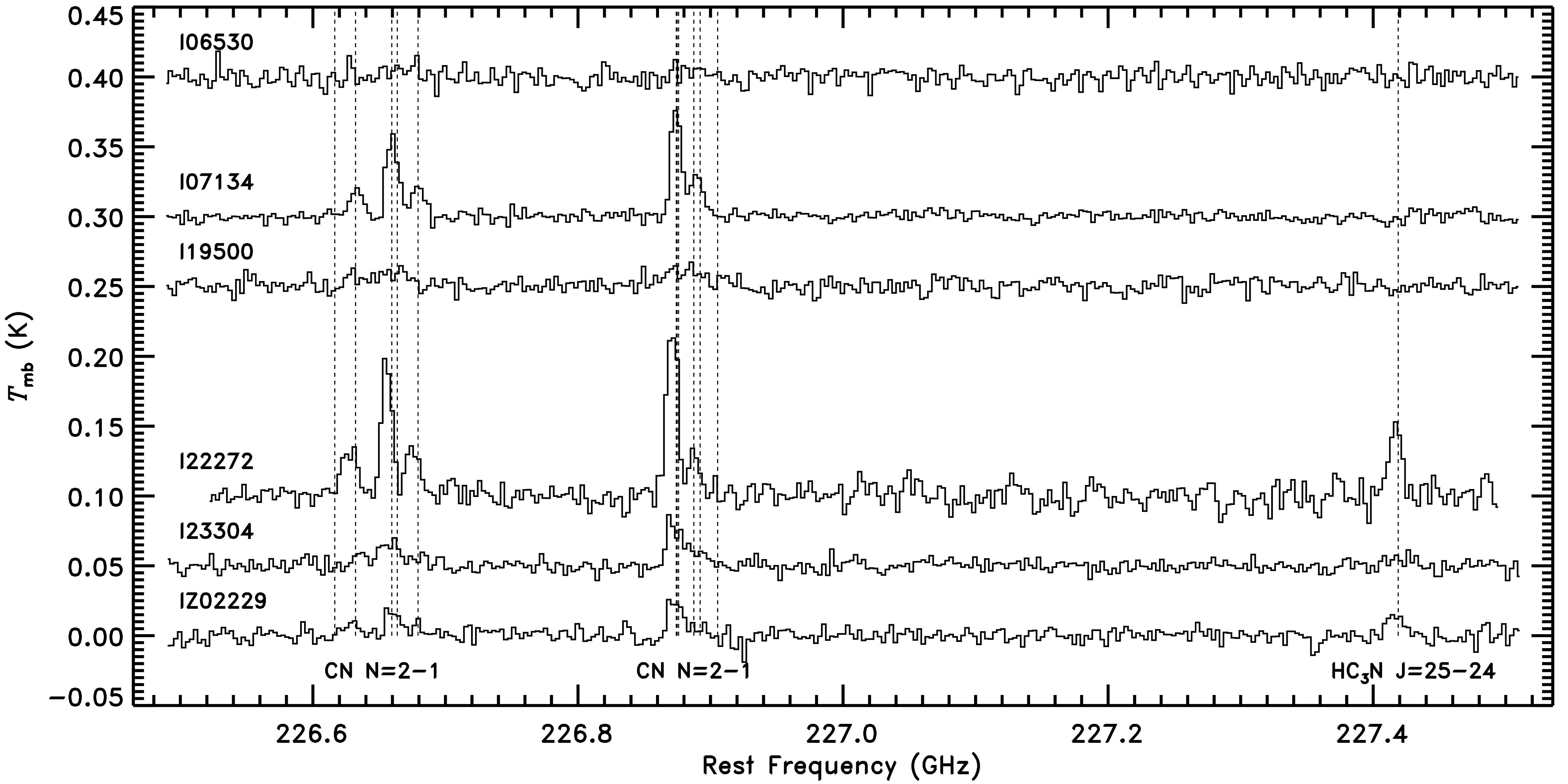}
\caption{}
\end{figure}

\renewcommand{\thefigure}{\arabic{figure} (Cont.)}
\addtocounter{figure}{-1}

\begin{figure}[htb]
\centering
\includegraphics[angle=0,scale=.21]{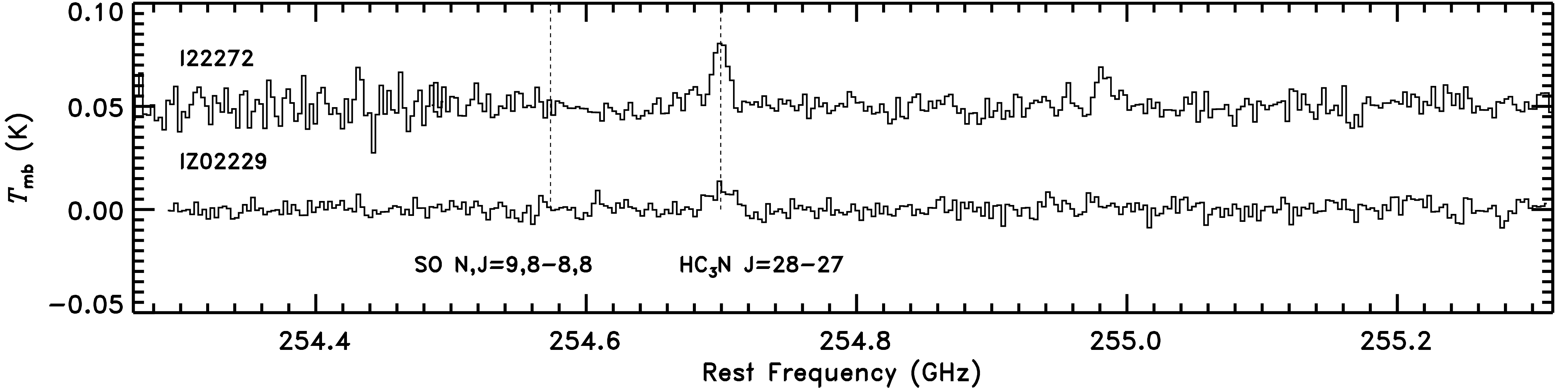}
\caption{}
\end{figure}

\renewcommand{\thefigure}{\arabic{figure} (Cont.)}
\addtocounter{figure}{-1}

\begin{figure}[htb]
\centering
\includegraphics[angle=0,scale=.21]{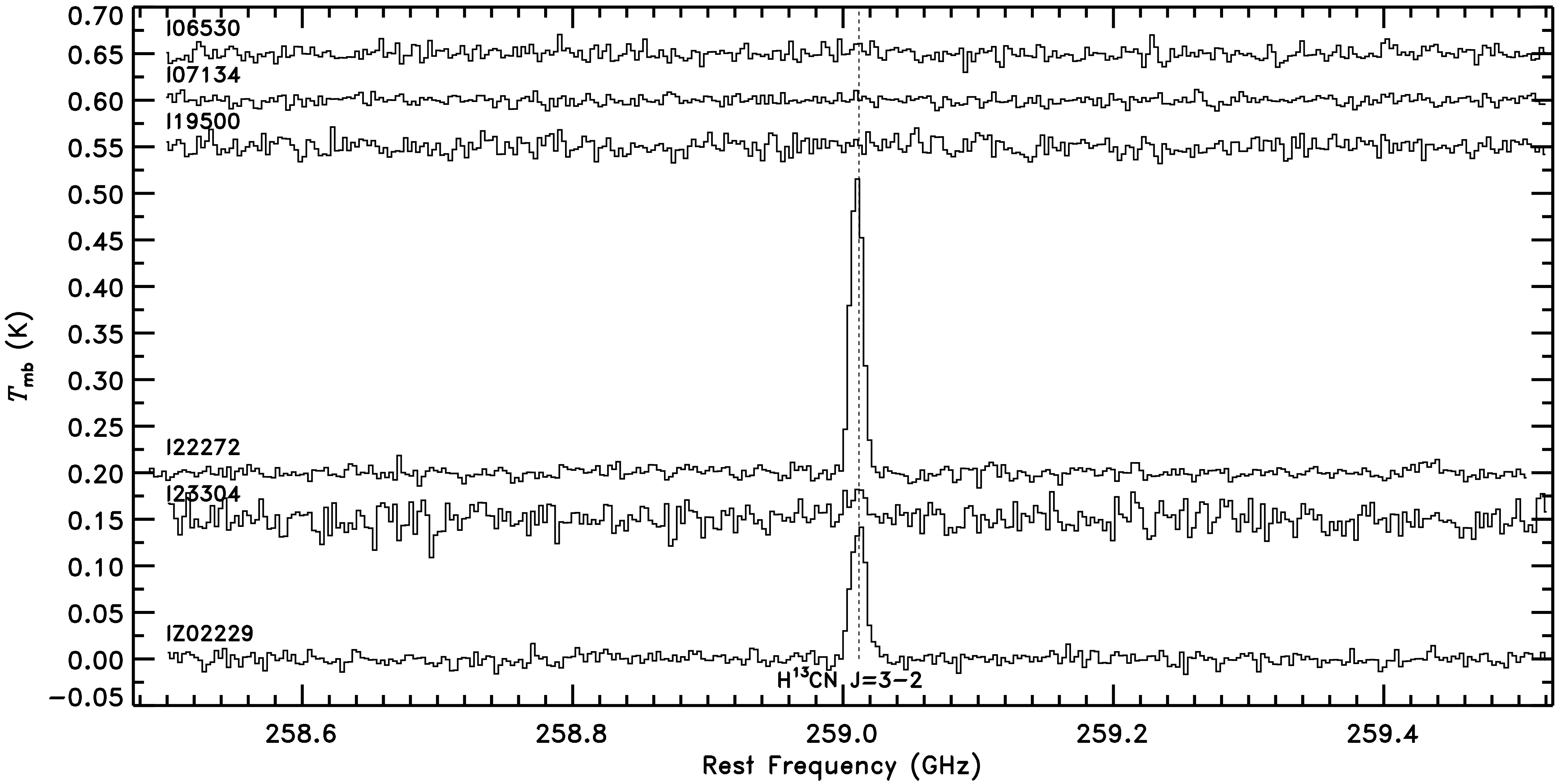}
\caption{}
\end{figure}

\renewcommand{\thefigure}{\arabic{figure} (Cont.)}
\addtocounter{figure}{-1}

\begin{figure}[htb]
\centering
\includegraphics[angle=0,scale=.21]{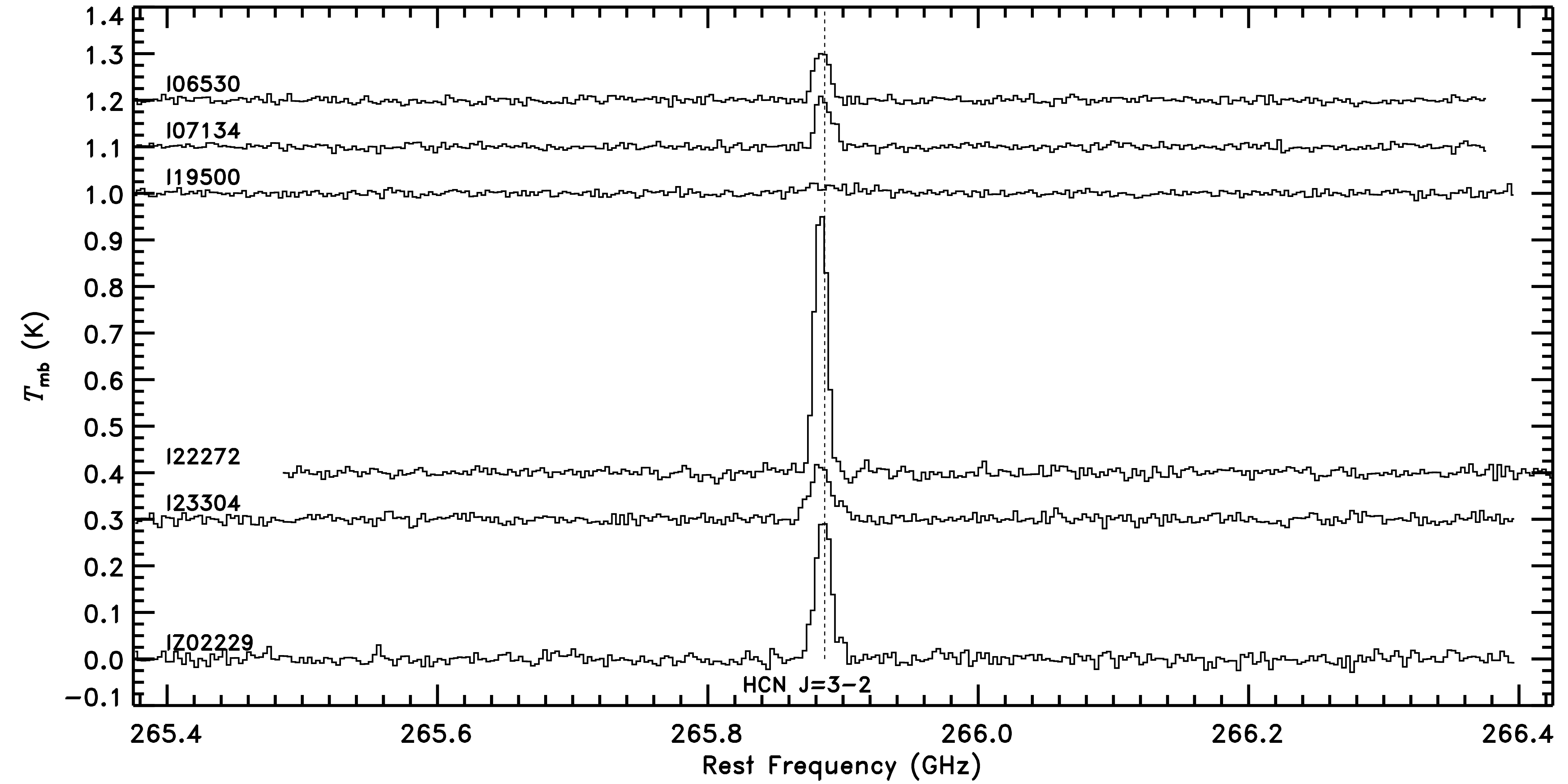}
\caption{}
\end{figure}

\renewcommand{\thefigure}{\arabic{figure} (Cont.)}
\addtocounter{figure}{-1}

\begin{figure}[htb]
\centering
\includegraphics[angle=0,scale=.21]{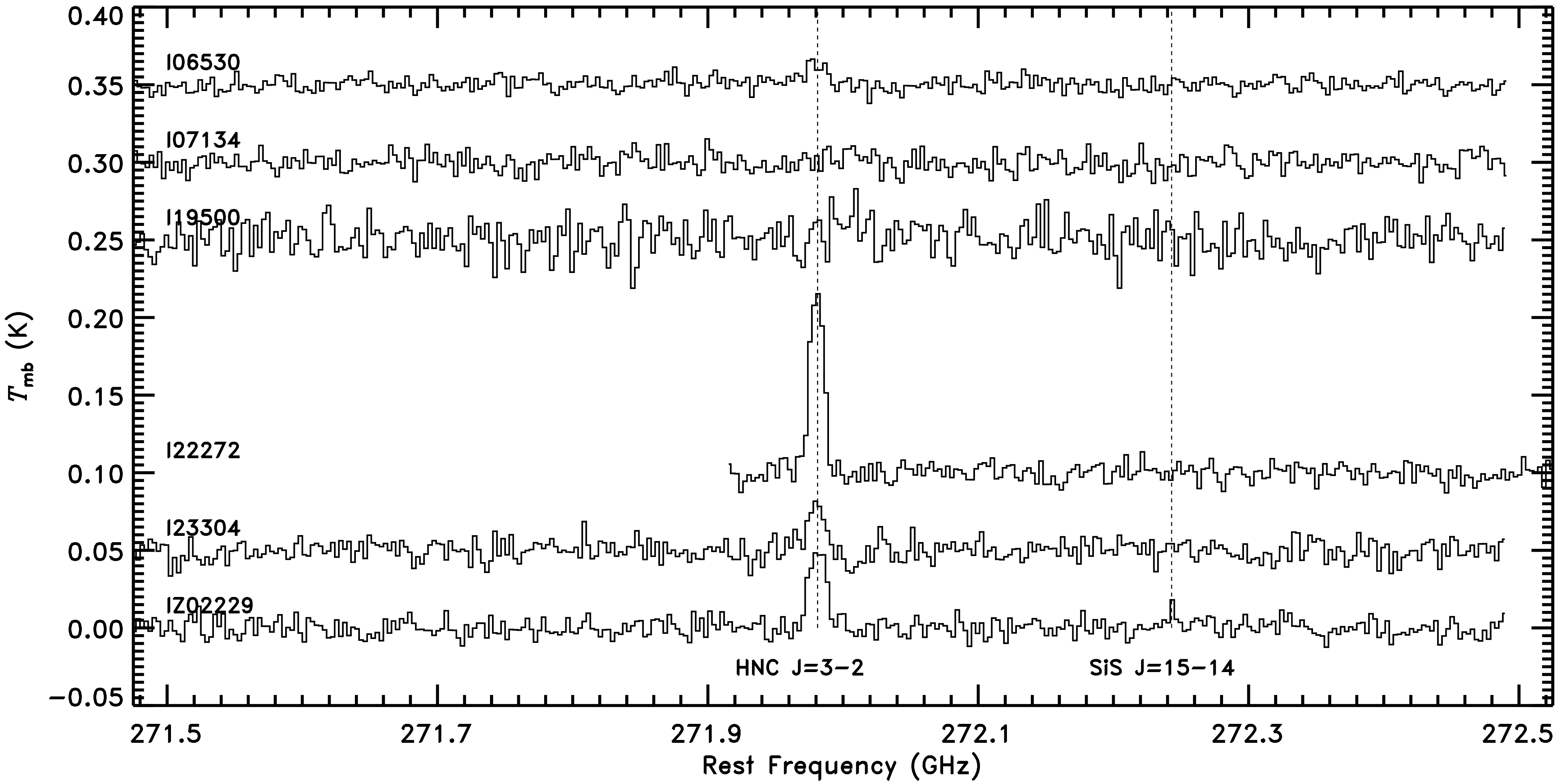}
\caption{}
\end{figure}

\renewcommand{\thefigure}{\arabic{figure}}

\begin{figure*}
\epsscale{.50}
\plotone{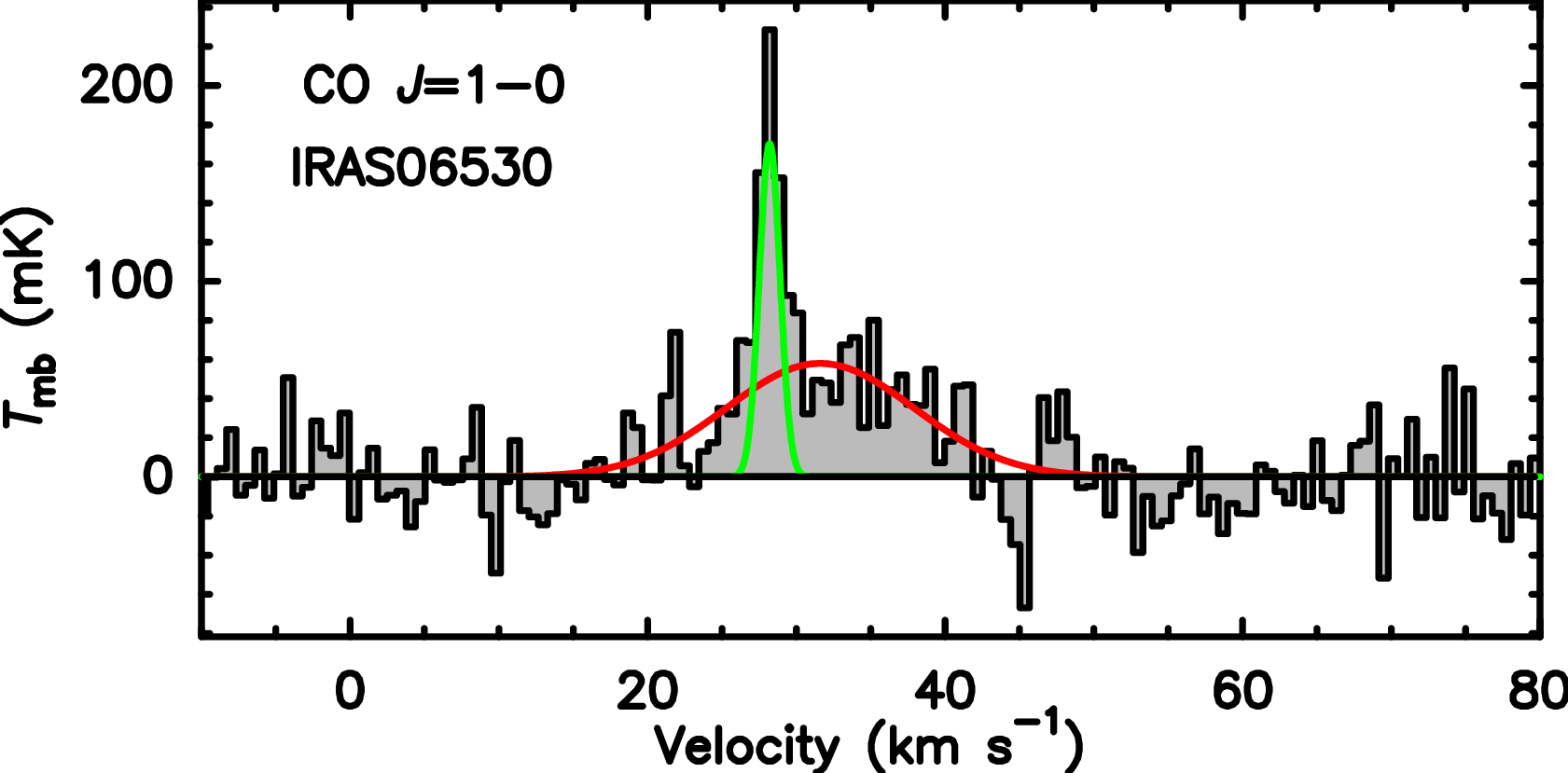}
\plotone{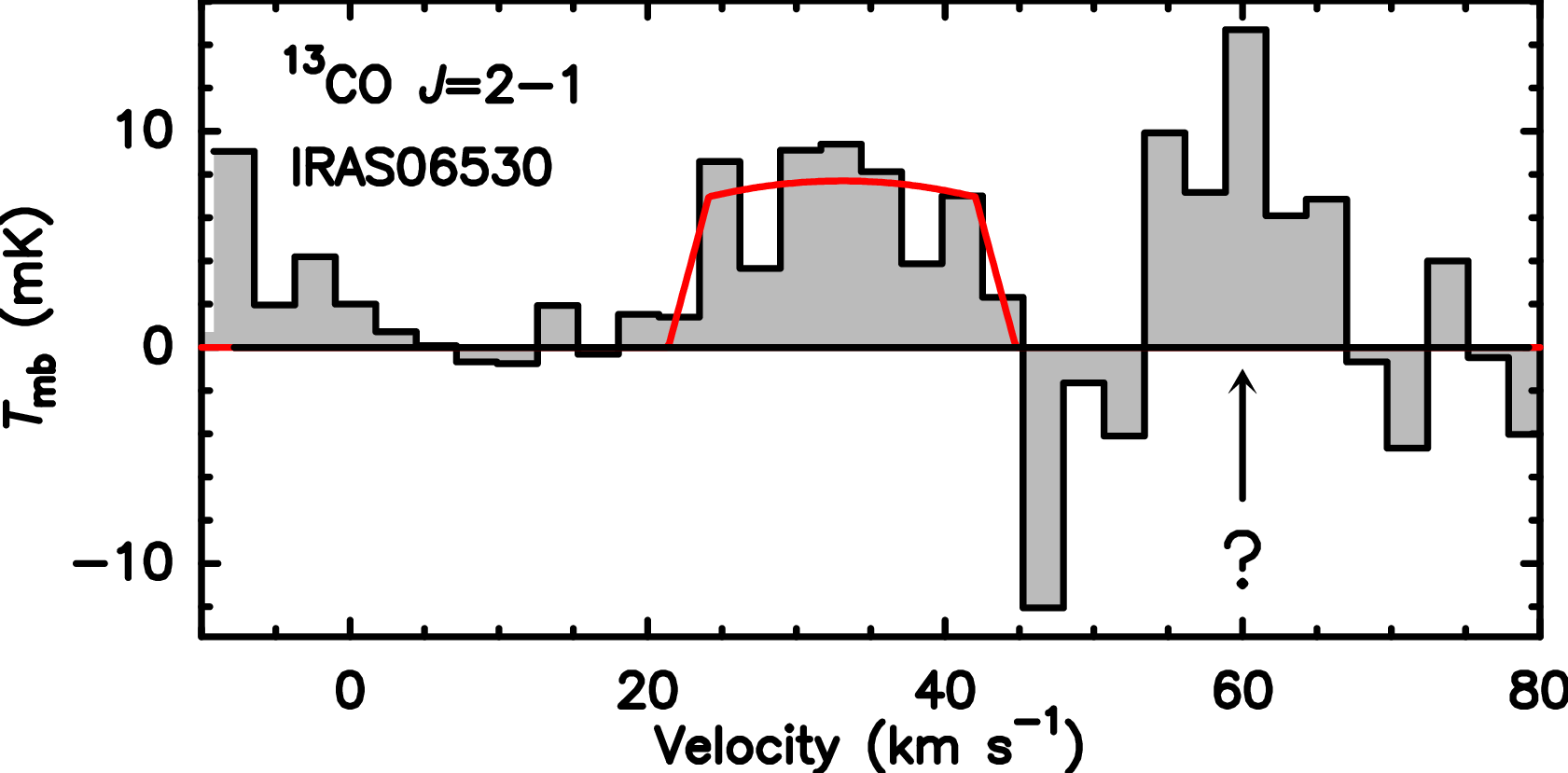}
\plotone{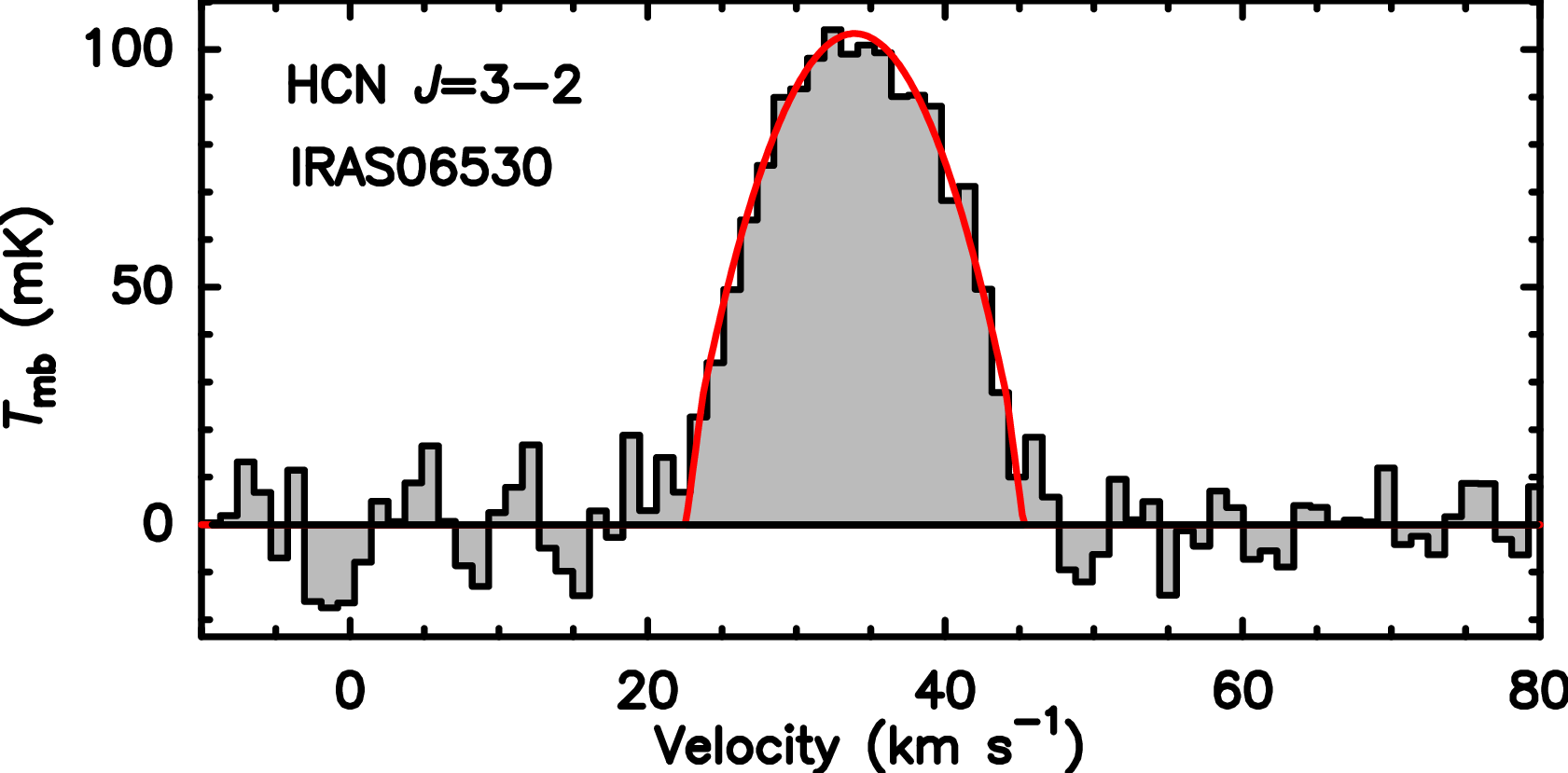}
\plotone{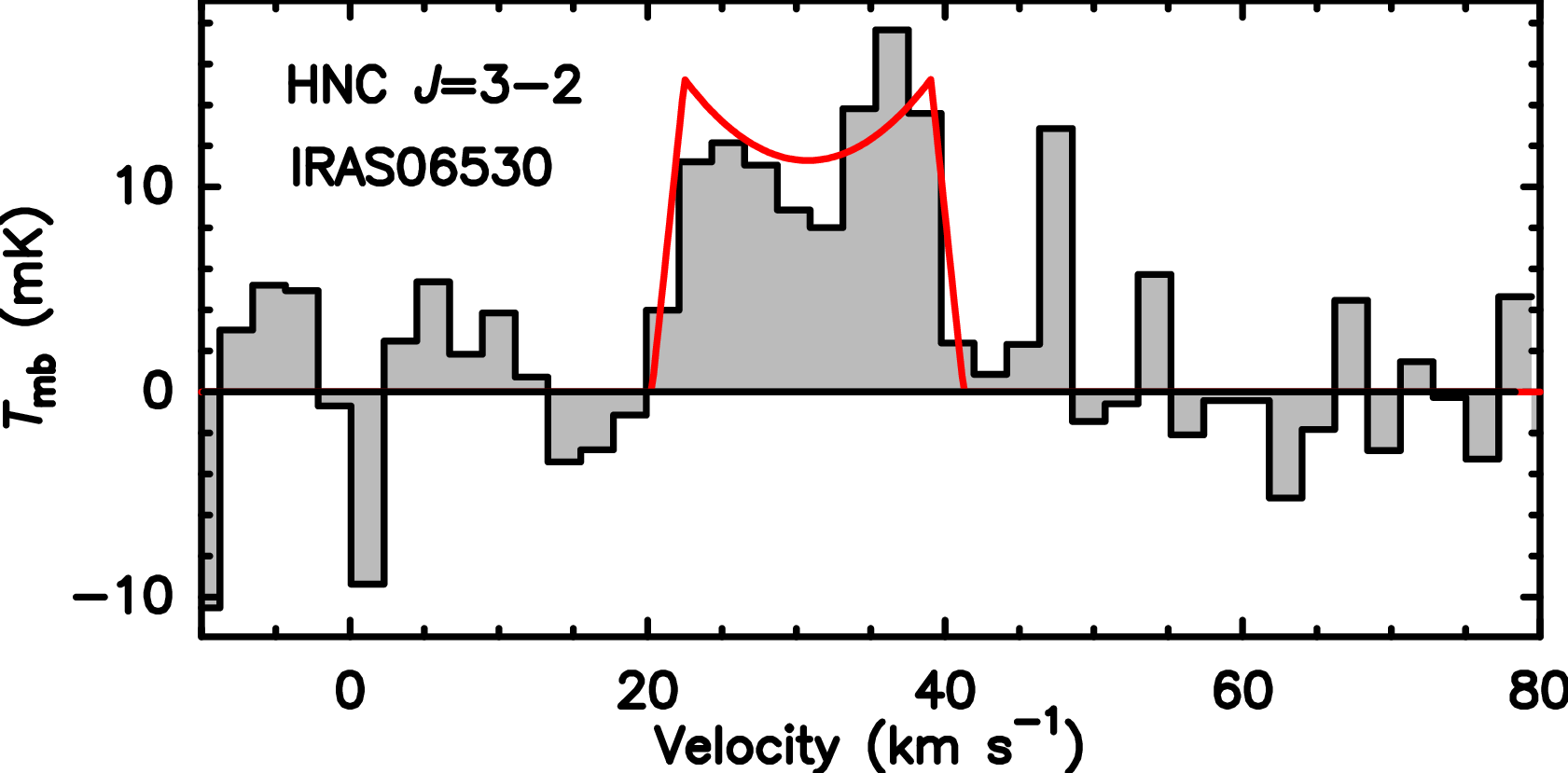}
\plotone{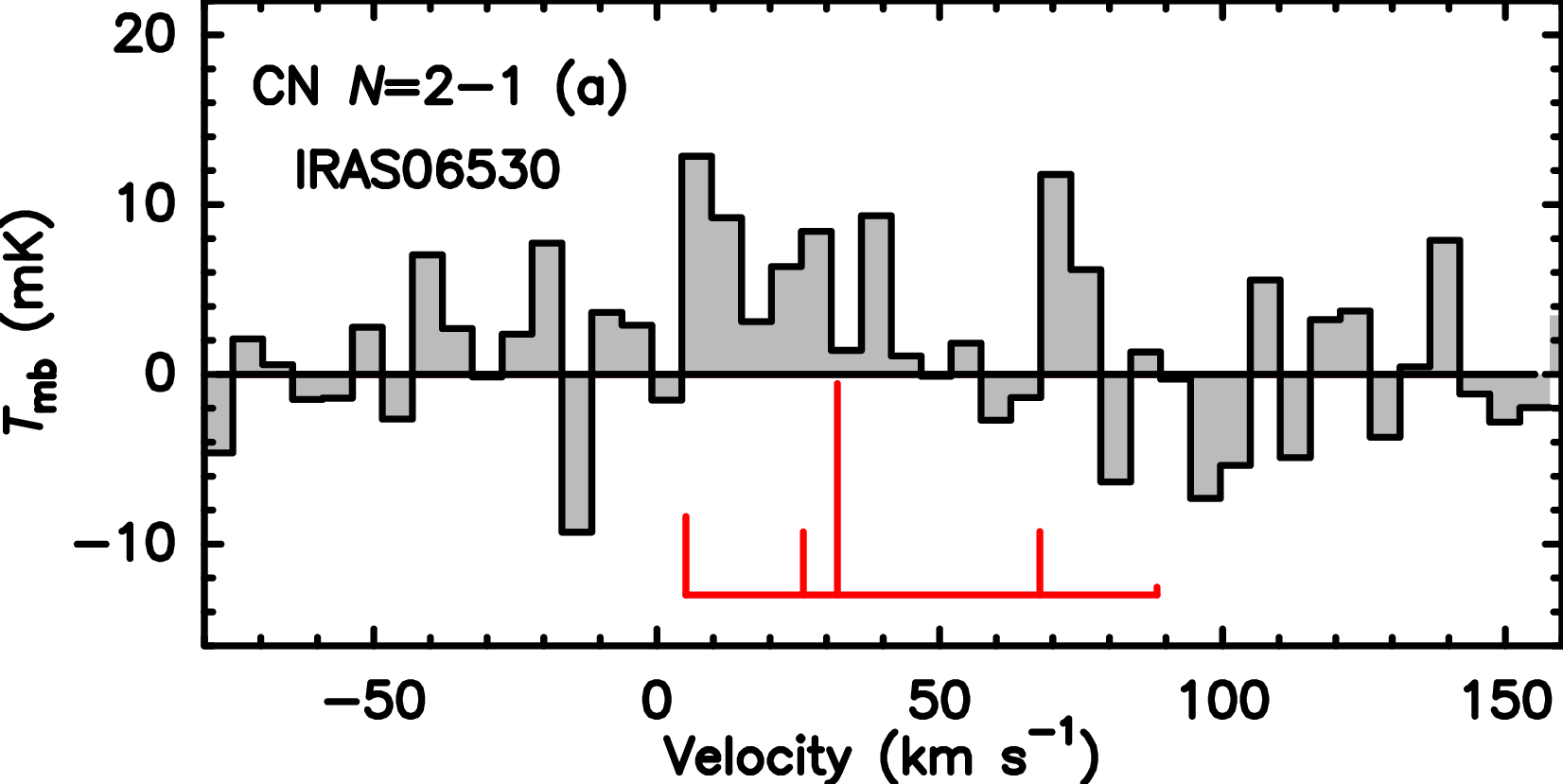}
\plotone{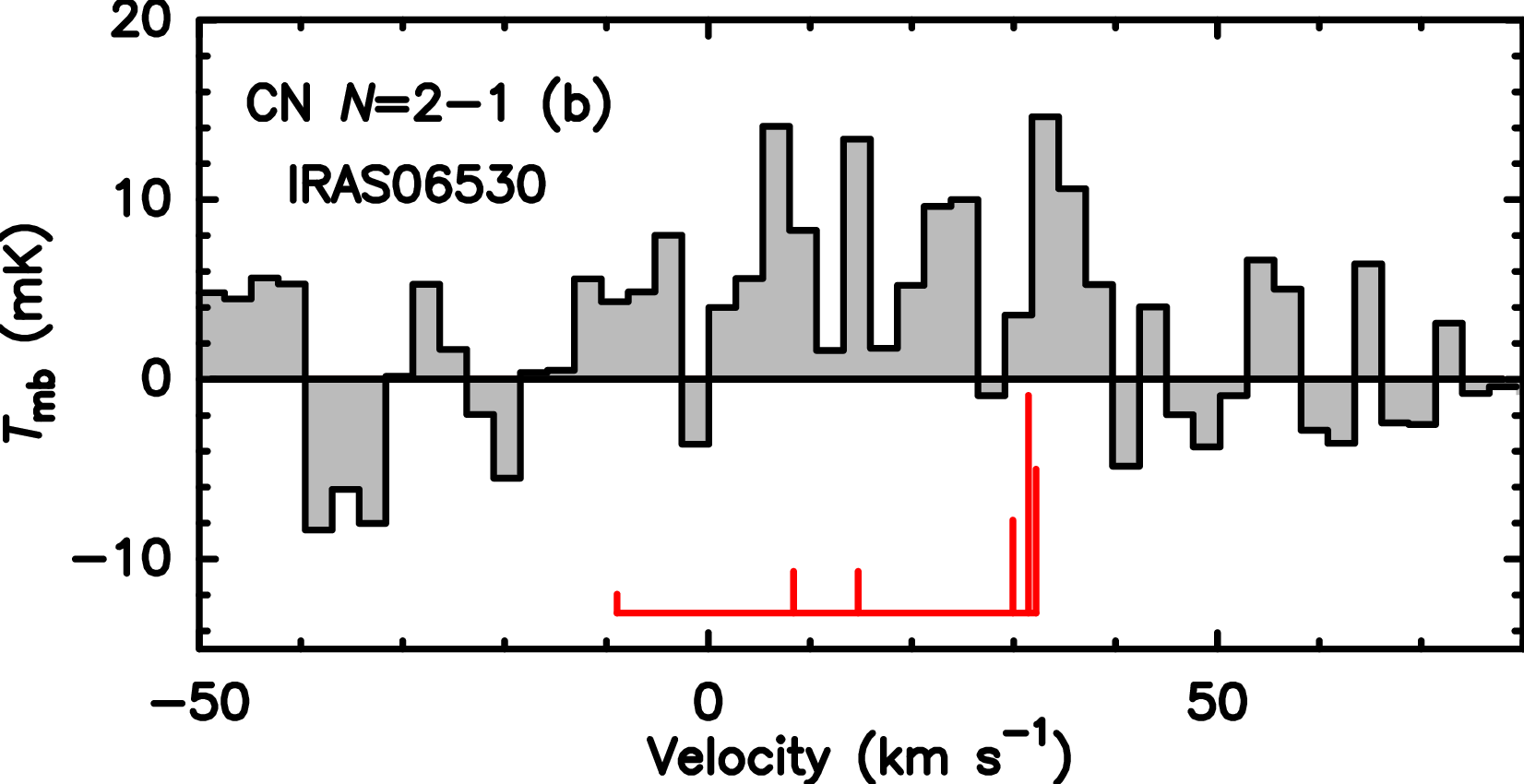}
\caption{The lines of CO, $^{13}$CO, HCN, HNC, and CN in IRAS\,06530-0213. 
The curves represent the stellar-shell fitting, 
except for the CO $J = 1 \rightarrow 0$ transition that is fitted with two Gaussian profiles. 
The vertical lines mark the positions and intrinsic relative intensities of hyperfine components. 
The velocity resolutions are listed in Tables~\ref{Table2}--\ref{Table7}.
For the CN $N = 2 \rightarrow 1$ transition, panel (a) shows the unsolved hyperfine structures of 
$J_{F} = 3/2_{1/2} \rightarrow 1/2_{3/2}$, 
$3/2_{3/2} \rightarrow 1/2_{3/2}$, 
$3/2_{5/2} \rightarrow 1/2_{3/2}$, 
$3/2_{1/2} \rightarrow 1/2_{1/2}$, and 
$3/2_{3/2} \rightarrow 1/2_{1/2}$ 
in the frequency range of 226.6--226.7\,GHz,
and panel (b) shows the unsolved hyperfine structures of 
$J_{F} = 5/2_{5/2} \rightarrow 3/2_{3/2}$,    
$5/2_{7/2} \rightarrow 3/2_{5/2}$, 
$5/2_{3/2} \rightarrow 3/2_{1/2}$, 
$5/2_{3/2} \rightarrow 3/2_{3/2}$, 
$5/2_{5/2} \rightarrow 3/2_{5/2}$, and 
$5/2_{3/2} \rightarrow 3/2,5/2$ 
in the frequency range of 226.8--226.9\,GHz.
}
\label{Figure3}
\end{figure*}

\begin{figure*}
\epsscale{.50}
\plotone{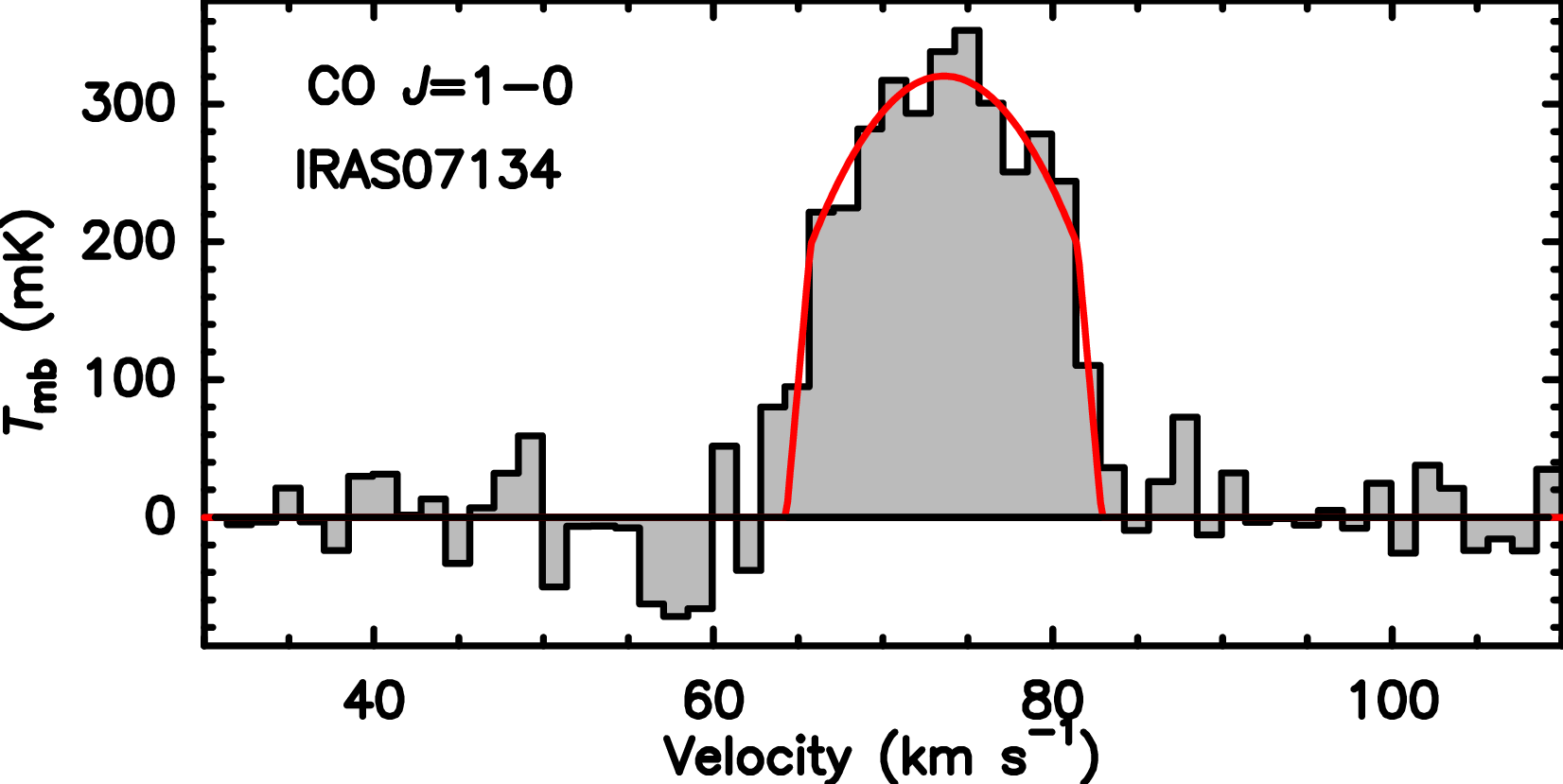}
\plotone{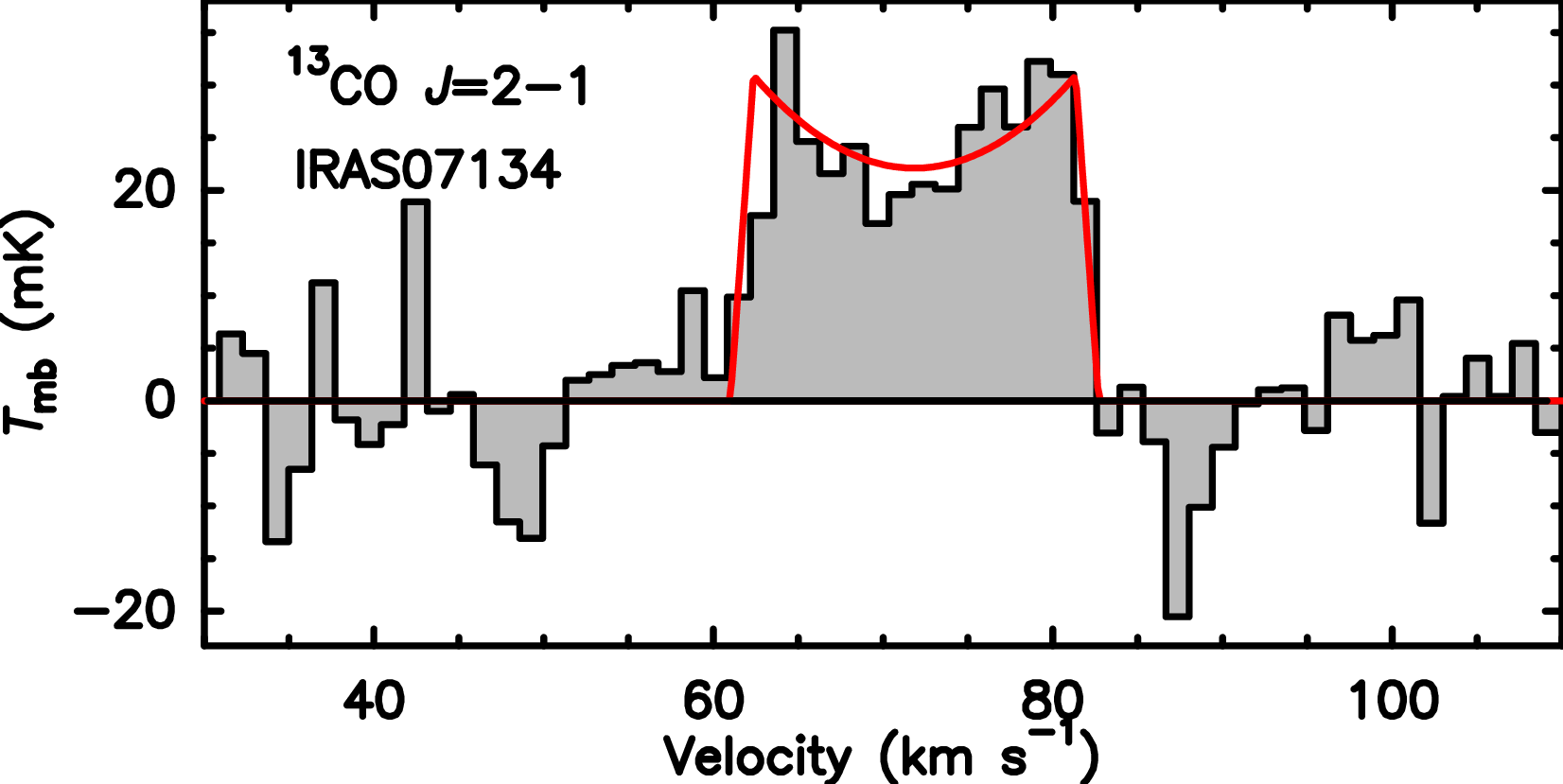}
\plotone{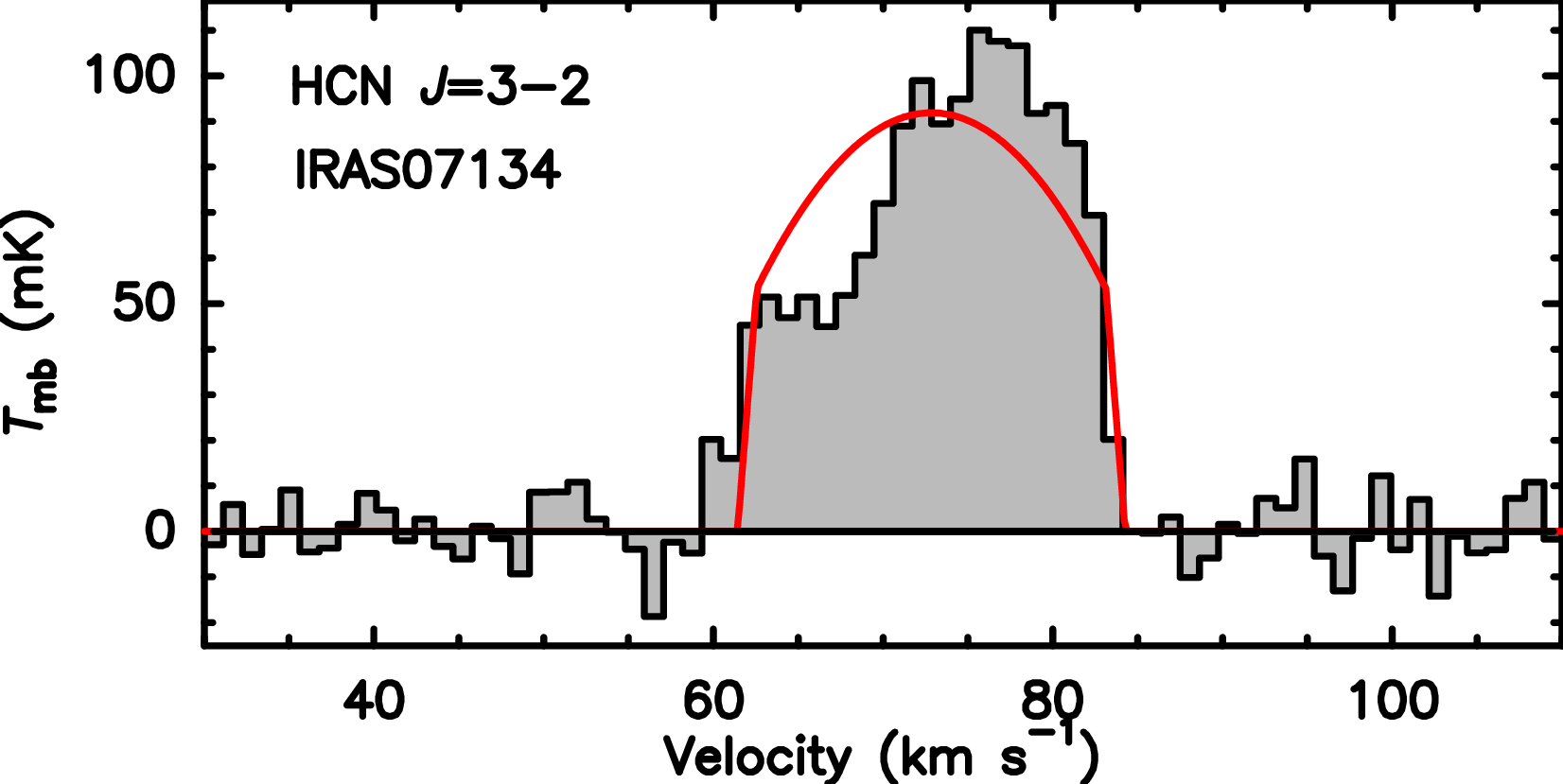}
\plotone{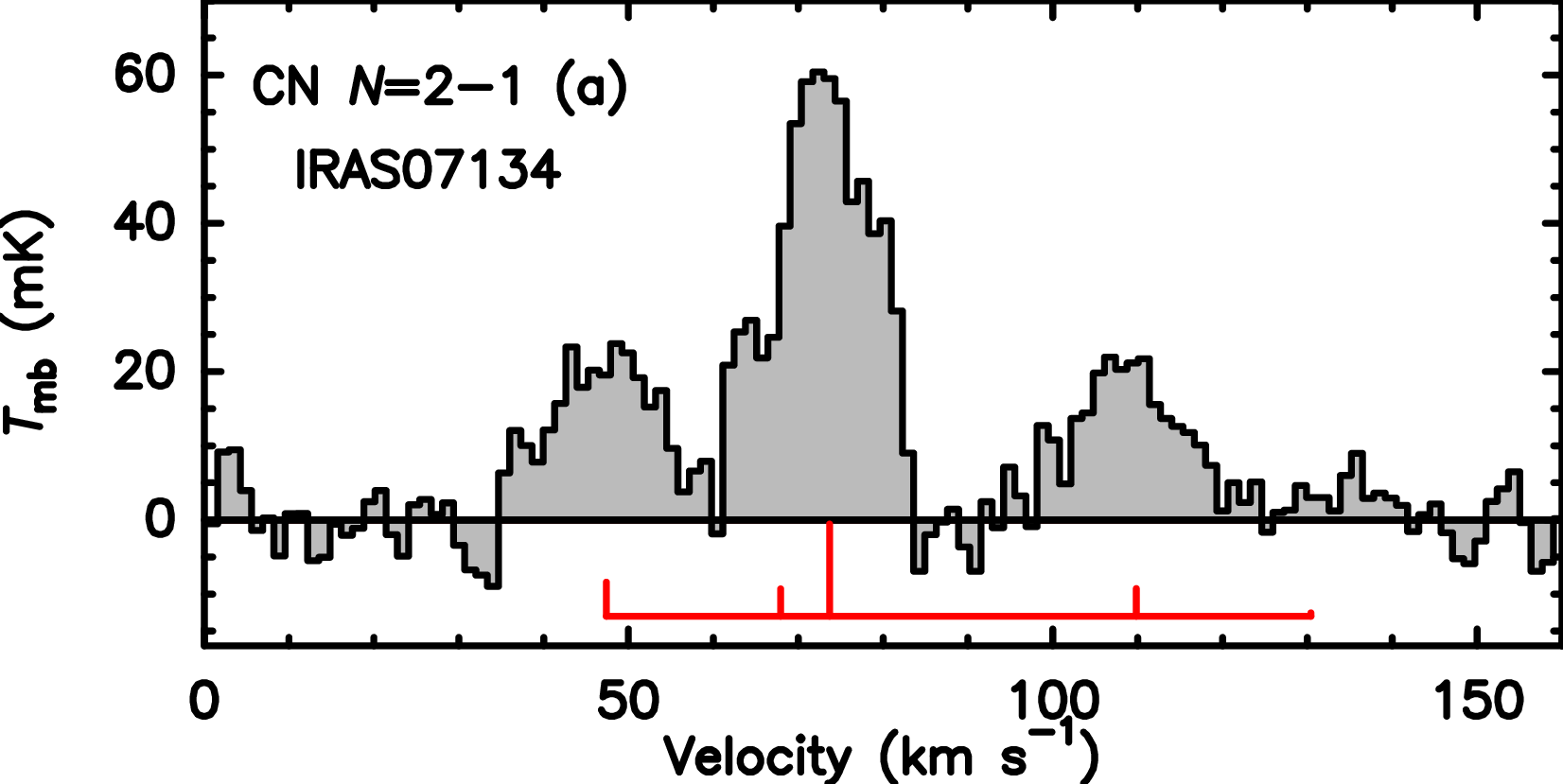}
\plotone{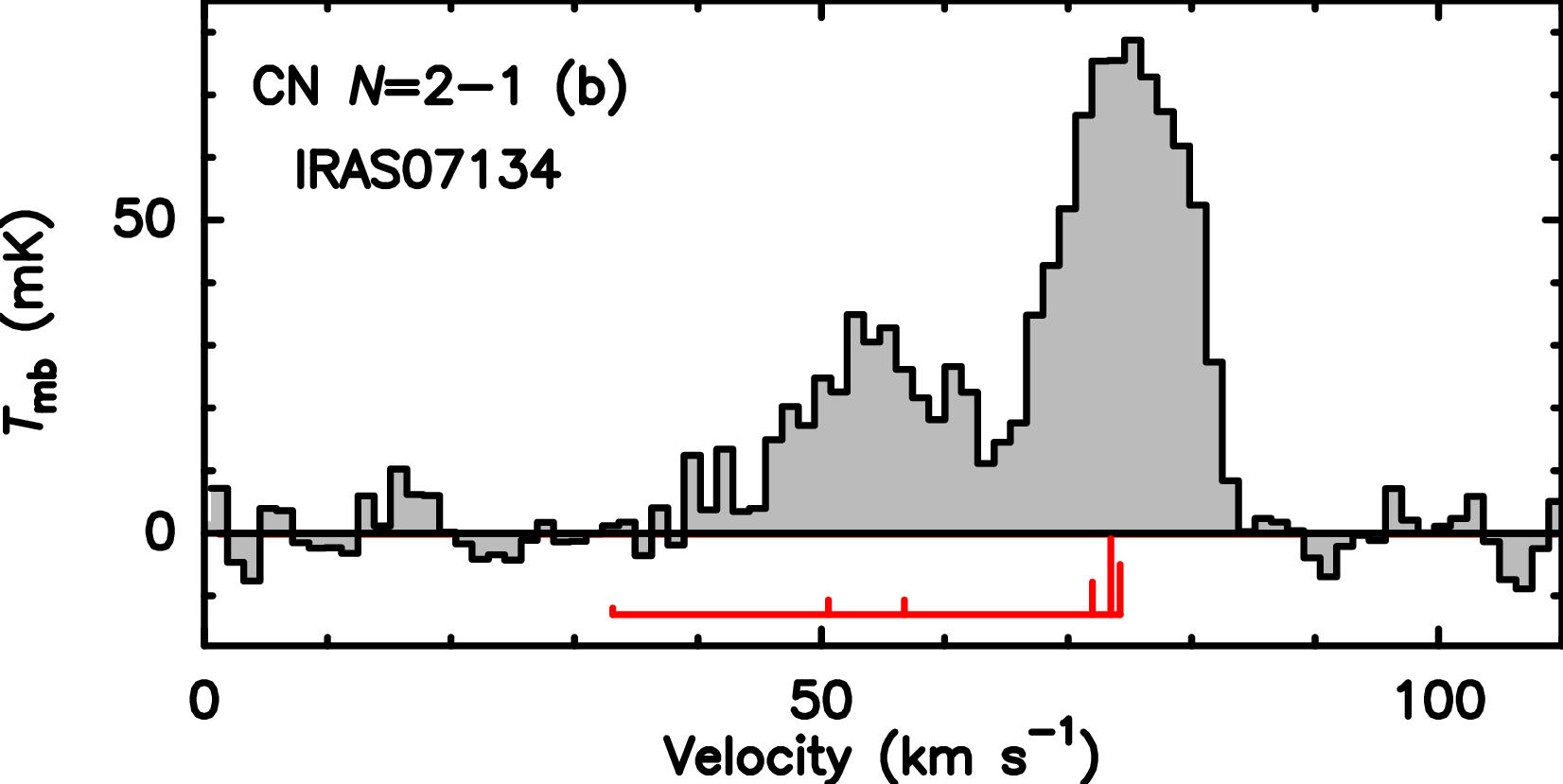}
\caption{Same as Fig.~\ref{Figure3}, but for CO, $^{13}$CO, HCN, and CN in IRAS\,07134+1005.
}
\label{Figure4}
\end{figure*}

\begin{figure*}
\epsscale{.50}
\plotone{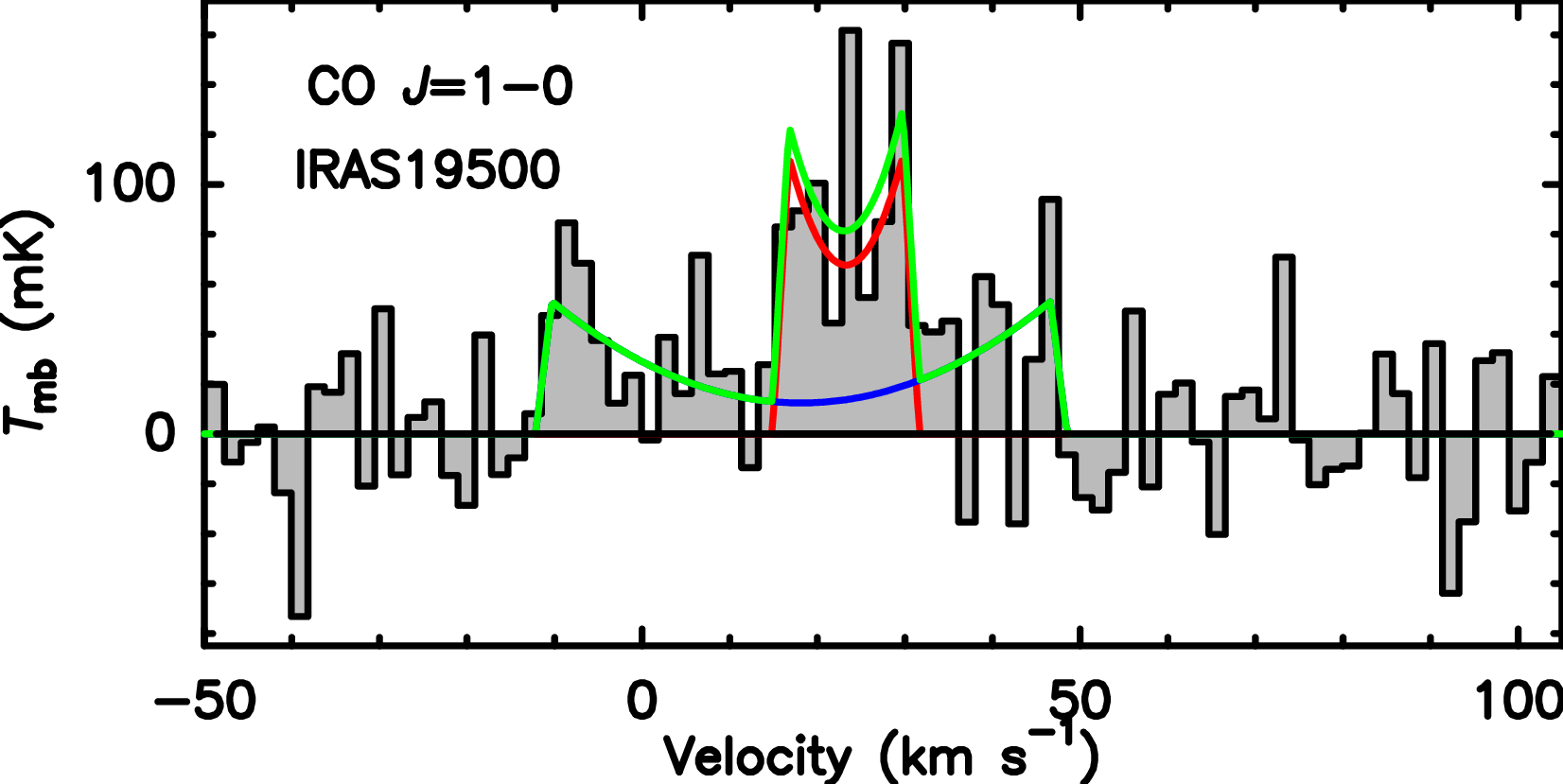}
\plotone{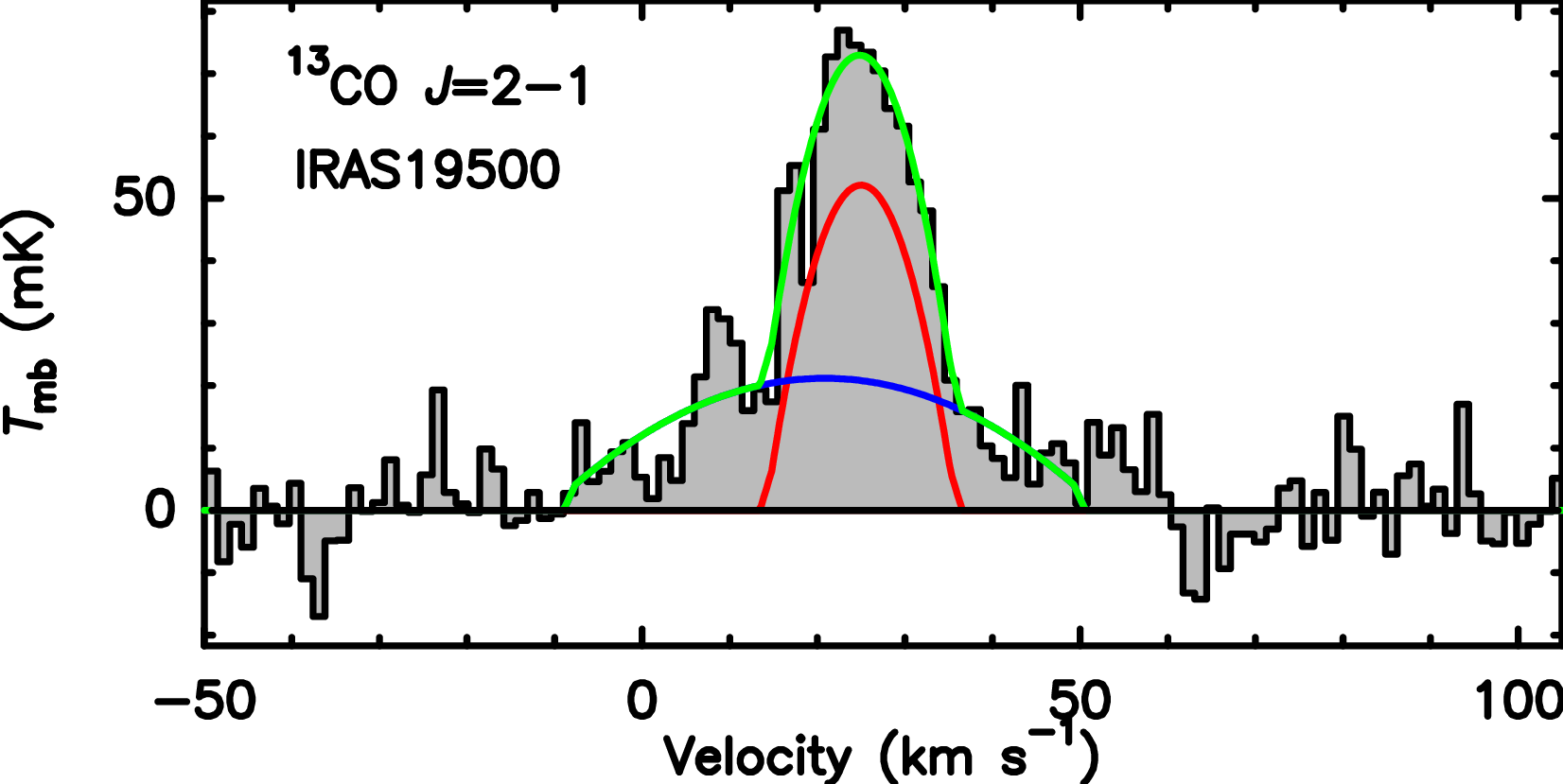}
\plotone{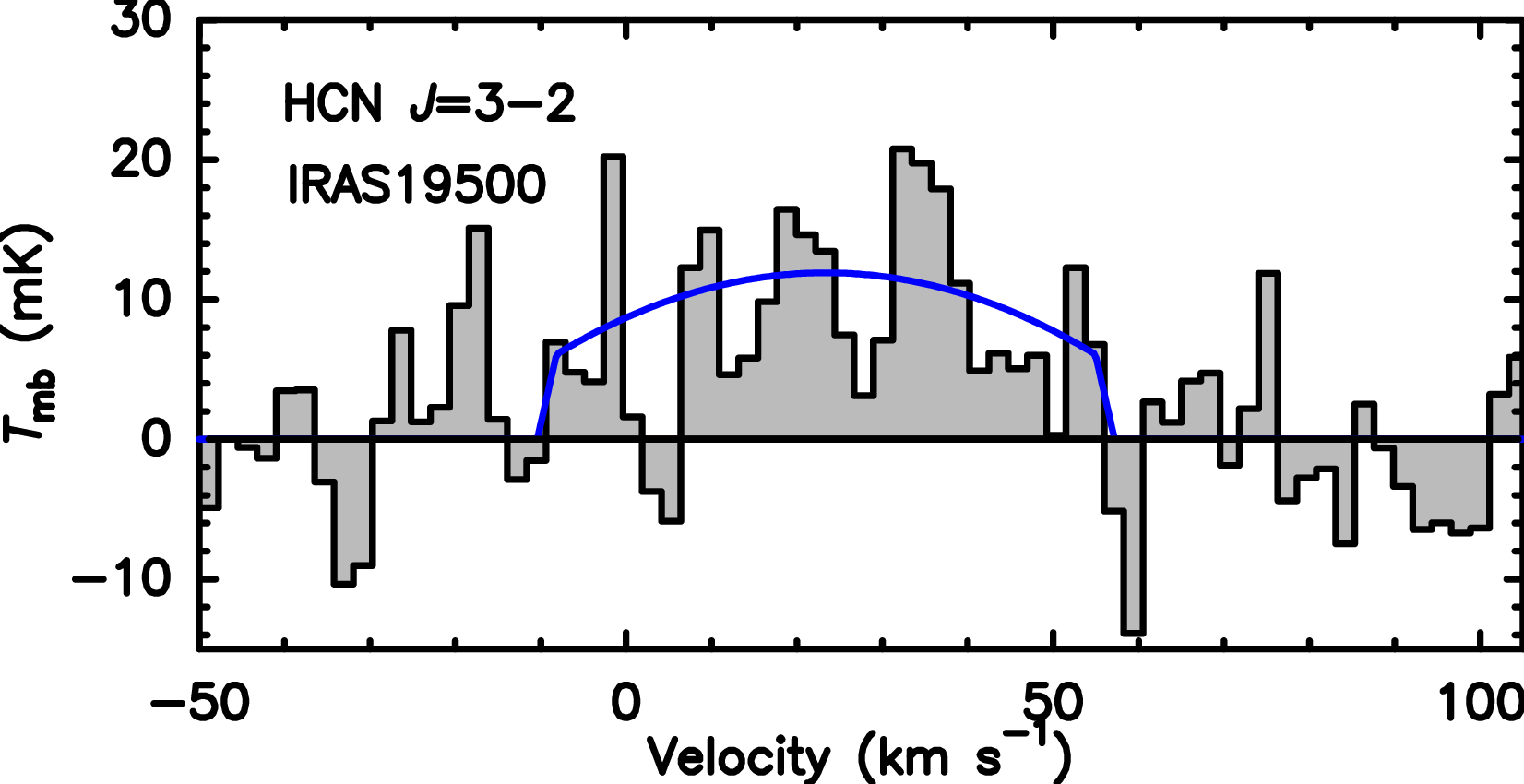}
\plotone{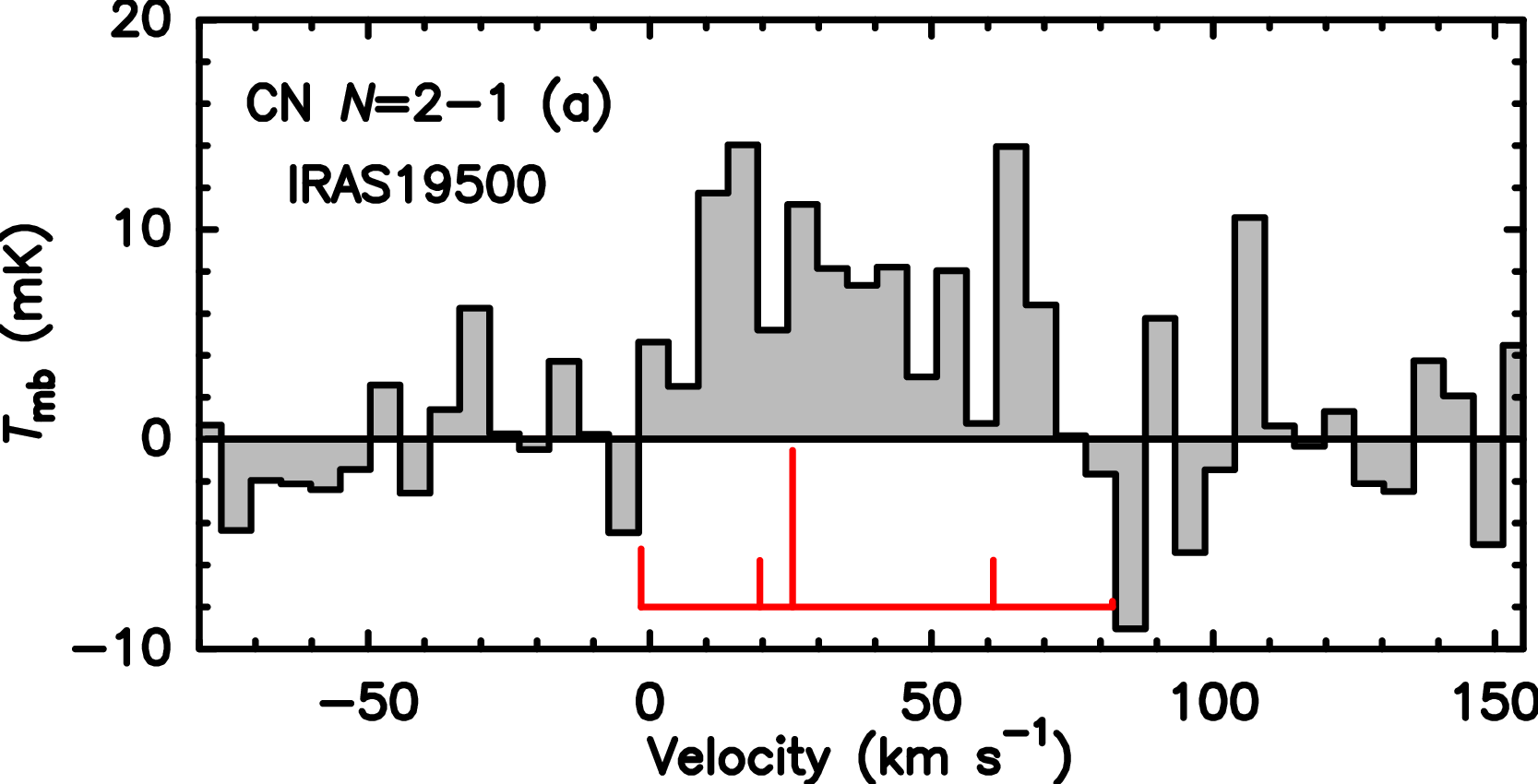}
\plotone{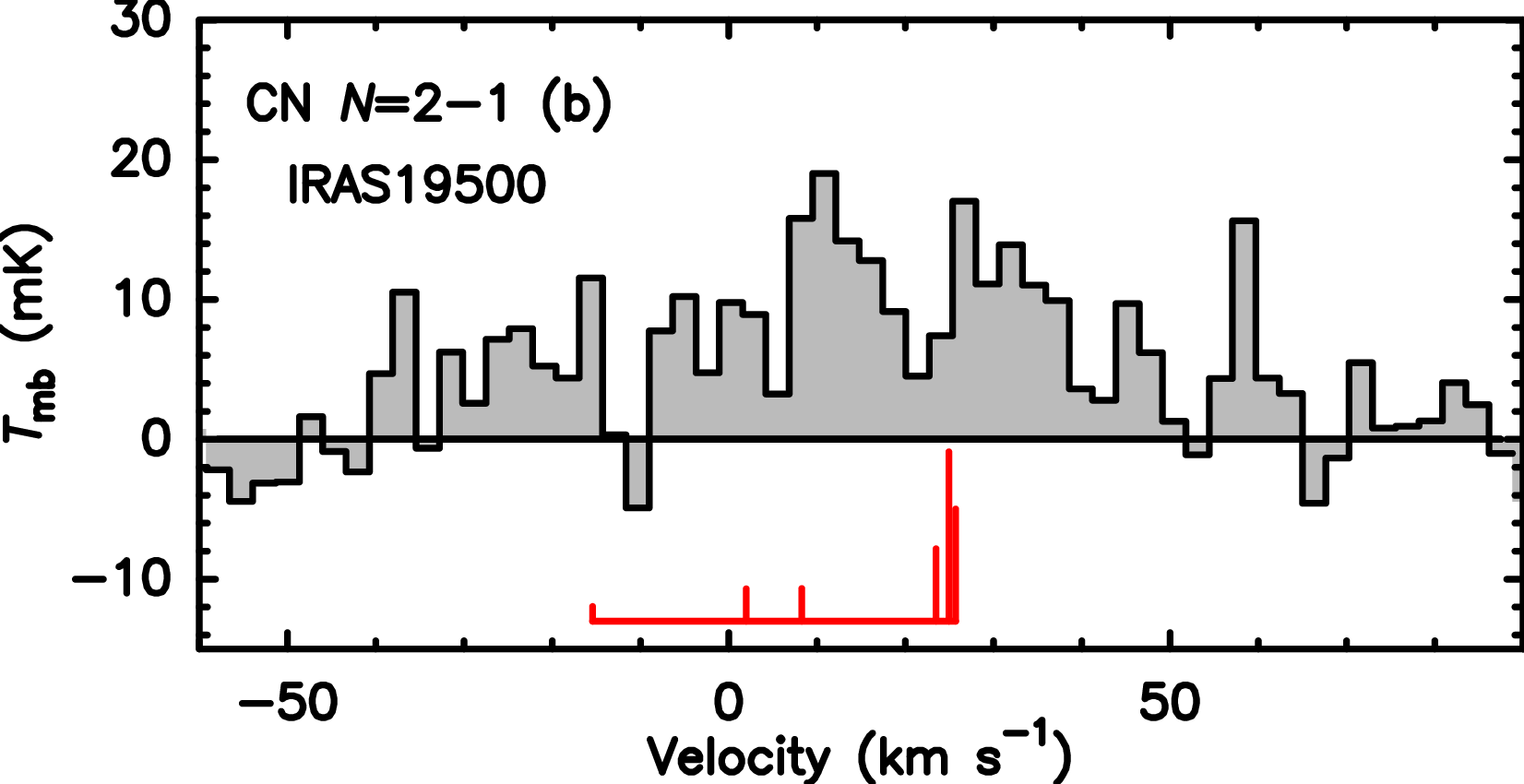}
\caption{Same as Fig.~\ref{Figure3}, but for CO, $^{13}$CO, HCN, HNC, and CN in IRAS\,19500-1709.}
\label{Figure5}
\end{figure*}

\begin{figure*}
\epsscale{.50}
\plotone{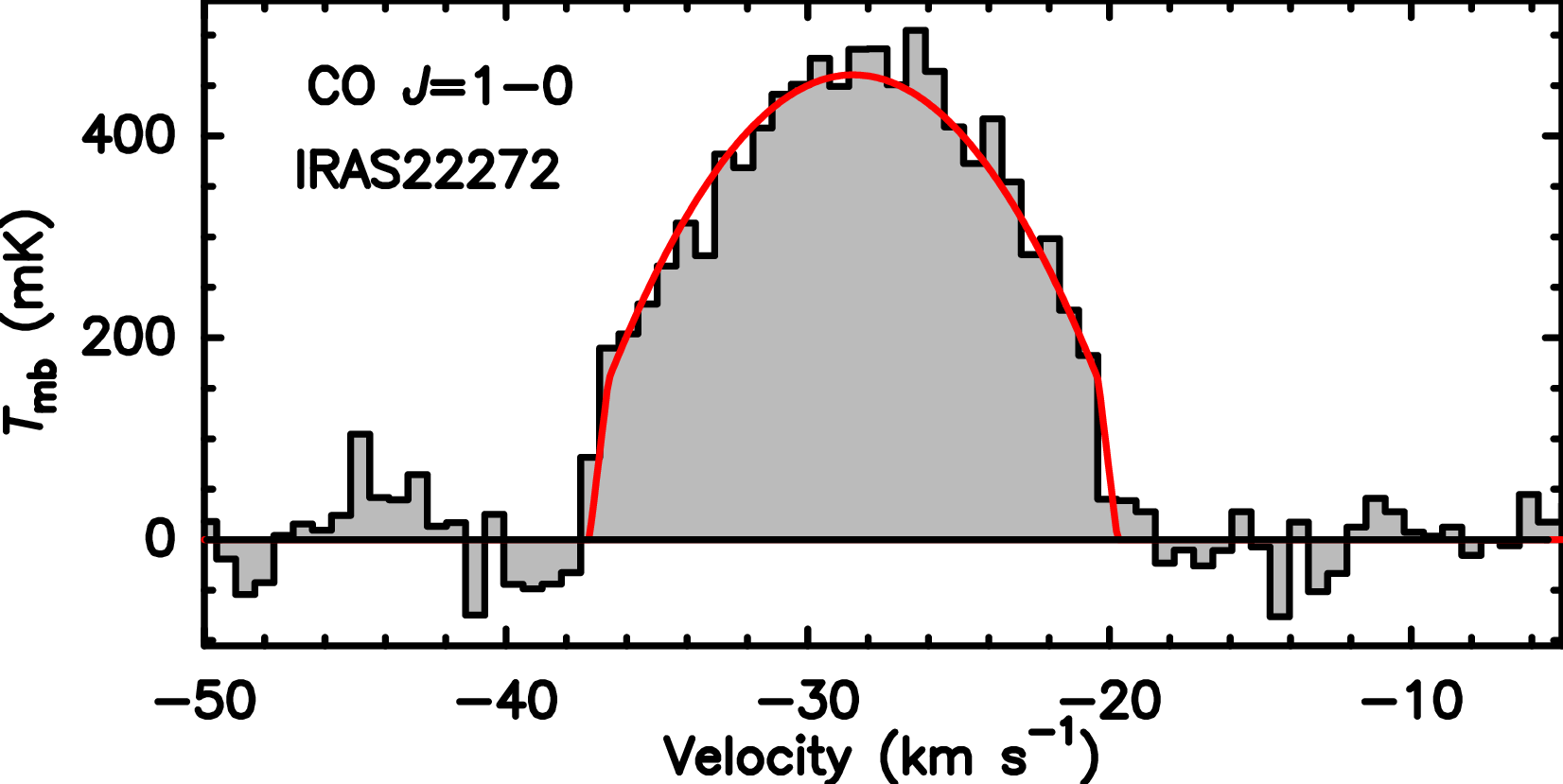}
\plotone{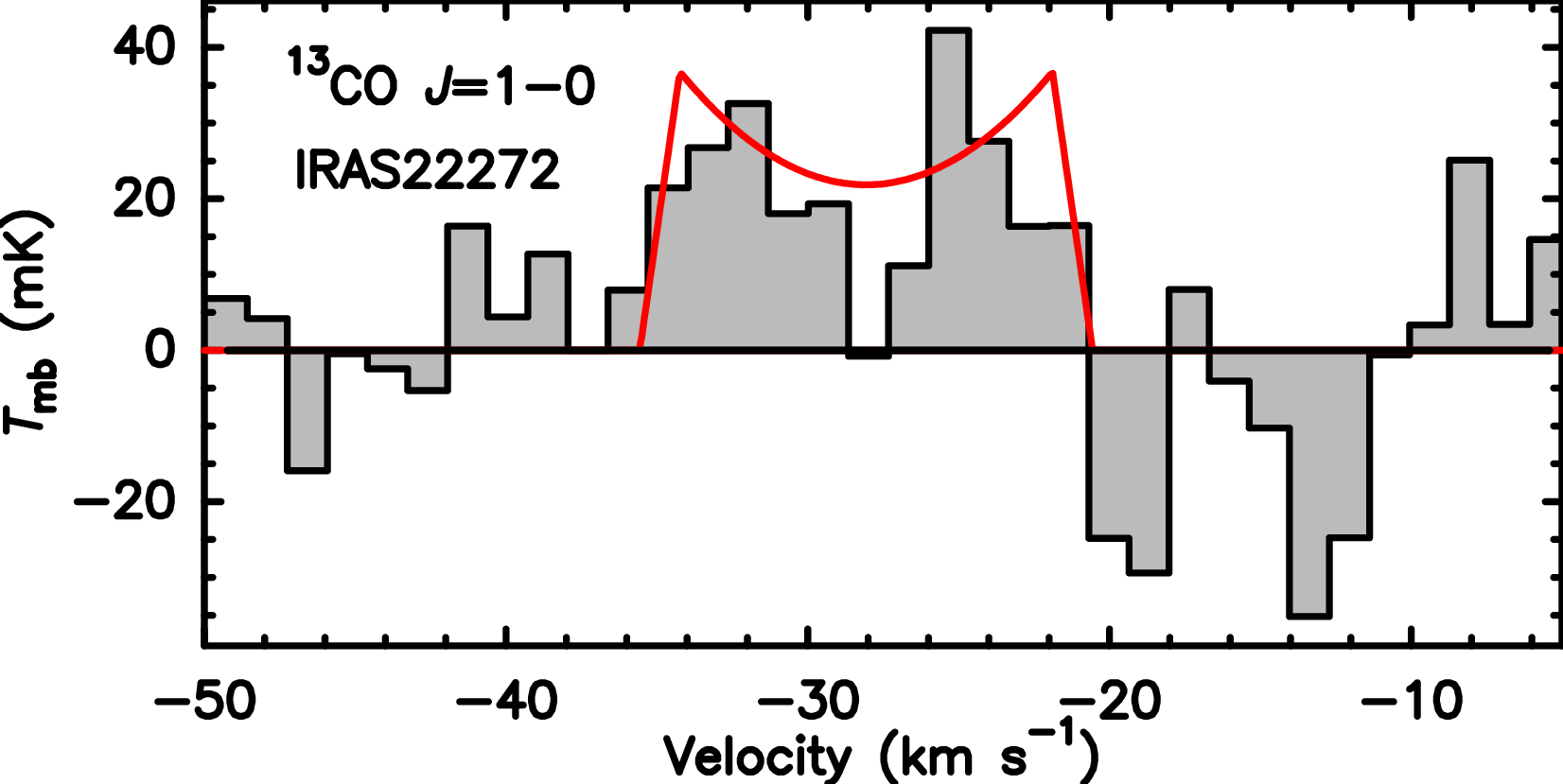}
\caption{Same as Fig.~\ref{Figure3}, but for CO in IRAS\,22272+5435.}
\label{Figure6}
\end{figure*}

\begin{figure*}
\epsscale{.50}
\plotone{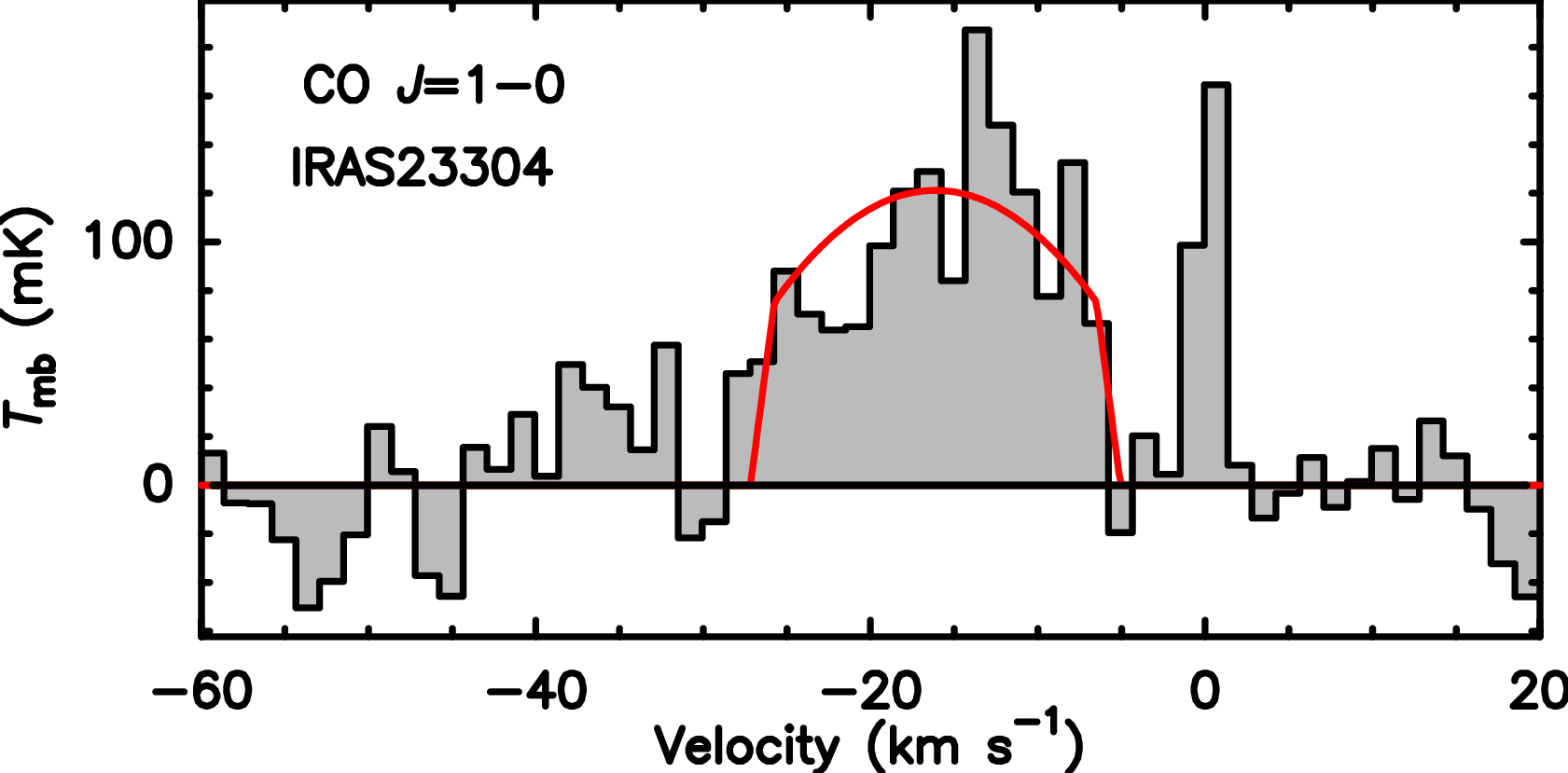}
\plotone{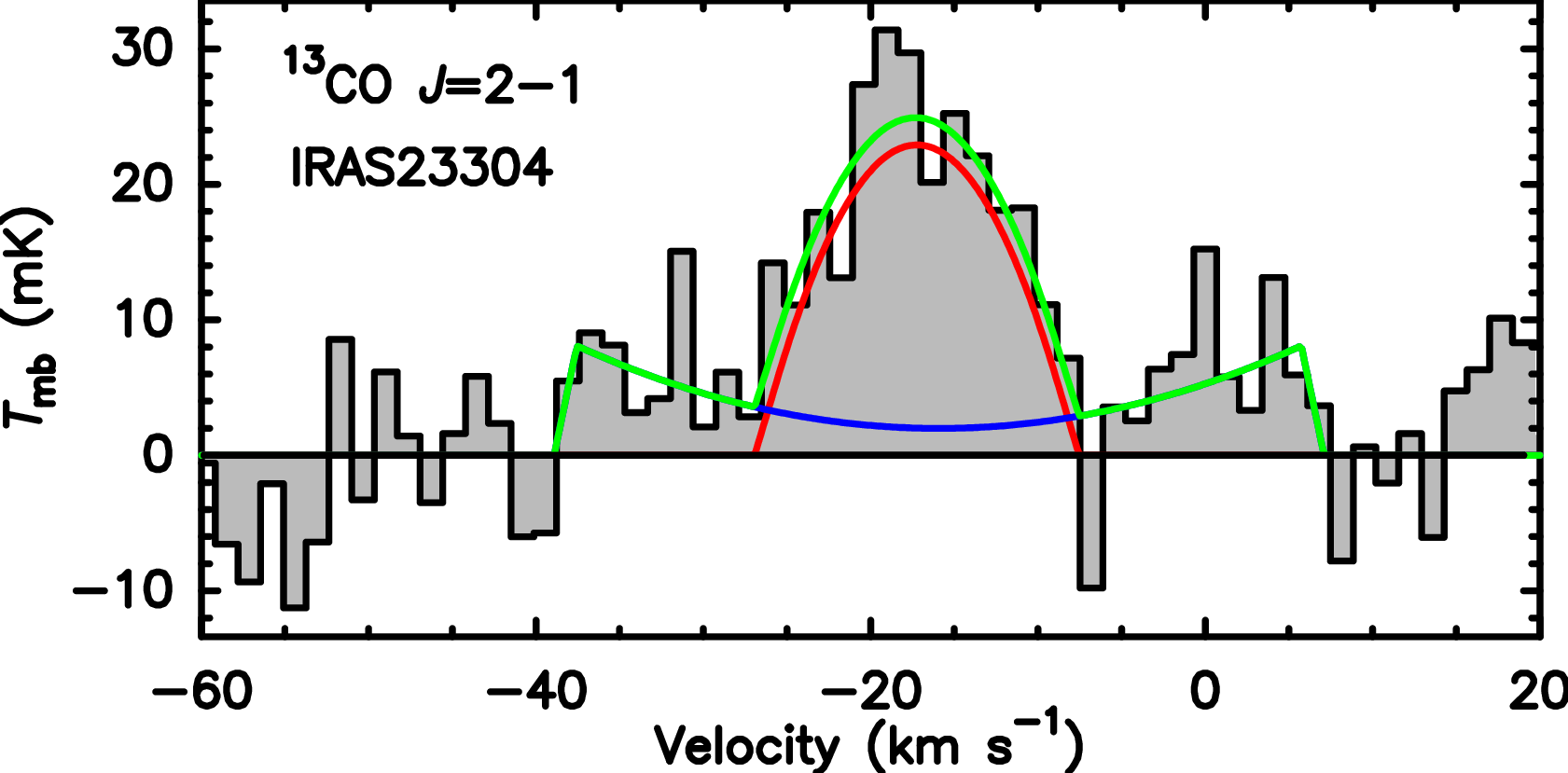}
\plotone{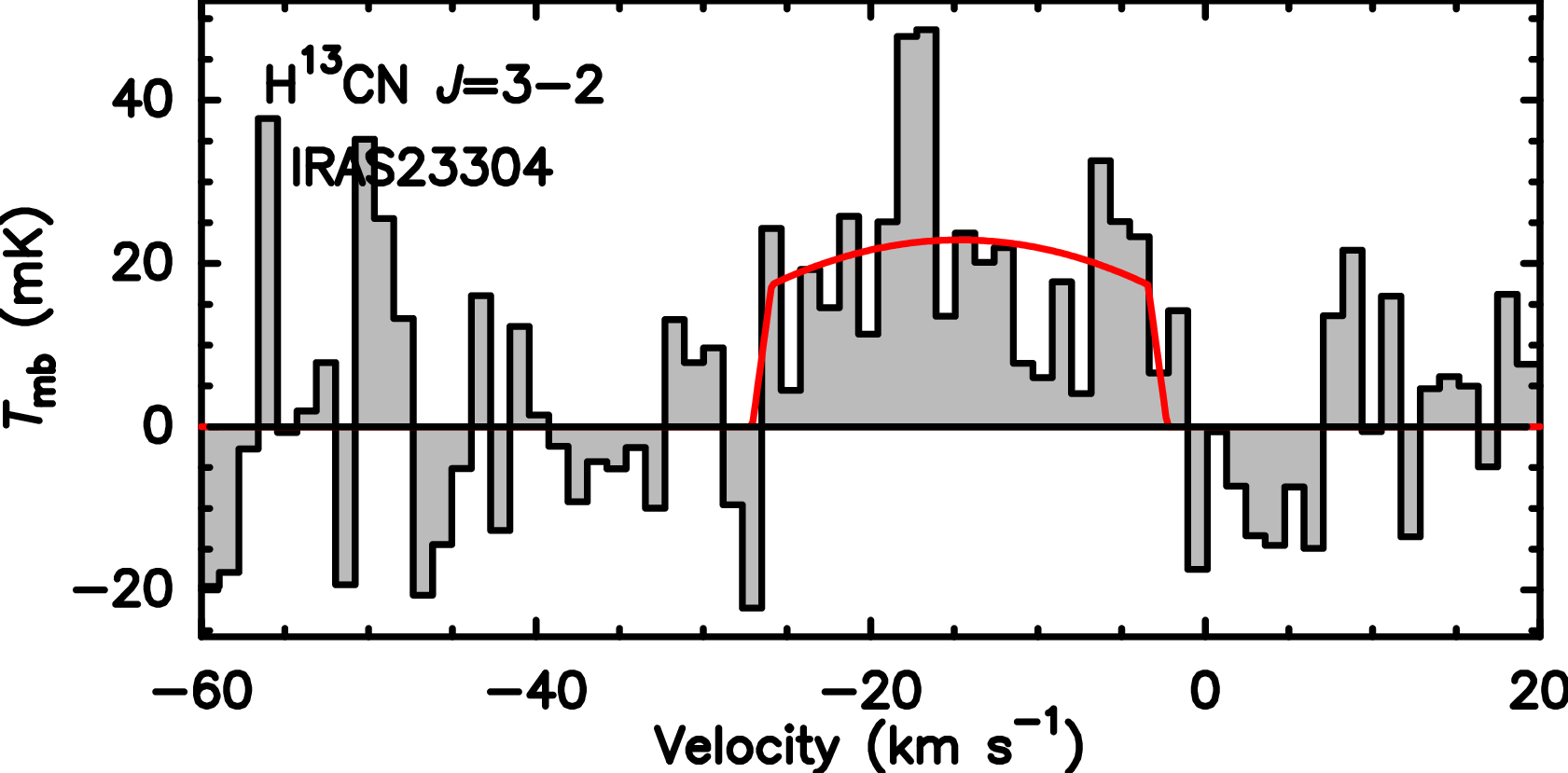}
\plotone{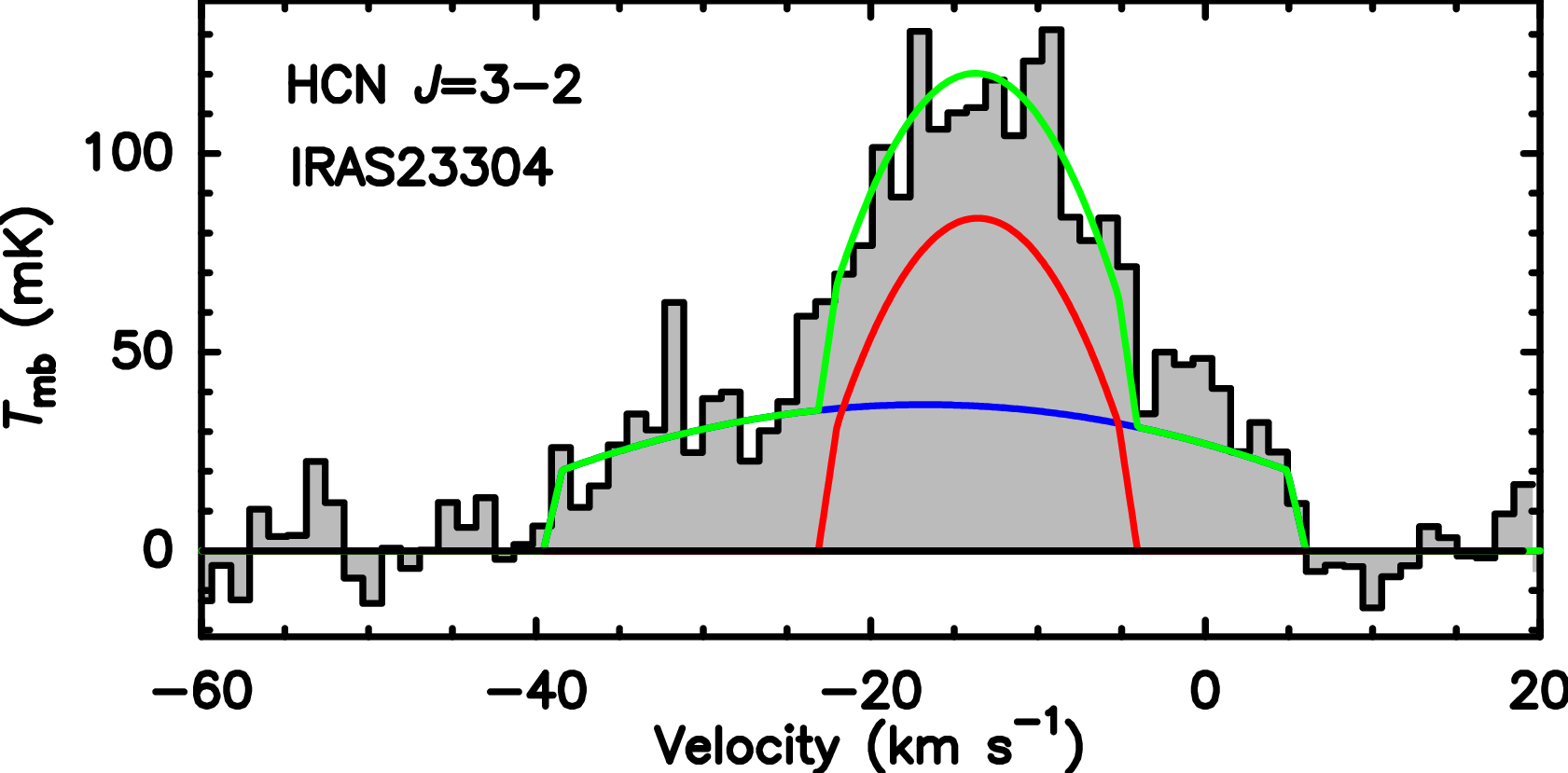}
\plotone{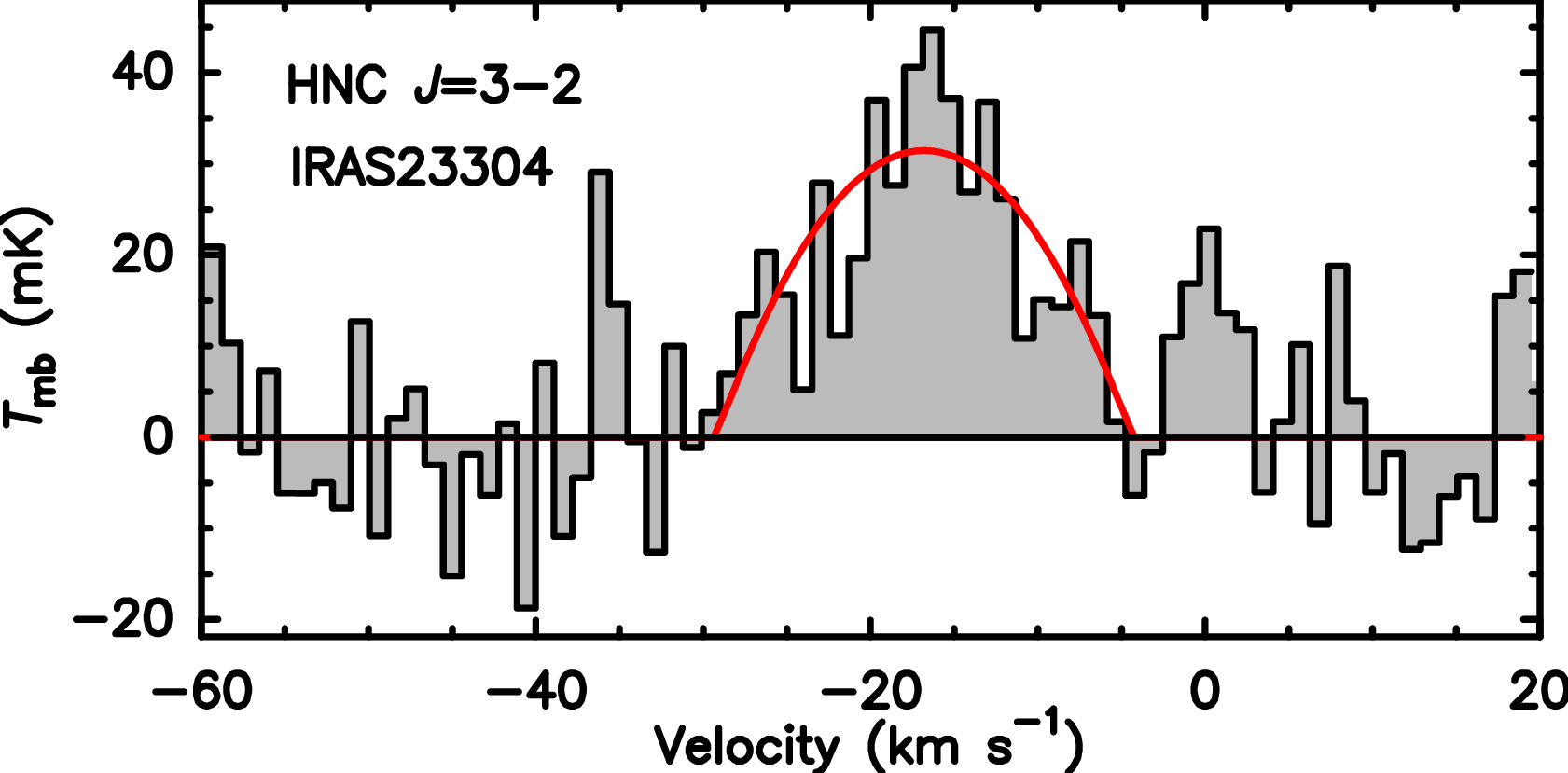}
\plotone{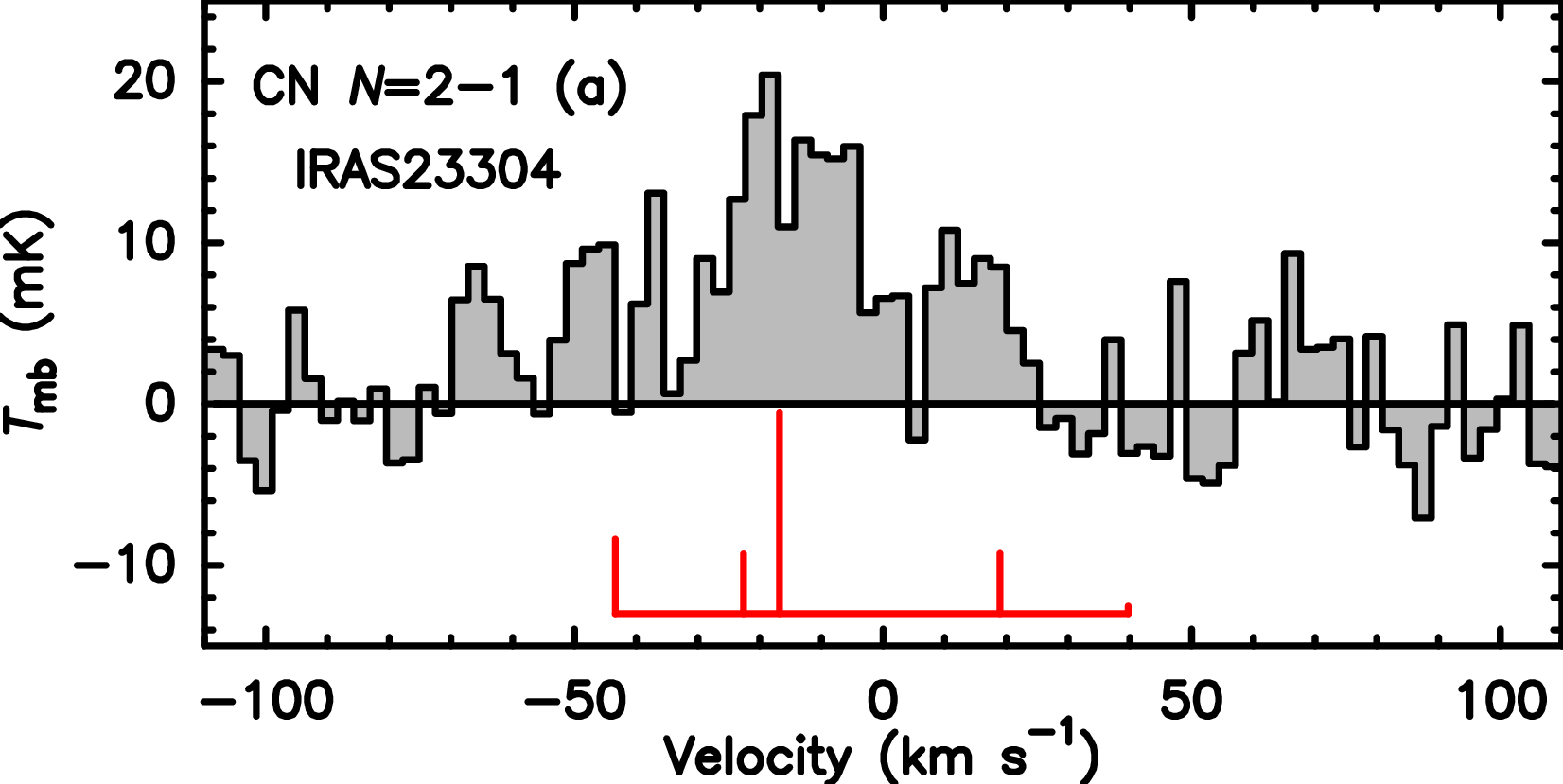}
\plotone{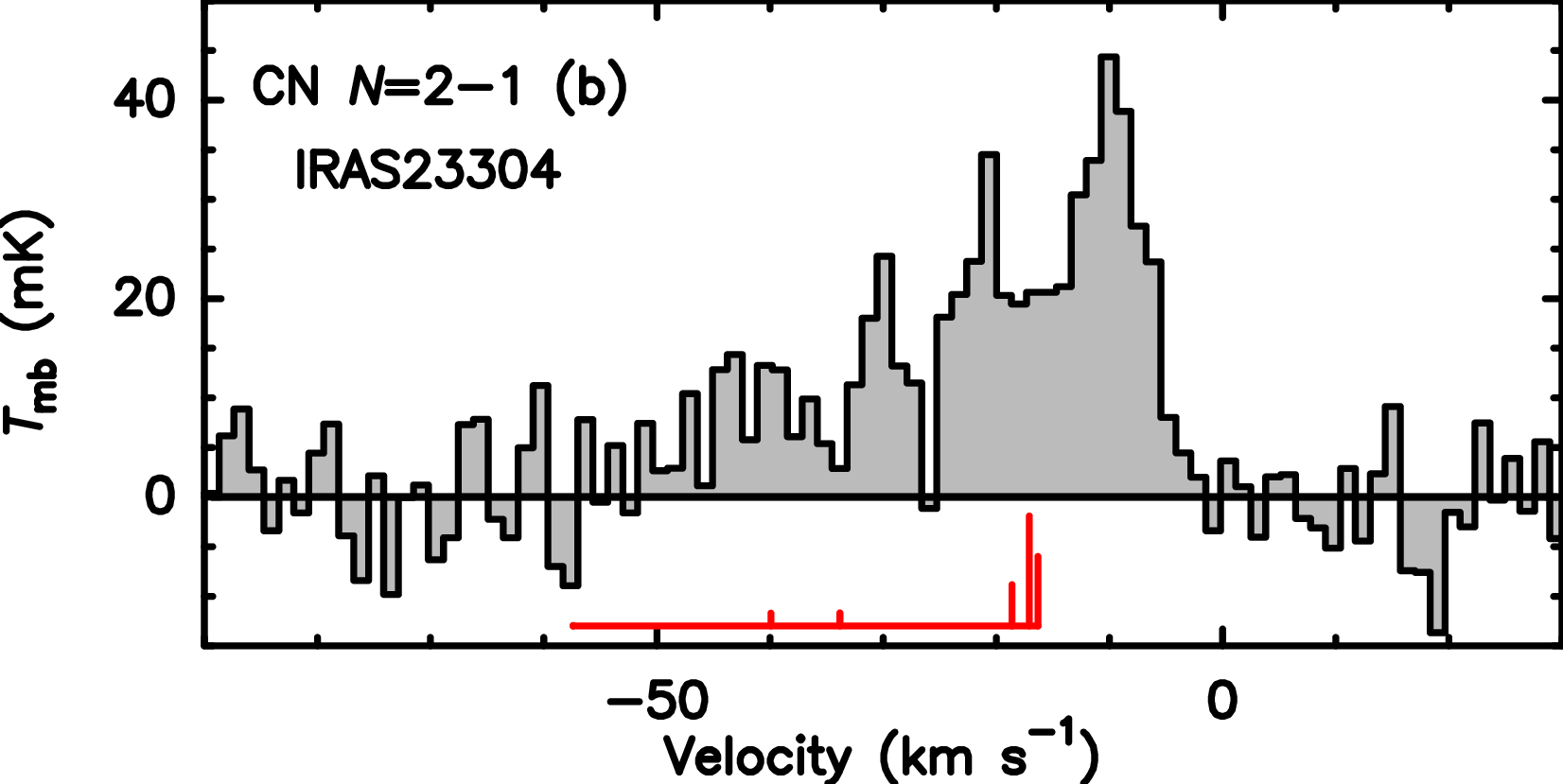}
\caption{Same as Fig.~\ref{Figure3}, but for CO, $^{13}$CO, HCN, H$^{13}$CN, HNC, and CN in IRAS\,23304+6147.}
\label{Figure7}
\end{figure*}

\begin{figure*}
\epsscale{.50}
\plotone{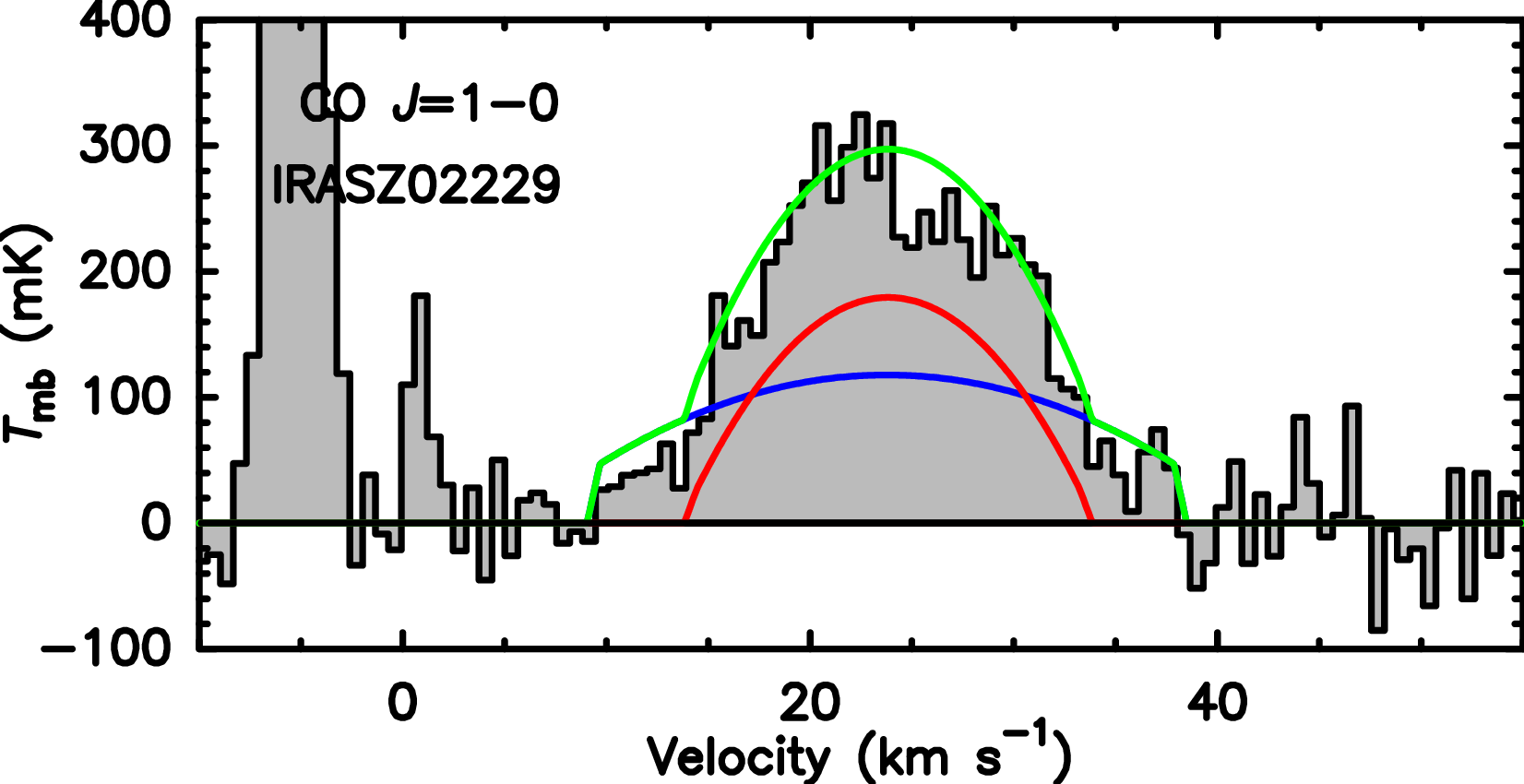}
\plotone{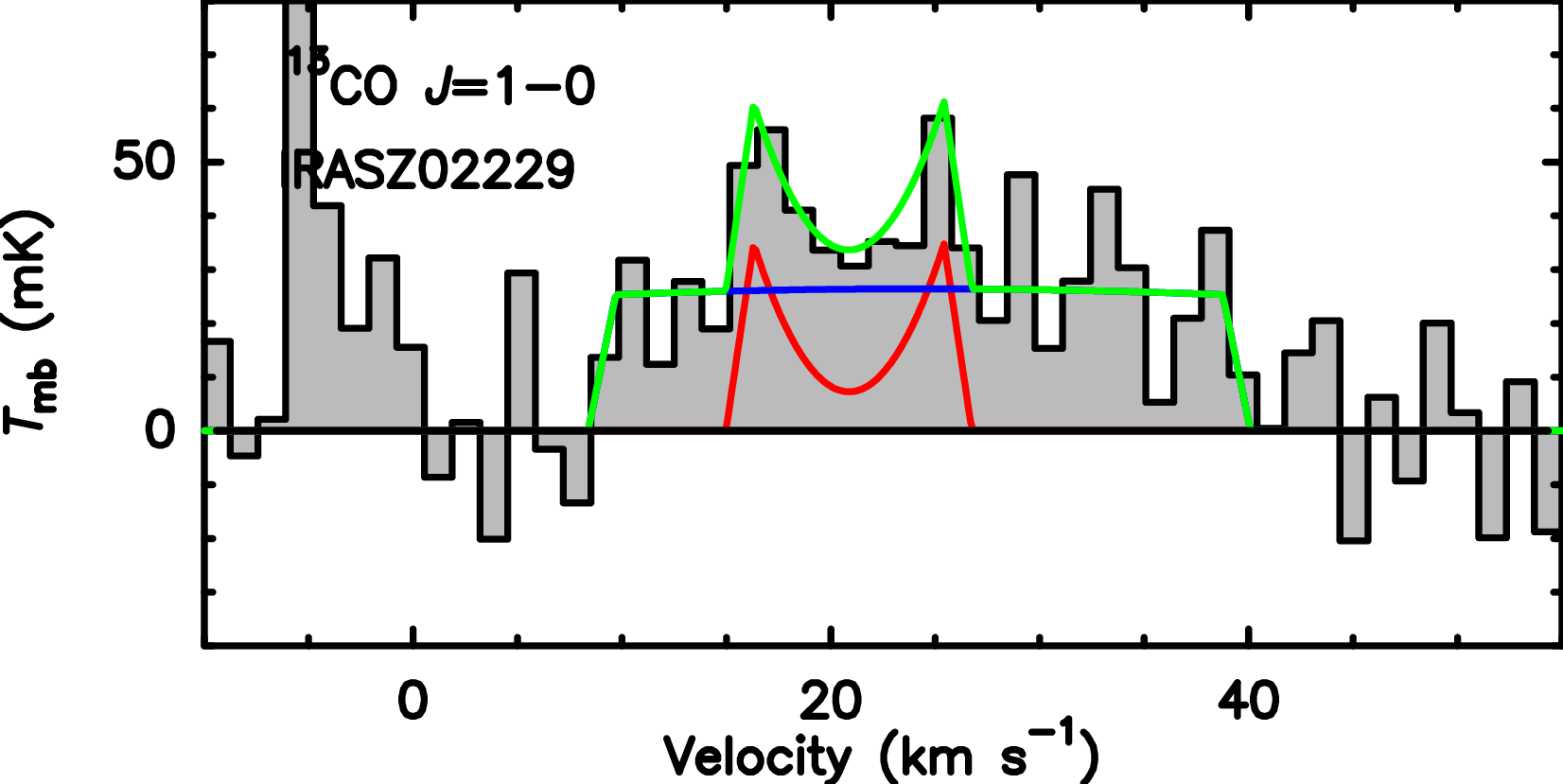}
\plotone{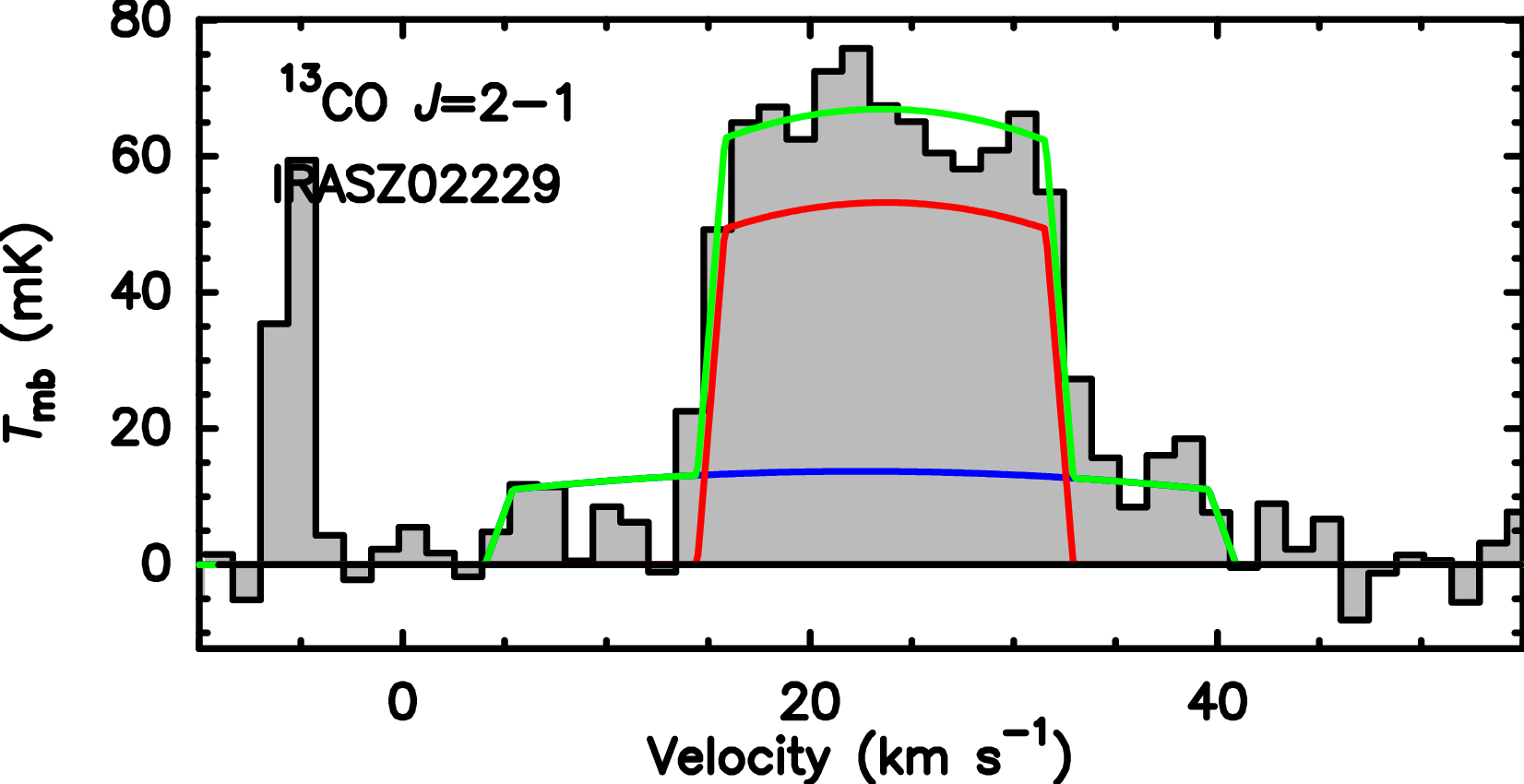}
\plotone{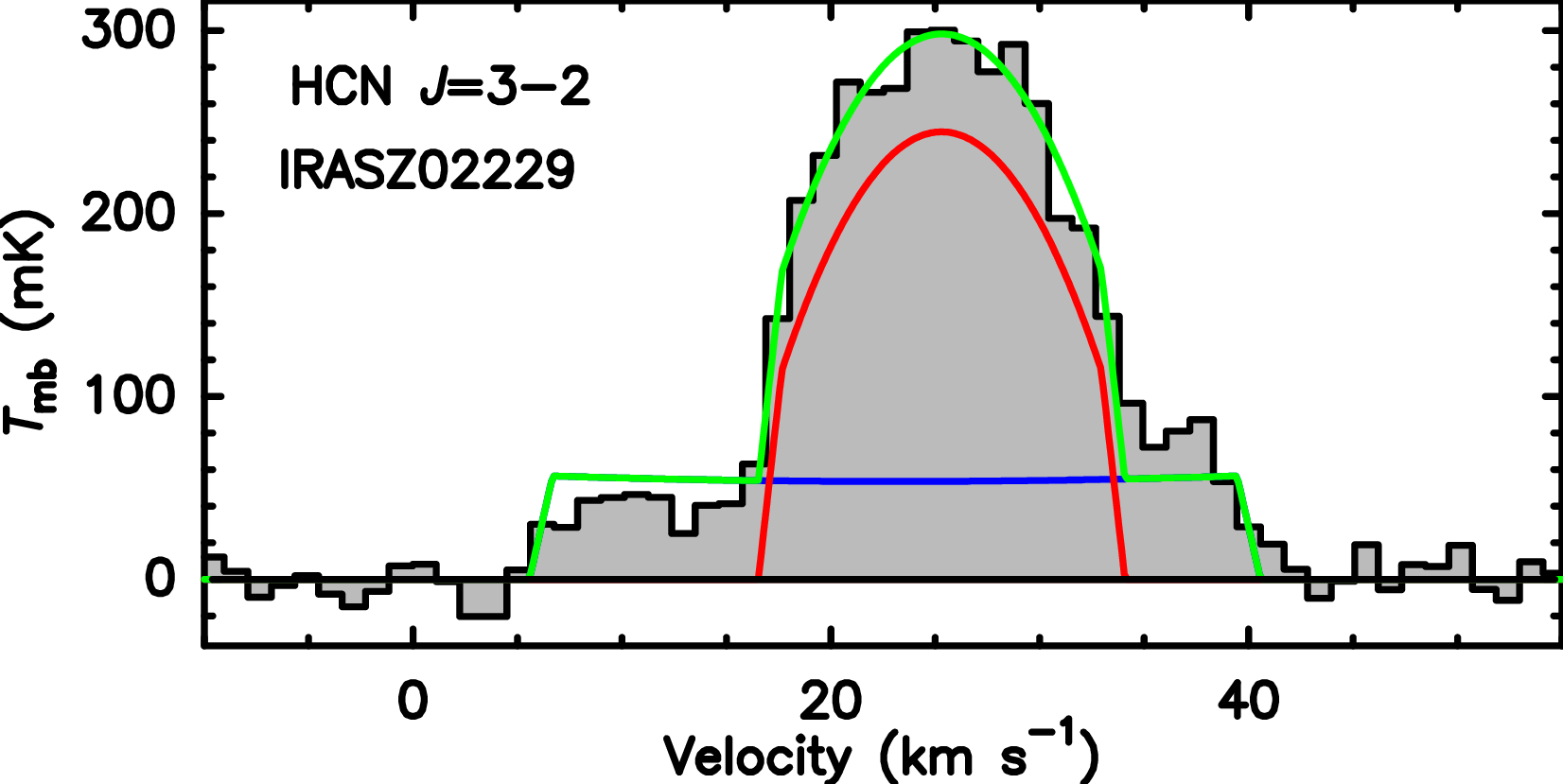}
\plotone{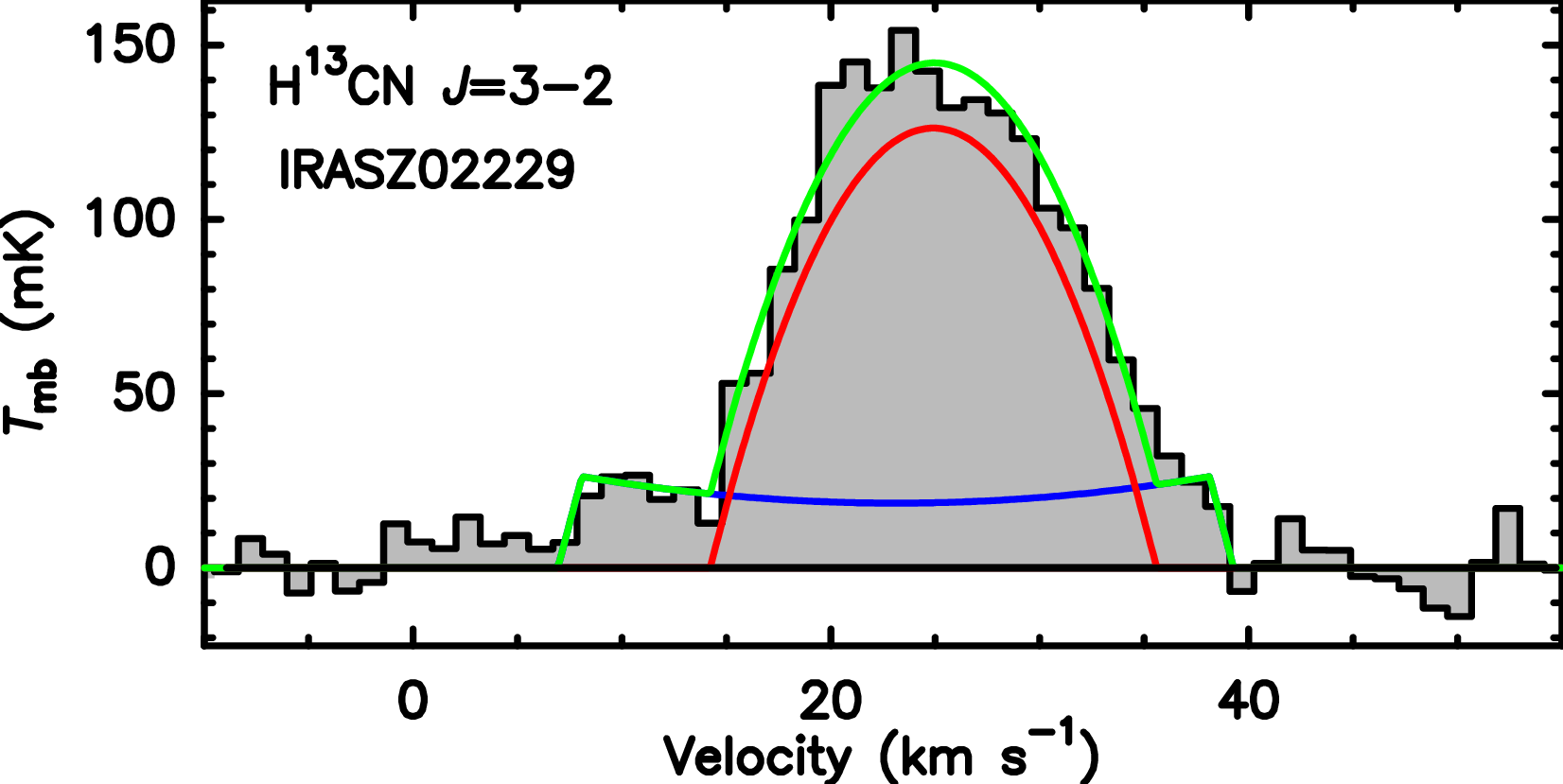}
\plotone{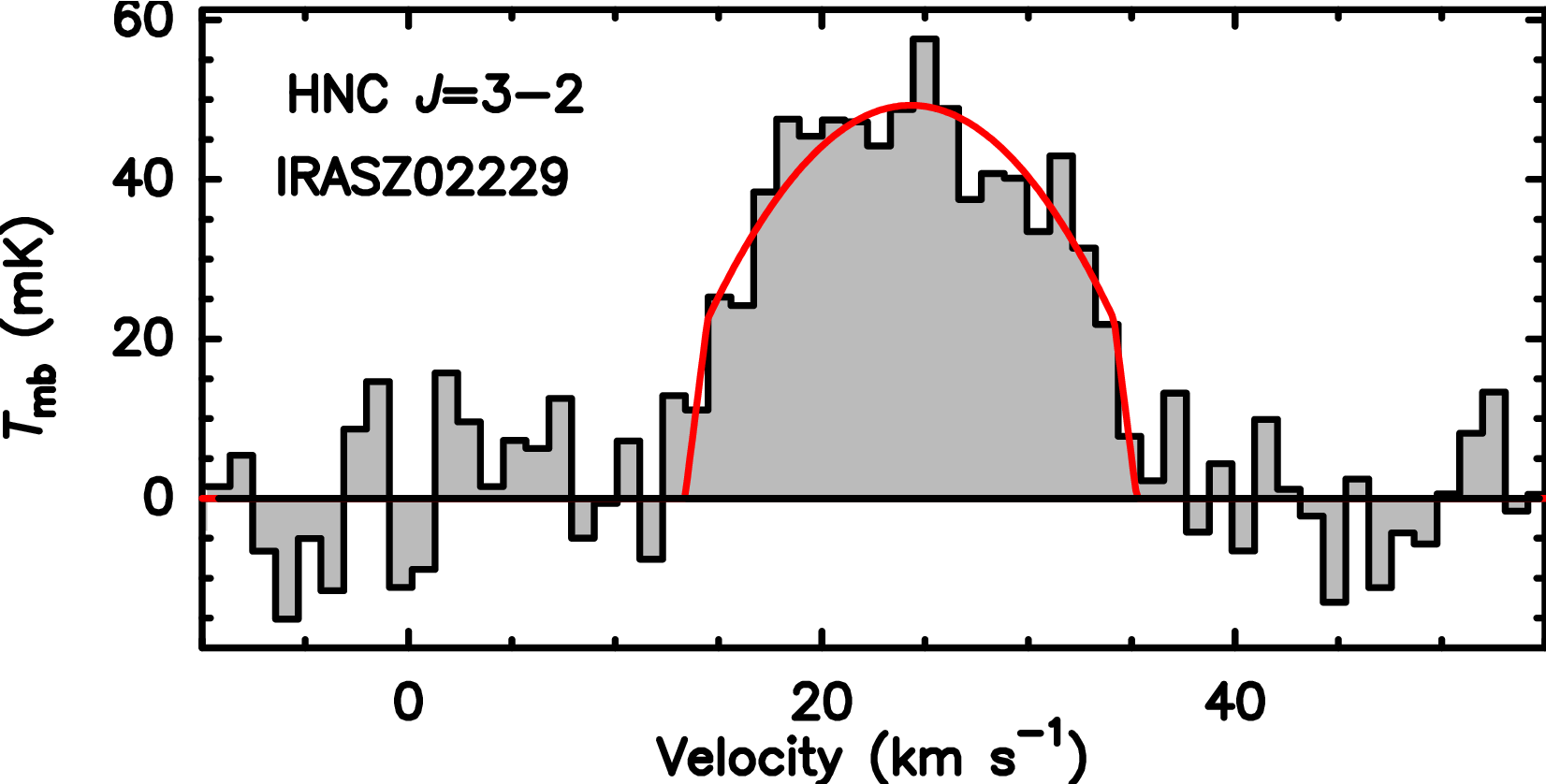}
\plotone{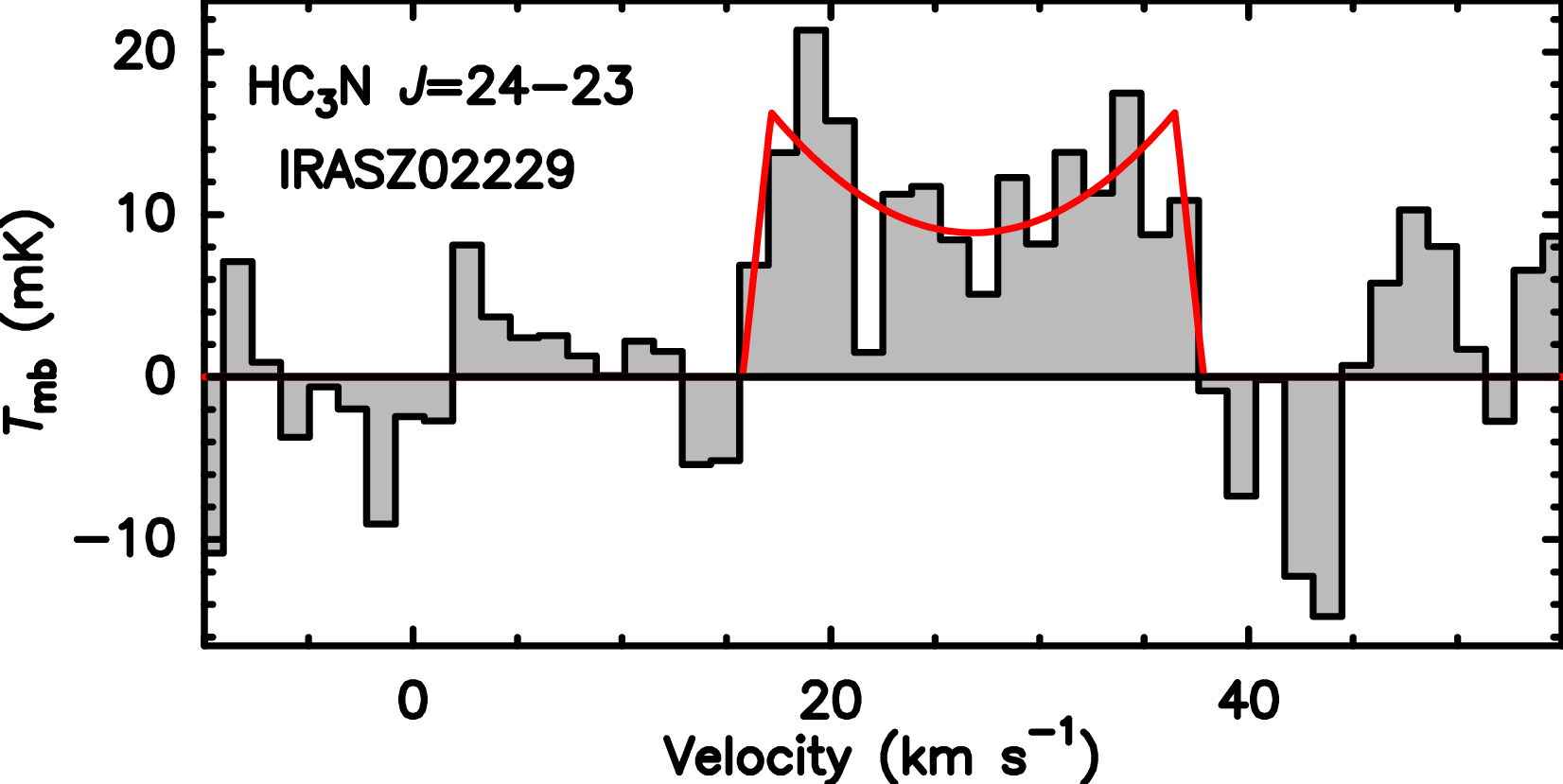}
\plotone{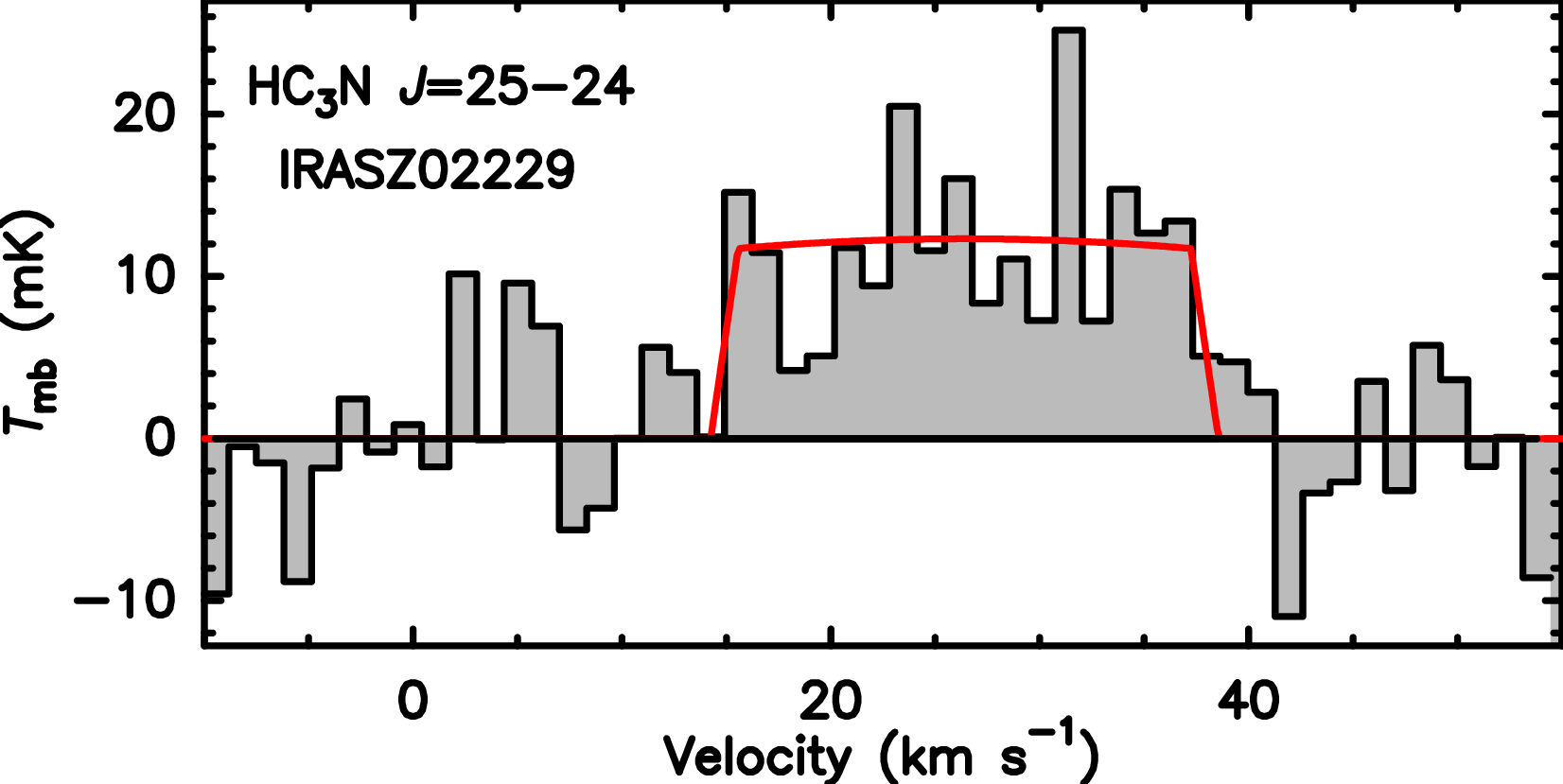}
\plotone{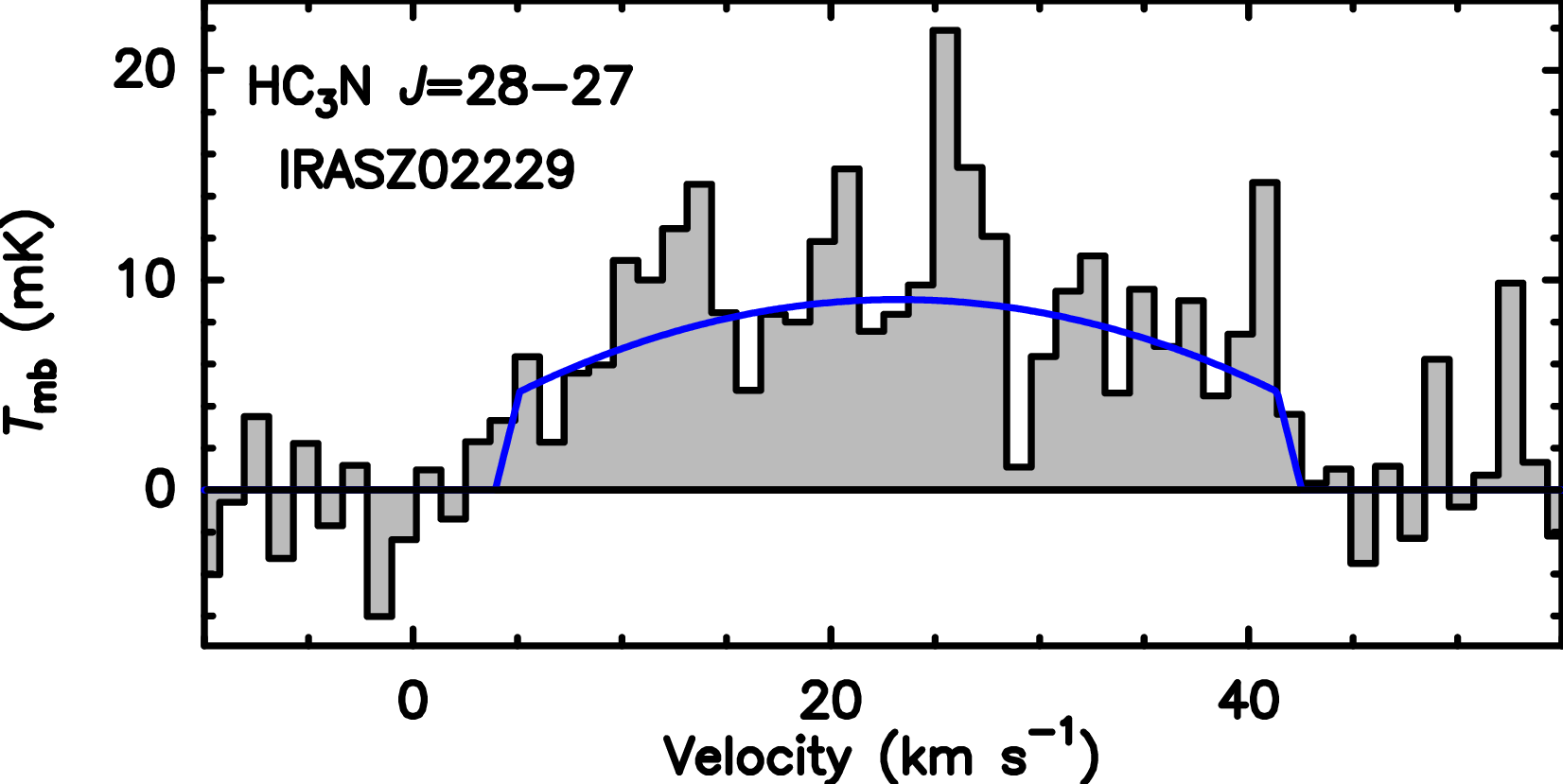}
\plotone{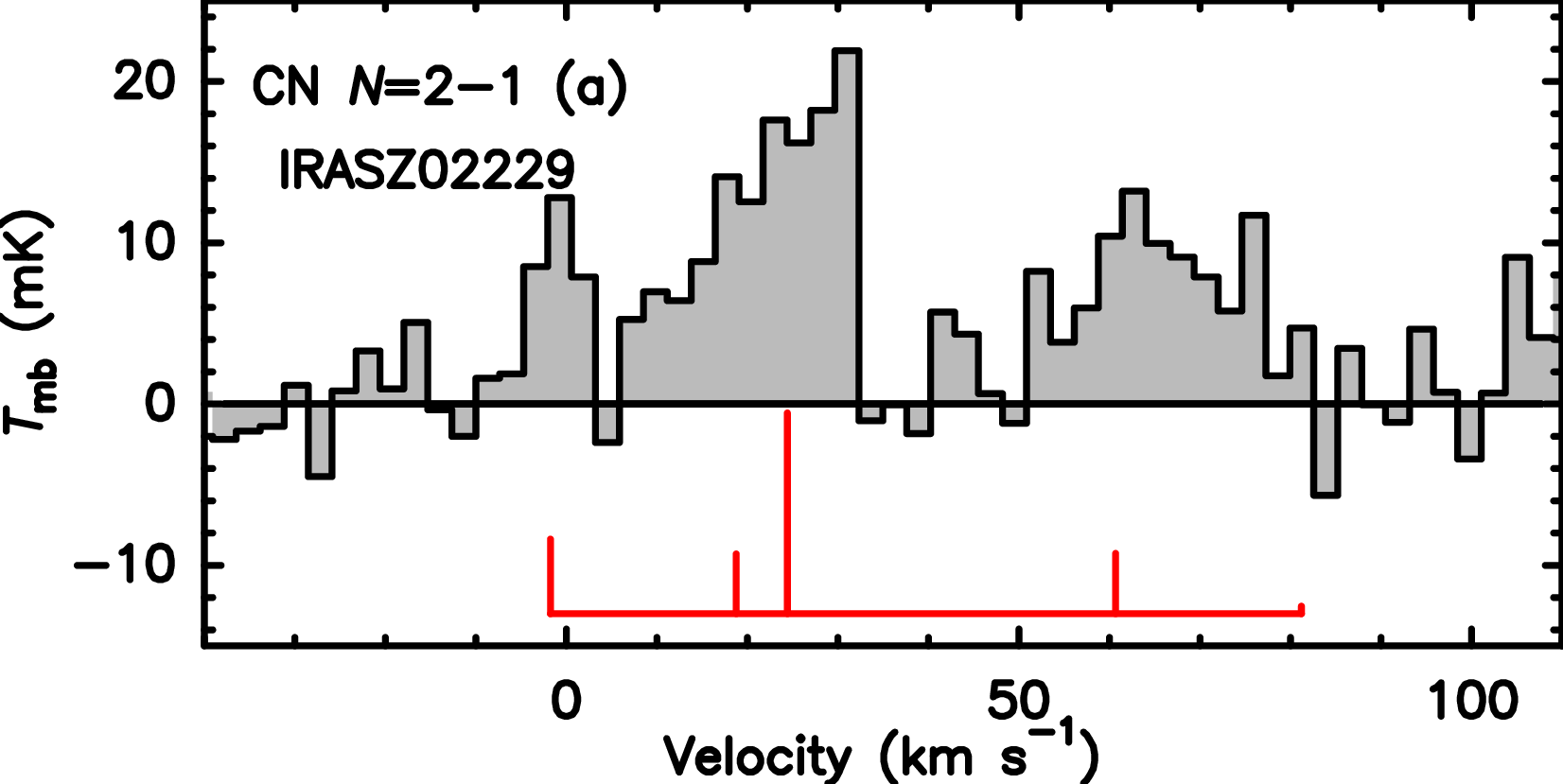}
\plotone{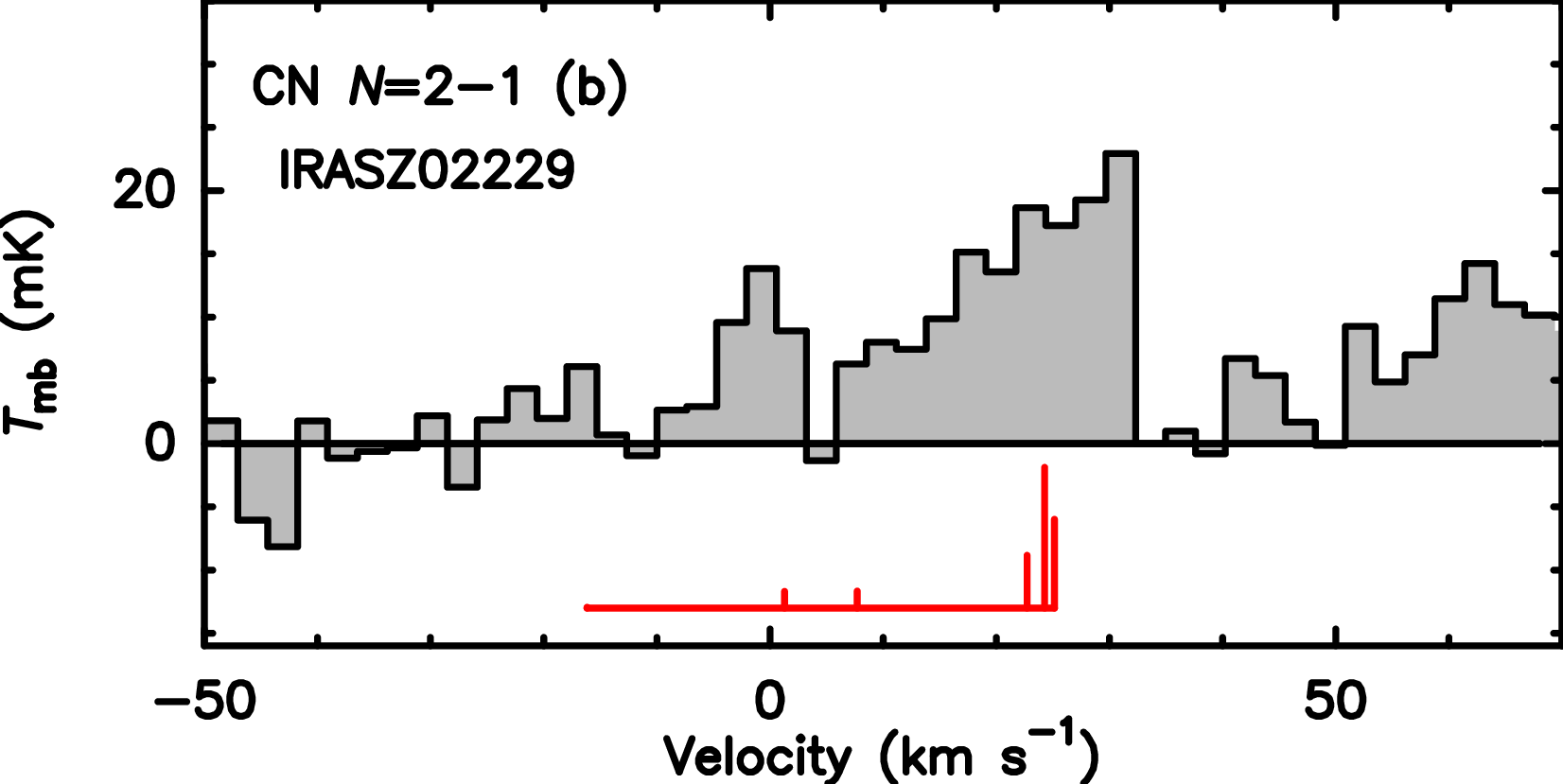}
\caption{Same as Fig.~\ref{Figure3}, but for CO, $^{13}$CO, HCN, H$^{13}$CN, HNC, HC$_{3}$N and CN in IRAS\,Z02229+6208.}
\label{Figure8}
\end{figure*}

\begin{figure*}
\plotone{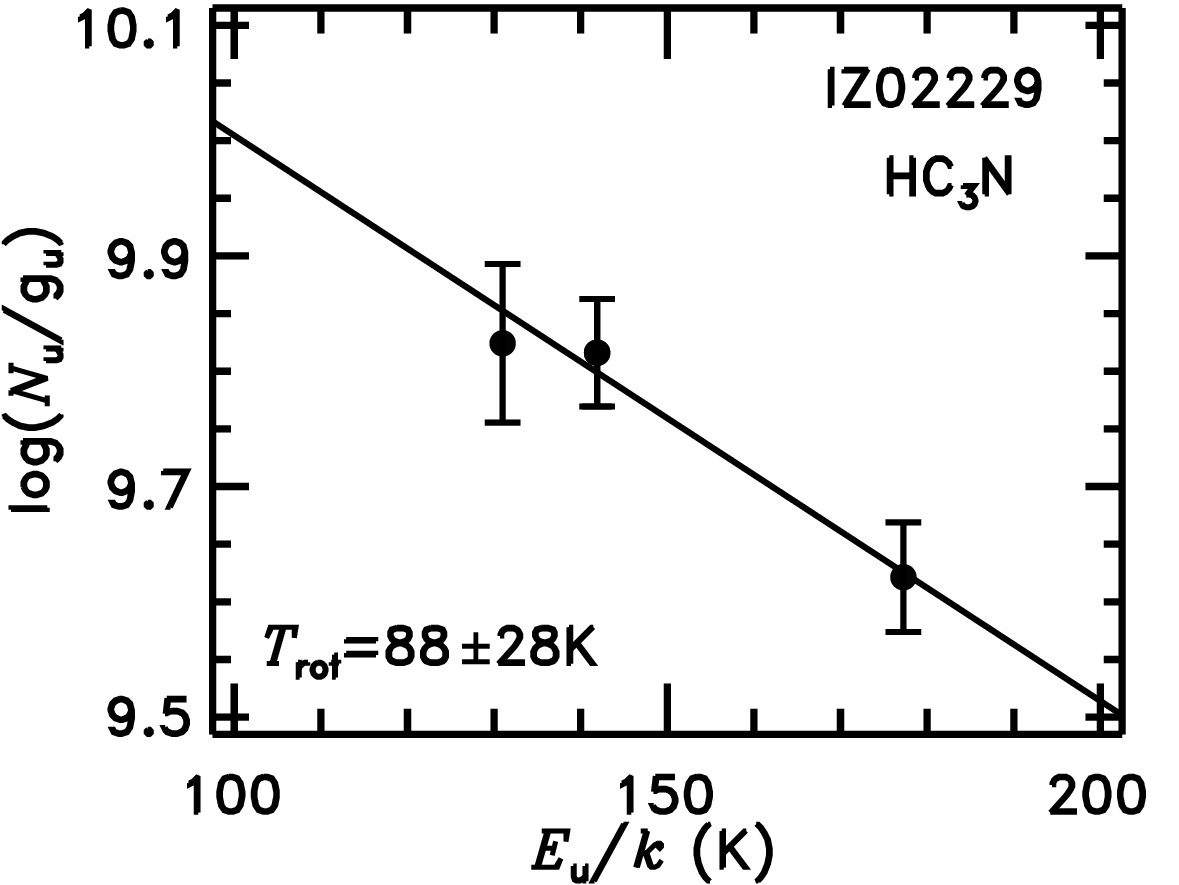}
\caption{
Rotational diagram of HC$_{3}$N in IRAS\,Z02229+6208.
}
\label{Figure9}
\end{figure*}

\begin{figure*}
\epsscale{1.2}
\plotone{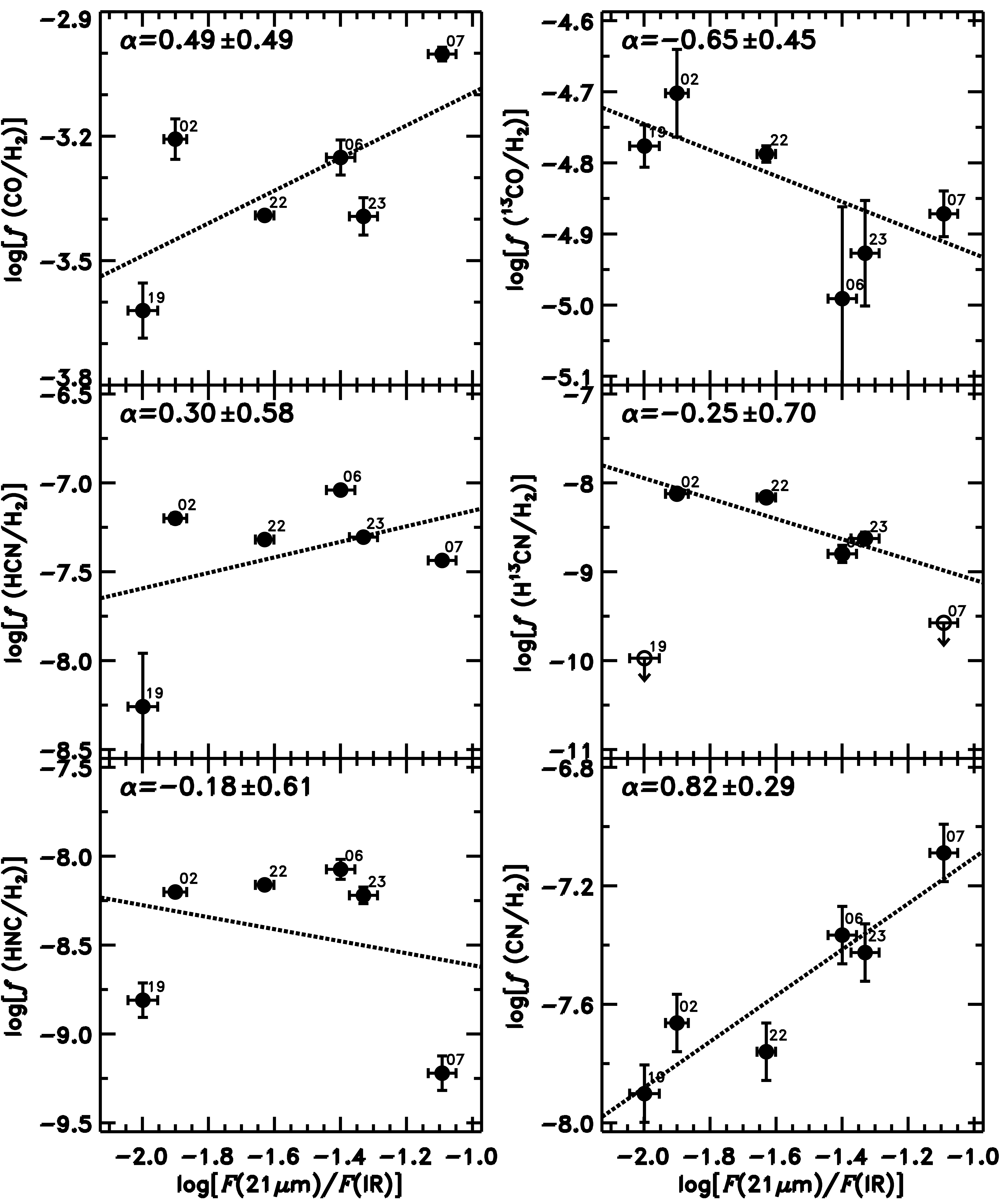}
\caption{Abundance ratios and densities versus intensities of the 21\,$\mu$m feature with respect to total IR emission. 
The dotted lines are the best linear fit. 
The correlation coefficient is marked at the upper-right corner of each panel. 
The numbers 02, 06, 07, 19, 22, and 23 on the plot are for IRAS\,Z02229+6208, IRAS\,06530-0213, IRAS\,07134+1005, IRAS\,19500-1709, IRAS\,22272+5435, and IRAS\,23304+6147, respectively.}
\label{Figure10}
\end{figure*}

\renewcommand{\thefigure}{\arabic{figure} (Cont.)}
\addtocounter{figure}{-1}

\begin{figure*}
\epsscale{1.2}
\plotone{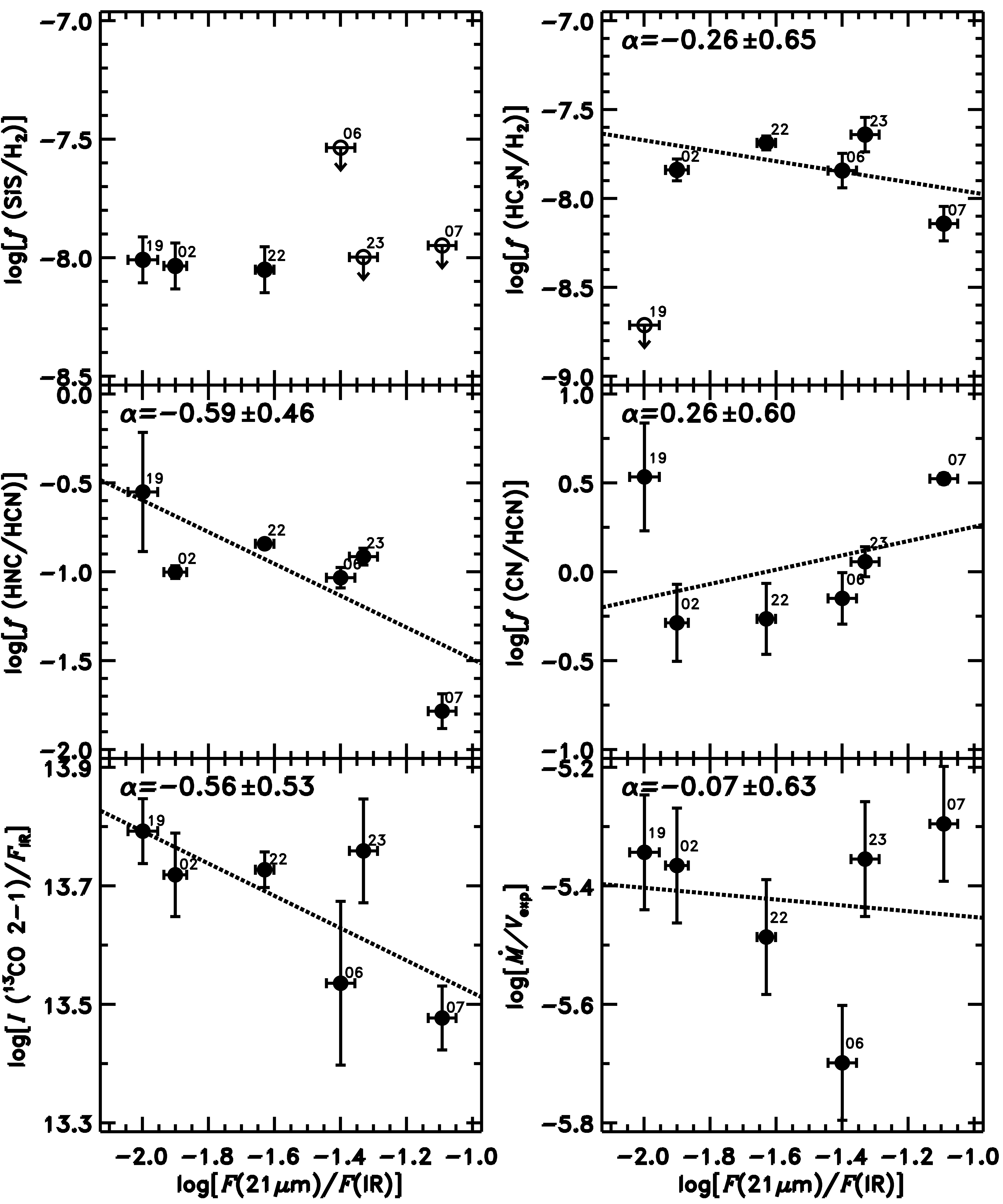}
\caption{}
\end{figure*}

\renewcommand{\thefigure}{\arabic{figure}}

\clearpage

\begin{deluxetable}{lcccccccccccc}[ht]
\rotate
\tablecaption{Stellar Parameters.
\label{Table1}}
\tabletypesize{\footnotesize}
\tablehead{
\colhead{Source name}   & \colhead{$F_{\rm 21 \mu m}$}         & \colhead{$F_{\rm IR}$}               & \colhead{$D$}    & \colhead{$V_{\rm exp}$}   & \colhead{$\dot{M}$}                           & \colhead{$M_{\rm dust}$}           & \colhead{$x_{i}$}   & \colhead{$x_{e}$}   & \colhead{$\theta_{\rm s}$} & \colhead{Duration}  \\
\colhead{}             & \colhead{(10$^{-12}$\,W\,m$^{-2}$)}  & \colhead{(10$^{-12}$\,W\,m$^{-2}$)}  & \colhead{(kpc)}  & \colhead{(km\,s$^{-1}$)}  & \colhead{(10$^{-5}$\,$M_{\odot}$\,yr$^{-1}$)}  & \colhead{(10$^{-4}$\,$M_{\odot}$)}  & \colhead{}      & \colhead{}      & \colhead{(\arcsec)}        & \colhead{(10$^{3}$yr)}             
}
\startdata
IRAS\,06530-0213    & 0.177 $\pm$ 0.005    & 4.44 $\pm$ 0.44       & 4.17 $\pm$ 1.22    & 14.0    & 2.8  &  4.9  & 0.02   & 0.15  & 7.9    &  4.3  \\
IRAS\,07134+1005    & 1.62 $\pm$ 0.06      & 17.3 $\pm$ 1.60       & 2.22 $\pm$ 0.10    & 10.0    & 5.1  &  3.7  & 0.03   & 0.09  & 6.0    &  1.9  \\
IRAS\,19500-1709    & 0.266 $\pm$ 0.002    & 26.52 $\pm$ 2.60      & 2.50 $\pm$ 0.19    & 11.0    & 5.0  & 18.9  & 0.05   & 0.25  & 15.4  &  6.5  \\
IRAS\,22272+5435    & 0.89 $\pm$ 0.05      & 35.8 $\pm$ 2.20       & 1.45 $\pm$ 0.06    & 9.80    & 3.2  &  4.8  & 0.05   & 0.22  & 15.2   &  3.7  \\
IRAS\,23304+6147    & 0.54 $\pm$ 0.045     & 8.90 $\pm$ 0.83       & 4.17 $\pm$ 0.52    & 15.5    & 6.8  & 10.7  & 0.03   & 0.16  & 10.9   &  5.6  \\
IRAS\,Z02229+6208   & 0.34 $\pm$ 0.02      & 25.6 $\pm$ 1.95       & 2.63 $\pm$ 0.42    & 14.8    & 6.4  &  6.9  & 0.02   & 0.15  & 10.4   &  3.6  \\
\enddata
\tablecomments{
The values of $F_{21\mu \rm m}$, $F_{\rm IR}$, expansion velocity ($V_{\rm exp}$), and inner and outer radius of the circumstellar shell ($r_{i}$ and $r_{e}$) are taken from \citet{Mishra15} and \citet{Mishra16}. 
The total dust mass ($M_{\rm dust}$) and the duration age from AGB stage are taken from \citet{Mishra15} and \citet{Mishra16}, 
which have been scaled to the Gaia\,DR3 distance. 
The average mass-loss rate $\dot{M}$ is deduced through the dust mass divided by the duration age under an assumption of the gas-to-dust mass ratio of 100. 
}
\end{deluxetable}

\begin{deluxetable}{lcccccccccc}
\tablecaption{Molecular transitions detected in IRAS\,06530-0213.
\label{Table2}}
\tabletypesize{\footnotesize}
\tablewidth{0pt}
\tablehead{
\colhead{Frequency}    & \colhead{Species}    & \colhead{Transition}     & \colhead{$V_{\rm exp}$}    & \colhead{$H$}    & \colhead{$T_{\rm c}$}    & \colhead{$\int T_{\rm mb}$d$V$}    & \colhead{RMS}     & \colhead{$V_{\rm LSR}$}     & \colhead{D$_{v}$}     & \colhead{HPBW}        \\
\colhead{(MHz)}        &                      & \colhead{Upper-Lower}    & \colhead{(km\,s$^{-1}$)}   &                  & \colhead{(mK)}           & \colhead{(mK\,km\,s$^{-1}$)}     & \colhead{(mK)}    & \colhead{(km\,s$^{-1}$)}    & \colhead{(km\,s$^{-1}$)}  & \colhead{(\arcsec)}        
}
\startdata
115271.2    & CO           & $J=1 \rightarrow 0$                               & 14.9 $\pm$ 1.1    & ...         & 58.0    & 917.8 $\pm$ 80.5    & 20.2   & 31.6        & 0.6    & 48.2  \\ 
110201.4    & $^{13}$CO    & $J=1 \rightarrow 0$                               & ?                 & ?           & $<$20   & $<$200               & 17.8   & ?           & 2.0    & 50.4  \\ 
220398.7    & $^{13}$CO    & $J=2 \rightarrow 1$                               & 10.3 $\pm$ 1.0    & -0.13       & 7.7     & 152.4 $\pm$ 39.2     & 5.5    & 33.1        & 2.7    & 34.3  \\ 
109782.2    & C$^{18}$O    & $J=1 \rightarrow 0$                               & ?                 & ?           & $<$20   & $<$200               & 17.8   & ?           & 2.0    & 50.6  \\
265886.4    & HCN          & $J=3 \rightarrow 2$                               & 10.8 $\pm$ 0.2    & -0.83       & 103.4   & 1619.0 $\pm$ 41.5    & 8.6    & 33.9        & 1.1    & 28.4  \\ 
259011.8    & H$^{13}$CN   & $J=3 \rightarrow 2$                               & ?                 & ?           & 10:     & 115.2$^{a}$          & 5.9    & ?           & 2.3    & 29.2  \\ 
271981.1    & HNC          & $J=3 \rightarrow 2$                               & 9.4 $\pm$ 0.4     & 0.45        & 11.3    & 244.0 $\pm$ 29.5     & 5.6    & 30.8        & 2.2    & 27.8  \\ 
219949.4    & SO           & $N_J=5_{6} \rightarrow 4_{5}$                     & ?                 & ?           & $<$5    & $<$50                & 3.6    & ?           & 2.7    & 34.3  \\
\enddata
\tablecomments{
\tablenotetext{a}{Obtained by integrating the main-beam temperature over the velocity coverage from 20--50\,km\,s$^{-1}$.}
\tablenotetext{b}{Obtained by integrating the main-beam temperature over the velocity coverage from -20--110\,km\,s$^{-1}$.}
\tablenotetext{c}{Obtained by integrating the main-beam temperature over the velocity coverage from -30--60\,km\,s$^{-1}$.}
}
\text{Symbol:}\\
\footnotesize{? means the values that cannot be derived from our observations because of insufficient sensitivity.}\\
\footnotesize{$\times$ means the values that cannot be obtained because of unsolved hyperfine structures.}\\ 
\footnotesize{: means uncertain measurements.}\\
\end{deluxetable}

\begin{deluxetable}{lcccccccccc}
\tablecaption{Same as Table~\ref{Table2}, but for IRAS\,07134+1005.
\label{Table3}}
\tabletypesize{\footnotesize}
\tablewidth{0pt}
\tablehead{
\colhead{Frequency}    & \colhead{Species}    & \colhead{Transition}     & \colhead{$V_{\rm exp}$}    & \colhead{$H$}    & \colhead{$T_{\rm c}$}    & \colhead{$\int T_{\rm mb}$d$V$}    & \colhead{RMS}     & \colhead{$V_{\rm LSR}$}     & \colhead{D$_{v}$}     & \colhead{HPBW}        \\
\colhead{(MHz)}        &                      & \colhead{Upper-Lower}    & \colhead{(km\,s$^{-1}$)}   &                  & \colhead{(mK)}           & \colhead{(mK\,km\,s$^{-1}$)}     & \colhead{(mK)}    & \colhead{(km\,s$^{-1}$)}    & \colhead{(km\,s$^{-1}$)}  & \colhead{(\arcsec)}       
}
\startdata
115271.2    & CO           & $J=1 \rightarrow 0$                               & 8.6 $\pm$ 0.1     & -0.46       & 320.4   & 4666.5 $\pm$ 175.3    & 87.8    & 73.6        & 1.4    & 48.2  \\ 
110201.4    & $^{13}$CO    & $J=1 \rightarrow 0$                               & ?                 & ?           & $<$20   & $<$200                & 15.9    & ?           & 2.0    & 50.4  \\ 
220398.7    & $^{13}$CO    & $J=2 \rightarrow 1$                               & 10.2 $\pm$ 0.2    & 0.46        & 22.1    & 518.8 $\pm$ 37.0      & 6.7     & 71.9        & 1.4    & 34.3  \\ 
109782.2    & C$^{18}$O    & $J=1 \rightarrow 0$                               & ?                 & ?           & $<$20   & $<$200                & 17.8    & ?           & 2.0    & 50.6  \\
265886.4    & HCN          & $J=3 \rightarrow 2$                               & 10.9 $\pm$ 0.2    & -0.47       & 91.9    & 1685.3 $\pm$ 41.1     & 8.0     & 72.9        & 1.1    & 28.4  \\ 
259011.8    & H$^{13}$CN   & $J=3 \rightarrow 2$                               & ?                 & ?           & $<$5    & $<$50                 & 3.4     & ?           & 2.3    & 29.2  \\ 
271981.1    & HNC          & $J=3 \rightarrow 2$                               & ?                 & ?           & 8:      & 45.1$^{a}$            & 7.9     & ?           & 1.1    & 27.8  \\ 
219949.4    & SO           & $N_J=5_{6} \rightarrow 4_{5}$                     & ?                 & ?           & $<$5    & $<$50                 & 7.7     & ?           & 1.4    & 34.3  \\
\enddata
\tablecomments{
\tablenotetext{a}{Obtained by integrating the main-beam temperature over the velocity coverage from 60--85\,km\,s$^{-1}$.}
\tablenotetext{b}{Obtained by integrating the main-beam temperature over the velocity coverage from 20--150\,km\,s$^{-1}$.}
\tablenotetext{c}{Obtained by integrating the main-beam temperature over the velocity coverage from 30--1000\,km\,s$^{-1}$.}
}
\end{deluxetable}

\begin{deluxetable}{lcccccccccc}
\tablecaption{Same as Table~\ref{Table2}, but for IRAS\,19500-1709.
\label{Table4}}
\tabletypesize{\footnotesize}
\tablewidth{0pt}
\tablehead{
\colhead{Frequency}    & \colhead{Species}    & \colhead{Transition}     & \colhead{$V_{\rm exp}$}    & \colhead{$H$}    & \colhead{$T_{\rm c}$}    & \colhead{$\int T_{\rm mb}$d$V$}    & \colhead{RMS}     & \colhead{$V_{\rm LSR}$}     & \colhead{D$_{v}$}     & \colhead{HPBW}        \\
\colhead{(MHz)}        &                      & \colhead{Upper-Lower}    & \colhead{(km\,s$^{-1}$)}   &                  & \colhead{(mK)}           & \colhead{(mK\,km\,s$^{-1}$)}     & \colhead{(mK)}    & \colhead{(km\,s$^{-1}$)}    & \colhead{(km\,s$^{-1}$)}  & \colhead{(\arcsec)}      
}
\startdata
115271.2    & CO           & $J=1 \rightarrow 0$                                & 7.4 $\pm$ 0.8     & 0.82        & 67.8     & 1284.5 $\pm$ 140.6    & 29.8    & 23.3        & 1.9    & 48.2  \\ 
            &              &                                                    & 29.5 $\pm$ 2.4    & 3.20        & 13.6     & 1652.0 $\pm$ 275.9    & 29.8    & 18.1        & 1.9    & 48.2  \\ 
110201.4    & $^{13}$CO    & $J=1 \rightarrow 0$                                & ?                 & ?           & $<$20    & $<$200                & 23.0    & ?           & 2.0    & 50.4  \\ 
220398.7    & $^{13}$CO    & $J=2 \rightarrow 1$                                & 10.9 $\pm$ 0.1    & -1.00       & 52.1     & 756.6 $\pm$ 45.8      & 6.8     & 25.0        & 1.4    & 34.3  \\
            &              &                                                    & 29.1 $\pm$ 1.2    & -0.84       & 21.2     & 887.2 $\pm$ 63.2      & 6.8     & 20.8        & 1.4    & 34.3  \\ 
109782.2    & C$^{18}$O    & $J=1 \rightarrow 0$                                & ?                 & ?           & $<$20    & $<$200                & 17.8    & ?           & 2.0    & 50.6  \\
265886.4    & HCN          &  $J=3 \rightarrow 2$                               & 32.7 $\pm$ 1.9    & -0.57       & 11.9     & 631.9:                & 5.6     & 23.4        & 2.3    & 28.4  \\ 
259011.8    & H$^{13}$CN   &  $J=3 \rightarrow 2$                               & ?                 & ?           & $<$5     & $<$50                 & 9.0     & ?           & 2.3    & 29.2  \\
\enddata
\tablecomments{
\tablenotetext{a}{Obtained by integrating the main-beam temperature over the velocity coverage from -15--60\,km\,s$^{-1}$.}
\tablenotetext{b}{Obtained by integrating the main-beam temperature over the velocity coverage from -30--110\,km\,s$^{-1}$.}
\tablenotetext{c}{Obtained by integrating the main-beam temperature over the velocity coverage from -30--50\,km\,s$^{-1}$.}
}
\end{deluxetable}

\begin{deluxetable}{lcccccccccc}
\tablecaption{Same as Table~\ref{Table2}, but for IRAS\,22272+5435.
\label{Table5}}
\tabletypesize{\footnotesize}
\tablewidth{0pt}
\tablehead{
\colhead{Frequency}    & \colhead{Species}    & \colhead{Transition}     & \colhead{$V_{\rm exp}$}    & \colhead{$H$}    & \colhead{$T_{\rm c}$}    & \colhead{$\int T_{\rm mb}$d$V$}    & \colhead{RMS}     & \colhead{$V_{\rm LSR}$}     & \colhead{D$_{v}$}     & \colhead{HPBW}            \\
\colhead{(MHz)}        &                      & \colhead{Upper-Lower}    & \colhead{(km\,s$^{-1}$)}   &                  & \colhead{(mK)}           & \colhead{(mK\,km\,s$^{-1}$)}     & \colhead{(mK)}    & \colhead{(km\,s$^{-1}$)}    & \colhead{(km\,s$^{-1}$)}  & \colhead{(\arcsec)}      
}
\startdata
115271.2    & CO           & $J=1 \rightarrow 0$                               & 8.4 $\pm$ 0.1    & -0.71       & 460.6    & 5938.5 $\pm$ 110.7     & 72.2   & -28.5    & 0.6    & 48.2  \\ 
110201.4    & $^{13}$CO    & $J=1 \rightarrow 0$                               & 6.8 $\pm$ 0.5    & 1.06        & 16.9     & 310.1 $\pm$ 66.6       & 14.0   & -28.1    & 1.3    & 50.4  \\ 
109782.2    & C$^{18}$O    & $J=1 \rightarrow 0$                               & ?                & ?           & $<$20    & $<$200                 & 17.8   & ?        & 2.0    & 50.6  \\
\hline
\hline
\enddata
\end{deluxetable}

\begin{deluxetable}{lcccccccccc}
\tablecaption{Same as Table~\ref{Table2}, but for IRAS\,23304+6147.
\label{Table6}}
\tabletypesize{\footnotesize}
\tablewidth{0pt}
\tablehead{
\colhead{Frequency}    & \colhead{Species}    & \colhead{Transition}     & \colhead{$V_{\rm exp}$}    & \colhead{$H$}    & \colhead{$T_{\rm c}$}    & \colhead{$\int T_{\rm mb}$d$V$}    & \colhead{RMS}     & \colhead{$V_{\rm LSR}$}     & \colhead{D$_{v}$}     & \colhead{HPBW}         \\
\colhead{(MHz)}        &                      & \colhead{Upper-Lower}    & \colhead{(km\,s$^{-1}$)}   &                  & \colhead{(mK)}           & \colhead{(mK\,km\,s$^{-1}$)}     & \colhead{(mK)}    & \colhead{(km\,s$^{-1}$)}    & \colhead{(km\,s$^{-1}$)}  & \colhead{(\arcsec)}       
}
\startdata
115271.2    & CO           & $J=1 \rightarrow 0$                               & 10.3 $\pm$ 0.6    & -0.44       & 121.2    & 2142.4 $\pm$ 210.9   & 27.5   & -16.2    & 1.4    & 48.2  \\ 
110201.4    & $^{13}$CO    & $J=1 \rightarrow 0$                               & ?                 & ?           & $<$20    & $<$200               & 11.3   & ?        & 2.0    & 50.4  \\ 
220398.7    & $^{13}$CO    & $J=2 \rightarrow 1$                               & 9.2 $\pm$ 0.6     & 0.86        & 24.9     & 328.8 $\pm$ 37.6     & 5.6    & -17.5    & 1.4    & 34.3  \\ 
            &              &                                                   & 21.8 $\pm$ 0.5    & 11.5        & 0.9      & 182.1 $\pm$ 42.6     & 5.6    & -16.5    & 1.4    & 34.3  \\ 
109782.2    & C$^{18}$O    & $J=1 \rightarrow 0$                               & ?                 & ?           & $<$20    & $<$200               & 17.8   & ?        & 2.0    & 50.6  \\
265886.4    & HCN          & $J=3 \rightarrow 2$                               & 9.0 $\pm$ 1.1     & -0.71       & 83.8     & 1150.1 $\pm$ 22.2    & 10.8   & -13.5    & 1.1    & 28.4  \\
            &              &                                                   & 22.2 $\pm$ 1.1    & -0.47       & 36.8     & 1379.3 $\pm$ 22.2    & 10.8   & -16.7    & 1.1    & 28.4  \\  
259011.8    & H$^{13}$CN   & $J=3 \rightarrow 2$                               & 11.8 $\pm$ 0.1    & -0.27       & 22.9     & 493.5 $\pm$ 75.9     & 14.3   & -14.7    & 1.2    & 29.2  \\
\enddata
\tablecomments{
\tablenotetext{a}{Obtained by integrating the main-beam temperature over the velocity coverage from -80--80\,km\,s$^{-1}$.}
\tablenotetext{b}{Obtained by integrating the main-beam temperature over the velocity coverage from -70--10\,km\,s$^{-1}$.}
\tablenotetext{c}{Obtained by integrating the main-beam temperature over the velocity coverage from -40--5\,km\,s$^{-1}$.}
}
\end{deluxetable}

\begin{deluxetable}{lcccccccccc}
\tablecaption{Same as Table~\ref{Table2}, but for IRAS\,Z02229+6208.
\label{Table7}}
\tabletypesize{\footnotesize}
\tablewidth{0pt}
\tablehead{
\colhead{Frequency}    & \colhead{Species}    & \colhead{Transition}     & \colhead{$V_{\rm exp}$}    & \colhead{$H$}    & \colhead{$T_{\rm c}$}    & \colhead{$\int T_{\rm mb}$d$V$}    & \colhead{RMS}     & \colhead{$V_{\rm LSR}$}     & \colhead{D$_{v}$}     & \colhead{HPBW}         \\
\colhead{(MHz)}        &                      & \colhead{Upper-Lower}    & \colhead{(km\,s$^{-1}$)}   &                  & \colhead{(mK)}           & \colhead{(mK\,km\,s$^{-1}$)}     & \colhead{(mK)}    & \colhead{(km\,s$^{-1}$)}    & \colhead{(km\,s$^{-1}$)}  & \colhead{(\arcsec)}    
}
\startdata
115271.2    & CO           & $J=1 \rightarrow 0$                               & 9.7 $\pm$ 0.2     & -0.91       & 179.5   & 2435.1 $\pm$ 186.3   & 42.8   & 23.9        & 0.6    & 48.2  \\ 
            &              &                                                   & 14.4 $\pm$ 0.1    & -0.63       & 117.9   & 2688.0 $\pm$ 357.8   & 42.8   & 23.8        & 0.6    & 48.2  \\ 
110201.4    & $^{13}$CO    & $J=1 \rightarrow 0$                               & 5.2 $\pm$ 1.1     & 4.97        & 7.3     & 202.1 $\pm$ 109.0    & 15.6   & 20.9        & 1.3    & 50.4  \\ 
            &              &                                                   & 15.2 $\pm$ 0.9    & -0.05       & 26.4    & 791.8 $\pm$ 196.9    & 15.6   & 24.3        & 1.3    & 50.4  \\ 
220398.7    & $^{13}$CO    & $J=2 \rightarrow 1$                               & 8.6 $\pm$ 0.3     & -0.09       & 53.2    & 885.0 $\pm$ 67.0     & 4.2    & 23.7        & 1.4    & 34.3  \\ 
            &              &                                                   & 17.8 $\pm$ 0.7    & -0.21       & 13.7    & 454.0 $\pm$ 105.6    & 4.2    & 22.5        & 1.4    & 34.3  \\ 
109782.2    & C$^{18}$O    & $J=1 \rightarrow 0$                               & ?                 & ?           & $<$20   & $<$200               & 17.8   & ?           & 2.0    & 50.6  \\
265886.4    & HCN          & $J=3 \rightarrow 2$                               & 8.2 $\pm$ 0.2     & -0.61       & 244.7   & 3197.8 $\pm$ 148.9   & 13.8   & 25.3        & 1.1    & 28.4  \\
\enddata
\tablecomments{
\tablenotetext{a}{Obtained by integrating the main-beam temperature over the velocity coverage from 5--40\,km\,s$^{-1}$.}
\tablenotetext{a}{Obtained by integrating the main-beam temperature over the velocity coverage from -10--90\,km\,s$^{-1}$.}
\tablenotetext{a}{Obtained by integrating the main-beam temperature over the velocity coverage from -20--50\,km\,s$^{-1}$.}
}
\end{deluxetable}

\clearpage

\begin{footnotesize}
\centering
\begin{longtable}{cccccc}
\caption{Column densities and fractional abundances.}
\label{Table8} \\ 
\hline
\hline
Object          &  Species         & $N$ (cm$^{-2}$)                      &    $f_{\rm X}$                    \\        
\hline
\endfirsthead
\hline
\hline
Object          &  Species         & $N$ (cm$^{-2}$)                      &    $f_{\rm X}$                    \\       
\hline
\endhead
\hline
\endfoot
IRAS 06530-0213   & CO               & (1.77 $\pm$ 0.16) $\times$ 10$^{17}$   & (5.61 $\pm$ 0.52) $\times$ 10$^{-4}$    \\          
                  & $^{13}$CO        & (7.31 $\pm$ 1.88) $\times$ 10$^{15}$   & (1.02 $\pm$ 0.26) $\times$ 10$^{-5}$    \\          
                  & H$^{13}$CN       & 2.56: $\times$ 10$^{12}$               & 1.59: $\times$ 10$^{-9}$                \\          
                  & HCN              & (3.26 $\pm$ 0.08) $\times$ 10$^{13}$   & (9.11 $\pm$ 0.23) $\times$ 10$^{-8}$    \\          
                  & HNC              & (4.20 $\pm$ 0.51) $\times$ 10$^{12}$   & (8.44 $\pm$ 1.02) $\times$ 10$^{-9}$    \\          
                  & CN               & 8.95: $\times$ 10$^{13}$               & 4.30: $\times$ 10$^{-8}$                \\          
                  & SO               & $<$1.66 $\times$ 10$^{13}$             & $<$4.83 $\times$ 10$^{-8}$              \\          
                  & SiO              & $<$1.94 $\times$ 10$^{12}$             & $<$5.68 $\times$ 10$^{-9}$              \\
\hline
\hline
\end{longtable}
\end{footnotesize}

\clearpage

\begin{deluxetable}{lccccccccc}
\tablecaption{Average fractional abundances.
\label{Table9}}
\tablewidth{0pt}
\tablehead{
\colhead{Species}     & \colhead{21\,$\mu$m objects$^{a}$} & RSD$^{b}$     & \colhead{CRL\,2688$^{c}$}           & \colhead{IRC+10216$^{d}$}      
}
\startdata
CO                         & (5.38 $\pm$ 2.61) $\times$ 10$^{-4}$  &  49$\%$      &  ...                           &  1.00$^{e}$ $\times$ 10$^{-3}$  \\
$^{13}$CO                  & (1.47 $\pm$ 0.36) $\times$ 10$^{-5}$  &  24$\%$      &  ...                           &  2.86$^{f}$ $\times$ 10$^{-5}$  \\
HCN                        & (4.90 $\pm$ 2.84) $\times$ 10$^{-8}$  &  58$\%$      &  1.54 $\times$ 10$^{-7}$       &  1.40 $\times$ 10$^{-5}$  \\
H$^{13}$CN                 & (4.59 $\pm$ 3.05) $\times$ 10$^{-9}$  &  66$\%$      &  5.34 $\times$ 10$^{-7}$       &  3.10 $\times$ 10$^{-7}$  \\
HNC                        & (4.97 $\pm$ 3.14) $\times$ 10$^{-9}$  &  63$\%$      &  3.01 $\times$ 10$^{-8}$       &  7.20 $\times$ 10$^{-8}$  \\
CN                         & (3.56 $\pm$ 2.54) $\times$ 10$^{-8}$  &  71$\%$      &  8.67 $\times$ 10$^{-6}$       &  2.18 $\times$ 10$^{-6}$  \\
SiO                        & 2.79: $\times$ 10$^{-9}$              &  ...         &  9.07 $\times$ 10$^{-9}$       &  1.30 $\times$ 10$^{-7}$  \\
SiS                        & (1.31 $\pm$ 0.79) $\times$ 10$^{-8}$  &  61$\%$      &  1.18 $\times$ 10$^{-7}$       &  1.05 $\times$ 10$^{-6}$  \\
\enddata
\tablecomments{
\tablenotetext{a}{The average values of fractional abundances of each species with a standard deviation.}
\tablenotetext{b}{The relative standard deviation (RSD) is the ratio between the standard deviation and the average value.}
\tablenotetext{c}{Taken from \cite{Qiu22}.}
\tablenotetext{d}{Taken from \cite{Woods03}.}
\tablenotetext{e}{Taken from \cite{Agundez09}.}
\tablenotetext{f}{Obtained by scaling the $f$(CO) value to the $^{12}$C/$^{13}$C ratio of 35 from \cite{Agundez12}.}
\tablenotetext{g}{In $M_{\odot}$\,yr$^{-1}$\,km\,s$^{-1}$.}
}
\end{deluxetable}

\begin{deluxetable}{lcccc}
\tablecaption{Lower limits of $^{12}$C/$^{13}$C and $^{16}$O/$^{18}$O isotopic abundance ratios. 
\label{Table10}}
\tablewidth{0pt}
\tablehead{
\colhead{Objects}     & \colhead{CO/$^{13}$CO}     & \colhead{CO/$^{13}$CO}     & \colhead{HCN/H$^{13}$CN}   & \colhead{CO/C$^{18}$O}  \\
\colhead{}            & \colhead{$J$=1--0}         & \colhead{$J$=2--1}         & \colhead{$J$=3--2}         & \colhead{$J$=1--0}        
}
\startdata
IRAS\,06530-0213      & $>$1.0$^{a}$               & 15.9 $\pm$ 6.3             & 14.1:                       & $>$5.1   \\
IRAS\,07134+1005      & $>$7.8$^{a}$               & 19.2 $\pm$ 3.5             & $>$11.2$^{c}$               & $>$23.3  \\
IRAS\,19500-1709      & $>$4.9$^{a}$               & 3.6 $\pm$ 0.6              & $>$4.2$^{c}$                & $>$14.7  \\
IRAS\,22272+5435      & 19.1 $\pm$ 4.1             & 9.2 $\pm$ 0.3$^{b}$        & 1.7 $\pm$ 0.1$^{b}$         & $>$29.7  \\
IRAS\,23304+6147      & $>$3.6$^{a}$               & 6.7 $\pm$ 2.3              & 5.1 $\pm$ 0.8               & $>$10.7  \\
IRAS\,Z02229+6208     & 12.0 $\pm$ 6.6$^{d}$       & 5.9 $\pm$ 1.4              & 1.8 $\pm$ 0.1$^{d}$         & $>$25.6  \\
                      & 3.4 $\pm$ 1.0$^{e}$        & ...                        & 2.8 $\pm$ 0.3$^{e}$         & ...      \\
\hline
\hline
\enddata
\tablecomments{
\tablenotetext{a}{The non-detection of the $^{13}$CO $J=1\rightarrow0$ line suggests that its
 integrated intensity is lower than 3$\sigma$.}
\tablenotetext{b}{Taken from \cite{Zhang20b}.}
\tablenotetext{c}{
The non-detection of the H$^{13}$CN $J=3\rightarrow2$ line suggests that its integrated intensity is lower than 3$\sigma$.}
\tablenotetext{d}{Derived from the broad line component.}
\tablenotetext{e}{Derived from the narrow line component.}
}
\end{deluxetable}

\begin{deluxetable}{lcccccccccccccc}
\tablecaption{
LINMIX fitting results between the fraction abundances of gas-phase molecules and the relative
intensities of the 21\,$\mu$m features.
\label{Table11}}
\tabletypesize{\footnotesize}
\tablewidth{0pt}
\tablehead{
              &                           & \colhead{CO}  & \colhead{$^{13}$CO}  & \colhead{HCN}   & \colhead{H$^{13}$CN}  & \colhead{HNC}  & \colhead{CN}    & \colhead{HC$_{3}$N}  & \colhead{HNC/HCN}  & \colhead{CN/HCN}  & \colhead{$^{13}$CO/$F_{\rm IR}$}  & \colhead{$\dot{M}$/$V_{\rm exp}$} 
}
\startdata
 a           & median                     & -2.70   & -5.11   & -6.46   & -10.23    & -8.95   & -6.33     & -8.26   & -2.39   &  0.66    &  13.25   & -5.50   \\
             & mean                       & -2.74   & -5.17   & -6.24   &  -7.68    & -9.16   & -6.33     & -8.29   & -2.55   &  0.46    &  13.46   & -5.49   \\
             & stdev                      &  0.89   &  0.46   &  2.37   &  18.60    &  3.18   &  0.70     &  2.72   &  2.06   &  2.69    &  1.20    &  0.90   \\
 b           & median                     &  0.39   & -0.18   &  0.65   &  -1.14    & -0.34   &  0.78     & -0.29   & -0.89   &  0.40    & -0.27    & -0.05   \\
             & mean                       &  0.38   & -0.22   &  0.74   &   0.88    & -0.45   &  0.78     & -0.25   & -1.00   &  0.24    & -0.13    & -0.04   \\
             & stdev                      &  0.55   &  0.29   &  1.48   &  13.27    &  1.95   &  0.45     &  1.77   &  1.26   &  1.68    &  0.76    &  0.56   \\
$\alpha$     & median                     &  0.67   & -0.84   &  0.53   &  -0.45    & -0.28   &  0.95     & -0.46   & -0.78   &  0.41    & -0.80    & -0.13   \\
             & mean                       &  0.49   & -0.65   &  0.38   &  -0.25    & -0.18   &  0.82     & -0.26   & -0.59   &  0.26    & -0.56    & -0.07   \\
\enddata
\tablecomments{ 
$\alpha$ is the linear correlation coefficient.
}
\end{deluxetable}


\end{CJK*}

\end{document}